%% file: main.tex
\documentclass{elsarticle}

\input{layout/theorems}

\usepackage{natbib}

\usepackage[utf8]{inputenc}
\usepackage{graphicx}
\usepackage{mathtools}
\usepackage{amsmath}
\usepackage{amsfonts}
\usepackage{mdframed}
\usepackage{amssymb}
\usepackage{ifsym}
\usepackage[british]{babel}
\usepackage{hyperref}
\usepackage{changepage}
\usepackage{tabularx}
\usepackage{paralist}
\usepackage{theoremref}
\usepackage{bbm}
\usepackage{subcaption}

\usepackage{todonotes}

\usepackage{floatrow}

\usepackage[T1]{fontenc}
\usepackage{lmodern}

\usepackage{pgfplots}
\usepackage{pbox}
\usepackage{epstopdf}

\usepackage[linesnumbered,ruled,vlined]{algorithm2e}

\usepackage{microtype}

\input{other/definitions.tex}

\let\emptyset\varnothing

\title{Solving Zero-Sum One-Sided Partially Observable Stochastic Games}
\author{Karel Hor\'ak$^1$}
\author{Branislav Bo\v{s}ansk\'y$^1$}
\author{Vojt\v{e}ch Kova\v{r}\'ik$^1$}
\author{Christopher Kiekintveld$^2$}

\address{$^1$ Artificial Intelligence Center, \\
Department of Computer Science, \\
Faculty of Electrical Engineering, \\
Czech Technical University in Prague \\
\texttt{\{karel.horak,branislav.bosasnsky,kovarvo1\}@fel.cvut.cz}\\~\\
$^2$ Computer Science Department,\\
University of Texas at El Paso,\\
\texttt{cdkiekintveld@utep.edu}
}

\begin{document}

\begin{abstract}
    \input{other/abstract.tex}

\end{abstract}


\maketitle

\input{chapters/introduction.tex}
\input{chapters/pomdp}
\input{chapters/onesided.tex}
\input{chapters/conclusion.tex}

\bibliographystyle{abbrvnat}
\bibliography{main.bib}

\appendix
\input{chapters/appendix}

\end{document}

%% file: layout/theorems.tex
\usepackage{amsthm}

\usepackage{thm-restate}

\theoremstyle{plain}
\newtheorem{theorem}{\protect\theoremname}
\newtheorem{lemma}{\protect\lemmaname}[section]
\newtheorem{proposition}[lemma]{\protect\propositionname}

\newtheorem{corollary}[lemma]{\protect\corollaryname}
\theoremstyle{definition}
\newtheorem{definition}[lemma]{\protect\definitionname}
\newtheorem{example}[lemma]{\protect\examplename}
\theoremstyle{remark}
\newtheorem*{remark}{\protect\remarkname}

\providecommand{\corollaryname}{Corollary}
\providecommand{\claimname}{Claim}
\providecommand{\definitionname}{Definition}
\providecommand{\lemmaname}{Lemma}
\providecommand{\notationname}{Notation}
\providecommand{\remarkname}{Remark}
\providecommand{\problemname}{Problem}
\providecommand{\propositionname}{Proposition}
\providecommand{\examplename}{Example}
\providecommand{\theoremname}{Theorem}

%% file: other/definitions.tex
\newcommand{\val}[0]{\mathrm{val}}
\newcommand{\E}[0]{\mathbb{E}}
\newcommand{\R}[0]{\mathbb{R}}
\newcommand{\Disc}[0]{\mathsf{Disc}}

\newcommand\numberthis{\addtocounter{equation}{1}\tag{\theequation}}

\DeclareMathOperator*{\argmax}{arg\,max}
\DeclareMathOperator*{\argmin}{arg\,min}

\renewcommand{\Pr}{\mathbb{P}}

\newcommand{\Supp}[0]{\mathsf{Supp}}

\newcommand{\linDS}[0]{\mathsf{lin}_{\Delta(S)}}

\newcommand{\uv}[0]{V_{\mathrm{LB}}^\Gamma}
\newcommand{\ov}[0]{V_{\mathrm{UB}}^\Upsilon}

\newcommand{\excess}[0]{\mathrm{excess}}

\makeatletter
\def\ps@pprintTitle{%
\let\@oddhead\@empty
\let\@evenhead\@empty
\def\@oddfoot{\reset@font\hfil\thepage\hfil}
\let\@evenfoot\@oddfoot
}
\makeatother

%% file: other/abstract.tex
Many security and other real-world situations are dynamic in nature and can be modelled as strictly competitive (or zero-sum) dynamic games.
In these domains, agents perform actions to affect the environment and receive observations -- possibly imperfect -- about the situation and the effects of the opponent's actions.
Moreover, there is no limitation on the total number of actions an agent can perform --- that is, there is no fixed horizon.
These settings can be modelled as partially observable stochastic games (POSGs).
However, solving \textit{general} POSGs is computationally intractable, so we focus on a broad subclass of POSGs called \emph{one-sided POSGs}.
In these games, only one agent has imperfect information while their opponent has full knowledge of the current situation.
We provide a full picture for solving one-sided POSGs: we 
(1) give a theoretical analysis of one-sided POSGs and their value functions, 
(2) show that a variant of a value-iteration algorithm converges in this setting,
(3) adapt the heuristic search value-iteration algorithm for solving one-sided POSGs,
(4) describe how to use approximate value functions to derive strategies in the game, and
(5) demonstrate that our algorithm can solve one-sided POSGs of non-trivial sizes and analyze the scalability of our algorithm in three different domains: pursuit-evasion, patrolling, and search games.

%% file: chapters/introduction.tex
\section{Introduction}

Non-cooperative game theory models the interaction of multiple agents in a joint environment. 
Rational agents perform actions in the environment to achieve their own, often conflicting goals.
The interaction of agents is typically very complex in real-world dynamic scenarios ---
the agents can perform sequences of multiple actions while only having partial information about the actions of others and the events in the environment.

Finding out (approximate) optimal strategies for agents in dynamic environment with imperfect information is a long-standing problem in Artificial Intelligence. Its applications range from recreational games, such as poker~\cite{moravcik2017-deepstack,brown2018-libratus}, to uses in security such as patrolling~\cite{Basilico2009,vorobeychik2014-icaps,basilico2016} and pursuit-evasion games~\cite{isler2005-peg,isler2008-os-peg,amigoni2012-peg}.

For tackling this problem, game theory can provide appropriate mathematical models and algorithms for computing (approximate) optimal strategies according to some game-theoretic solution concept. 
Among all existing game-theoretic models suitable for modelling dynamic interaction with imperfect information, \emph{partially observable stochastic games (POSGs)} are one of the most general ones.
POSGs model situations where all players have only partial information about the state of the environment, agents perform actions and receive observations, and the length of the interaction among agents is not a priori bounded.
As such, the expressive possibilities of POSGs are broad. In particular, they can model all considered security scenarios as well as recreational games. 

Despite having high expressive power, POSGs have limited applications due to the complexity of computing (approximate) optimal strategies.
There are two main reasons for this.
First, the imperfect information provides challenges for sequential decision-making even in the single-agent case -- partially observable Markov decision processes (POMDPs).
Theoretical results show that various exact and approximate problems in POMDPs are undecidable~\cite{madani1999-undecidability}.
Focused research effort has yielded several approximate algorithms with convergence guarantees~\cite{smith2004-hsvi,kurniawati2008sarsop} scalable even to large POMDPs~\cite{silver2010monte}.
The main step when solving a POMDP is to reason about \emph{belief states} -- probability distributions over possible states.
Note that an agent can easily deduce a belief state in a POMDP since the environment changes only as a result of the agent's actions or because of the environment's stochasticity (which is known).
In POSGs, however, the presence of another agent(s) changing the environment generates another level of complexity.
Suppose all players have partial information about the environment. In that case, each player needs to reason not only about their belief over environment states, but also about opponents' beliefs, their beliefs over beliefs, and so on. 
This issue is called the problem with \emph{nested beliefs}~\cite{macdermed2013-magii} and cannot be avoided in general unless we pose additional assumptions on the game model.
This is primarily because, in general POSGs, the choice of the optimal action (strategy) of a player depends on these nested beliefs.
To avoid this issue, we will focus on a subclass of POSGs that does not suffer from the problem of nested beliefs while still being expressive enough to contain many existing real-world games and scenarios. 

One such sub-class of POSGs are two-player concurrent-move games where one player is assumed to have full knowledge about the environment and only one player has partial information.
In this case, the player with partial information (player 1 from now on) does not have to reconstruct the belief of the opponent (player 2) since player 2 always has full information about the true state of the environment. 
Similarly, player 2 can always reconstruct the belief of player 1 by using the full information about the environment, which includes information about the action-history of player 1.
The game is played over stages where both players independently choose their next action (i.e., albeit player 2 has full knowledge about the current history and state, he does not know the action player 1 is about to play in the current stage).
The state of the game with the joint action of the players determines the next state and the next observation generated for player 1.
We term this class of games as \emph{one-sided POSG}.
While this class of games has appeared before in the literature (e.g., in~\cite{sorin2003-stochastic-incomplete} as \emph{Level-1 stochastic games}, or in~\cite{chatterjee2005semiperfect} as \emph{semiperfect-information} stochastic games\footnote{In this work, however, the authors assumed that the game is turn-taking. In contrast, we consider a more general case where at each timestep, both players choose simultaneously next action to be played.}) we are the first to focus on designing a practical algorithm for computing (approximately) optimal strategies. 

Despite the seemingly-strong assumption on the perfect information for player 2, the studied class of one-sided POSG has broad application possibilities, especially in security.
In particular, this model subsumes patrolling games~\cite{Basilico2009,vorobeychik2014-icaps,basilico2016} or pursuit-evasion games~\cite{isler2005-peg,isler2008-os-peg,amigoni2012-peg}.
In many security-related problems, the defender is protecting an area (or a computer network) against the attacker that wants to attack it (e.g., by intruding into the area or infiltrating the network). 
The defender does not have full information about the environment since he does not know which actions the attacker performed (e.g., which hosts in the computer network have been compromised by the attacker).
At the same time, it is difficult for the defender to exactly know what information the attacker has since the attacker can infiltrate the system or use insider information, and can thus have substantial knowledge about the environment. 
Hence, as the worst-case assumption, the defender can assume that the attacker has full knowledge about the environment.
From this perspective, one-sided POSG can be used to compute robust defense strategies.
We restrict to the strictly competitive (or zero-sum) setting. In this case, the defender has guaranteed expected outcome when using such robust strategies even against attackers with less information.
Finally, we use the standard assumption that payoffs are computed as discounted sums of immediate rewards. However, our approach could be generalized to the non-discounted version to some extent (by proceeding similarly to \cite{horak2018-ijcai}). 

Our main contribution is the description of the first practical algorithm for computing (approximate) optimal solution for two-player zero-sum one-sided POSG with discounted rewards.\footnote{Parts of this work appeared in conference publications~\cite{horak2017-aaai}. This submission is significantly extended from the published works by (1) containing all the proofs and all the technical details regarding the algorithm, (2) full description of the procedure for extracting strategies computed by the algorithm, and (3) new experiments with improved implementation of the algorithm. Finally, we acknowledge that a modification of presented algorithm has been provided in \cite{horak2019-ijcai,horak2019-cose} where a compact representation of belief space was proposed for a specific cybersecurity domain and demonstrate that proposed algorithm can scale even beyond experiments however at the cost of losing theoretical guarantees.}
The contribution is threefold: (1) the theoretical contribution proving that our proposed algorithm has guarantees for approximating the value of any one-sided POSG, (2) showing how to extract strategies from our algorithm and use them to play the game, (3) implementation of the algorithm and experimental evaluation on a set of games.
The theoretical work is a direct extension of the theory behind the single-player case (i.e., POMDPs).
In POMDPs, an optimal strategy in every step depends on the player's belief over environment states and on the outcomes achievable in each state. In other words, we have a \emph{value function} which takes a belief $b$ and returns the optimal expected value that can be achieved under $b$ (by following an optimal strategy in both the current decision point and those encountered afterwards). 
\begin{figure}[ptb]
  \centering
  \includegraphics[width=0.98\linewidth]{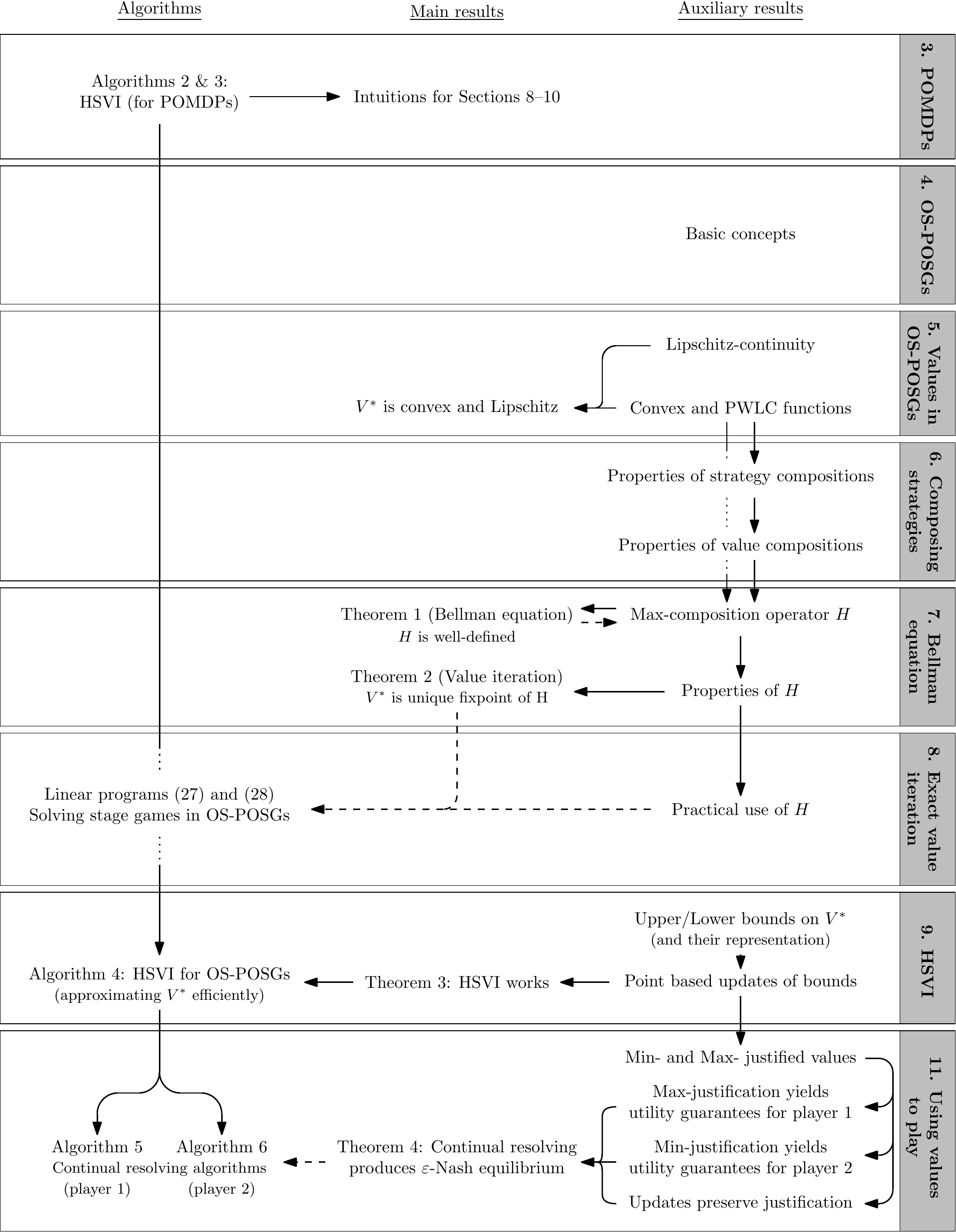}
  \caption{Outline of the theoretical results presented in the paper.}
  \label{fig:map}
\end{figure}

Figure~\ref{fig:map} visualizes the outline and key results provided in each section of the paper. 
After reviewing related work (Section~\ref{sec:rw}) we state relevant technical background for POMDPs~(Section~\ref{sec:pomdps}). 
We then formally define one-sided POSG (Section~\ref{sec:osposg:model}) and restate some known results \cite{sorin2003-stochastic-incomplete} regarding the characteristics of the value function (convexity) and show that the value function can be computed using a recursive formula (Section~\ref{sec:osposg:value}).
We then observe that each strategy can be decomposed into the distribution that determines the very next action and the strategy for the remainder of the game and that this structure is mirrored on the level of value functions (Section~\ref{sec:osposg:composing}).
With these tools, we derive a Bellman equation for one-sided POSG a prove that the iterative application of the corresponding operator $H$ is guaranteed to converge to the optimal value function $V^*$ (Section~\ref{sec:osposg:bellman}).
To get a baseline method of computing $V^*$, we show that the operator $H$ can be computed using a linear program (Section~\ref{sec:osposg:vi}).
To get a method with better scaling properties, we design novel approximate algorithms that aim at approximating $V^*$ (Section~\ref{sec:osposg:hsvi}).
Namely, we follow the heuristic search value iteration algorithm (HSVI)~\cite{smith2004-hsvi,smith2005-hsvi} that uses two functions to approximate the value function, an upper bound function and a lower bound function.
By decreasing the gap between these approximations, the algorithm approximates the optimal expected value for relevant belief points.
We show that a similar approach can also work in one-sided POSG and that, while the overall idea remains, most of the technical parts of the algorithm have to be adapted for one-sided POSG. 
We identify and address these technical challenges in order to formally prove that our HSVI algorithm for one-sided POSG converges to optimal strategies. 
As defined, the HSVI algorithm primarily approximates optimal value for a given game. 
To extract strategies that reach computed values in expectation, we provide an additional online algorithm (based on ideas from online game-playing algorithms with imperfect information but finite horizon~\cite{moravcik2017-deepstack}) that generates actions from (approximate) optimal strategies according to the computed approximated value functions (Section~\ref{sec:osposg:playing}). 
Finally, we experimentally evaluate the proposed algorithm on a set of different games, show scalability for these games, and provide deep insights into the performance for each specific part of the algorithm~(Section~\ref{sec:osposg:experiments}).
We demonstrate that our implementation of the algorithm is capable of solving non-trivial games with as much as 4\,500 states and 120\,000 transitions.

\section{Related Work}\label{sec:rw}
General, domain-independent algorithms for solving\footnote{Or even approximating an optimal solution to a given error.} (subclasses of) partially observable stochastic games with infinite horizon are not commonly studied. 
As argued in the introduction, the problem of nested beliefs is one of the reasons.
One way of tackling this issue is by using history-dependent strategies.
One of the few such approaches is the bottom-up dynamic programming for constructing relevant finite-horizon policy trees for individual players while pruning-out dominated strategies~\cite{hansen2004dynamic,kumar2009dynamic}.
However, while the history-dependent policies can cope with the necessity of considering the nested beliefs, the number of the strategies is doubly exponential in the horizon of the game (i.e., the number of turns in the game), which greatly limits the scalability and applicability of the algorithm.

We take another approach and restrict to subclasses of POSGs, where the problem of nested beliefs does not appear. 
Besides the works focused directly on one-sided POSGs, there are other works that consider specific subclasses of POSGs.
For example, Ghosh et al.~\citeyear{Ghosh2004} study zero-sum POSGs with public actions and observations. 
The authors show that the game has a well-defined value and present an algorithm that exploits the transformation of such a model into a game with complete information.
In one-sided POSG, however, the actions are not publicly observable since the imperfectly-informed player lacks the information about their opponent's action.
Compared to existing works studying one-sided POSG~\cite{sorin2003-stochastic-incomplete,chatterjee2005semiperfect}, our work is the first to provide a practical algorithm that can be directly used to solve games of non-trivial sizes. 

Our algorithm focuses on the \emph{offline problem} of (approximately) solving a given one-sided POSG. 
However, a part of our contribution is the extraction of the strategy that reaches the computed value.
On the other hand, \emph{online algorithms} focus on computing strategies that will be used while playing the game.
For a long time, no online algorithms for dynamic imperfect-information games provided guarantees on the (near-)optimality of the resulting strategies.
While several new algorithms with theoretical guarantees emerged~\cite{lisy2015online,moravcik2017-deepstack,sustr2019monte} in recent years, they only considered limited-horizon games and produced history-dependent strategies.
Using such online algorithms for POSGs is thus only possible with very limited lookahead or when using a heuristic evaluation function.
Our approach is fully domain-independent and avoids considering complete histories and the use of evaluation functions while nevertheless being able to consider strategies with horizon of 100 turns or more.
Finally, note that the recent work \cite{sustr2020sound} has shown that online algorithms which seem to be consistent with some Nash equilibrium strategy might fail to be ``sound'' (i.e., there will be a way to exploit them).
Fortunately, our algorithm is provably $\epsilon$-sound in this sense, since (the proof of) \thref{thm:equilibrium} shows that it is always guaranteed to get at least the equilibrium value minus $\epsilon$.

%% file: chapters/pomdp.tex
\section{Partially Observable MDPs}
\label{sec:pomdps}
Partially observable Markov decision processes (POMDPs)~\citep{astrom1965-pomdp,sondik1978-pomdp,pineau2003-pbvi,smith2004-hsvi,smith2005-hsvi,spaan2005-perseus,bonet2009-rtdpbel,somani2013-despot} are a standard tool for single-agent decision making in stochastic environment under uncertainty about the states.
From the perspective of partially observable stochastic games, POMDPs can be seen as a variant of POSG that is only played by a single player.

\begin{definition}[Partially observable Markov decision process]
\thlabel{def:pomdp}
    A \emph{partially observable Markov decision process} is a tuple $(S,A,O,T,R)$ where
    \begin{compactitem}
        \item $S$ is a finite set of states,
        \item $A$ is a finite set of actions the agent can use,
        \item $O$ is a finite set of observations the agent can observe,
        \item $T(o, s' \mid s, a)$ is a probability to transition to $s'$ while generating observation $o$ when the current state is $s$ and agent uses action $a$,
        \item $R(s, a)$ is the immediate reward of the agent when using action $a$ in state $s$.
    \end{compactitem}
\end{definition}

In POMDPs, the agent starts with a known belief $b^{\mathrm{init}} \in \Delta(S)$ that characterizes the probability $b^{\mathrm{init}}(s)$ that $s$ is the initial state.
The play proceeds similarly as in POSGs, except that there is only one decision-maker involved:
The initial state $s^{(1)}$ is sampled from the distribution $b^{\mathrm{init}}$.
Then, in every stage $t$, the agent decides about the current action $a^{(t)}$ and receives reward $R(s^{(t)}, a^{(t)})$ based on the current state of the environment $s^{(t)}$.
With probability $T(o^{(t)}, s^{(t+1)} \mid s^{(t)}, a^{(t)})$ the system transitions to $s^{(t+1)}$ and the agent receives observation $o^{(t)}$.
The decision process is then repeated.
Although many objectives have been studied in POMDPs, in this section we discuss only discounted POMDPs with infinite-horizon, i.e., the objective is to optimize $\sum_{t=1}^\infty \gamma^{t-1} r_t$ for a discount factor $\gamma \in (0, 1)$.

A strategy $\sigma: (A_1 O)^* \rightarrow A_1$ in POMDPs is traditionally called a \emph{policy} and assigns a deterministic action to each observed history $\omega \in (A_1 O)^*$ of the agent.\footnote{As usual, we take $X^*$ to denote the set of all finite sequences over $X$. For a set $Y$ of sequences, $YZ$ denotes the set of sequences obtained by concatenating a single element of $Z$ to some sequence from $Y$. (Combining this notation yields, e.g., $axbyc \in (AX)^*A$ for $a,b,c \in A$ and $x,y \in X$.)}
Since the agent is the only decision-maker within the environment, and the probabilistic characterization of the environment is known, the player is able to infer his belief $\Pr_{b^{\mathrm{init}}}[s^{(t+1)} \mid (a^{(i)} o^{(i)})_{i=1}^t ]$ (i.e., how likely it is to be in a particular state after a sequence of actions and observations $(a^{(i)} o^{(i)})_{i=1}^t$ has been used and observed).
This belief can be defined recursively
\begin{equation}
    \tau(b,a,o)(s') = \eta \sum_{s \in S} b(s) \cdot T(o,s' \mid s,a)
\end{equation}
where $\eta$ is a normalizing term, and $\tau(b,a,o) \in \Delta(S)$ is the updated belief of the agent when his current belief was $b$ and he played and observed $(a,o)$.
\cite{sondik1971-thesis} has shown that the belief of the agent is a sufficient statistic, and POMDPs can therefore be translated into \emph{belief-space MDP}.
In theory, standard methods for solving MDPs can be applied, and POMDPs can be solved, e.g., by iterating
\begin{equation}
    V^{t+1}(b) = [HV^t](b) = \max_{a \in A} \left[ \sum_{s \in S} b(s) \cdot R(s,a) + \gamma \sum_{o \in O} \Pr_{b}[o \mid a] \cdot V^t(\tau(b,a,o)) \right] \ \text{.} \label{eq:pomdp:bellman}
\end{equation}
Since $H$ is a contraction, the repeated application of Equation~\eqref{eq:pomdp:bellman} converges to a unique convex value function $V^*: \Delta(S) \rightarrow \R$ of the POMDP.
However, since the number of beliefs is infinite, it is impossible to apply this formula to approximate $V^*$ directly.

\paragraph{Exact value iteration}
The value iteration can be, however, rewritten in terms of operations with so-called $\alpha$-vectors~\citep{sondik1978-pomdp}.
An $\alpha$-vector can be seen as a linear function $\alpha: \Delta(S) \rightarrow \R$ characterized by its values $\alpha(s)$ in the vertices $s \in S$ of the belief simplex $\Delta(S)$.
We thus have $\alpha(b) = \sum_{s \in S} b(s) \cdot \alpha(s)$.

Assume that $V^t$ is a piecewise-linear and convex function where $V^t(b) = \max_{\alpha \in \Gamma^t} \alpha(b)$ for a finite set of $\alpha$-vectors $\Gamma^t$.
We can then form a new (finite) set $\Gamma^{t+1}$ of $\alpha$-vectors to represent $V^{t+1}$ from Equation~\eqref{eq:pomdp:bellman} by considering all possible combinations of $\alpha$-vectors from the set $\Gamma^t$:
\begin{flalign}
    \Gamma^{t+1} = \Big\lbrace \ \alpha : \Delta(S) \to \R \ \Big| \ \alpha(s) = R(s,a) + \gamma \!\!\!\!\!\!\!\! \sum_{(o,s') \in O \times S} \!\!\!\!\!\!\!\! T(o, s' \,|\, s, a) \alpha^o(s') \hspace{-32em} \nonumber && \\
    && \textnormal{for some } a \in A \textnormal{ and } \alpha^o \in \Gamma^t, \ o \in O \ \Big\rbrace \ \text{.}
\end{flalign}
As $|\Gamma^{t+1}| = |A| \cdot |\Gamma^t|^{|O|}$, this exact approach suffers from poor scalability.
Several techniques have been proposed to reduce the size of sets $\Gamma^t$~\citep{littman1996-thesis,zhang2001speeding}, however, this still does not translate to an efficient algorithm.

\bigskip

In the remainder of this section, we present two scalable algorithms for solving POMDPs that are relevant to this thesis.
First, we present RTDP-Bel that uses discretized value function and applies Equation~\eqref{eq:pomdp:bellman} directly.
Second, we present heuristic search value iteration (HSVI)~\citep{smith2004-hsvi,smith2005-hsvi} that inspires our methods for solving POSGs.

\paragraph{RTDP-Bel}
The RTDP-Bel algorithm~\citep{bonet1998-rtdpbel} is based on RTDP~\citep{BBS95} and has been originally framed in the context of Goal-POMDPs.
Goal-POMDPs do not discount rewards (i.e., they set $\gamma=1$ in Equation~\eqref{eq:pomdp:bellman}). However, the agent is incentivized to reach the goal state $g$ as his reward for every transition before reaching the goal is negative (i.e., it represents the cost).
The RTDP-Bel also applies to discounted POMDPs as discounting can be modelled within the Goal-POMDP framework as a fixed probability $1-\gamma$ of reaching the goal state during every transition~\citep{bonet2009-rtdpbel}.

RTDP-Bel adapts RTDP to partially observable domains by using a grid-based approximation of $V^*$ and using a hash-table to store the values, where $V^*(b) \sim \widehat{V}(\lfloor K \cdot b \rfloor)$ for some fixed parameter $K \in \mathbb{N}$.
This approximation, however, loses the theoretical properties of RTDP.
The algorithm need not converge as the values of the discretized value function may oscillate.
Moreover, there is no guarantee that the values stored in the hash-table will provide a bound on the values of $V^*$~\cite[p.~3, last paragraph of Section~3]{bonet2009-rtdpbel}.
Despite the lack of theoretical properties, RTDP-Bel has been shown to perform well in practice.
The RTDP-Bel algorithm performs a sequence of trials (see Algorithm~\ref{alg:rtdp-bel}) that updates the discretized value function $\widehat{V}$.
\begin{algorithm}
\caption{A single trial of the RTDP-Bel algorithm.}\label{alg:rtdp-bel}
\DontPrintSemicolon
\small
$b \gets b^{\mathrm{init}}$; ~$s \sim b$ \;
\While{$b(g) < 1$}{
  $Q(b,a) \gets \sum_{s \in S} b(s) R(s,a) + \sum_{o \in O} \Pr_b[o \mid a] \cdot \widehat{V}(\lfloor K \cdot \tau(b,a,o) \rfloor)$ \;
  $a^* \gets \argmax_{a \in A} Q(b,a)$ \;
  $\widehat{V}(\lfloor K \cdot b \rfloor) \gets Q(b,a^*)$ \;
  $(o,s') \sim T(o,s' \mid s, a^*)$;\ \ \ 
  $b \gets \tau(b, a^*, o)$;\ \ \ 
  $s \gets s'$ \;
}
\end{algorithm}

\begin{figure}
  \centering
  \begin{subfigure}{0.32\linewidth}
    \centering\includegraphics[width=0.92\linewidth]{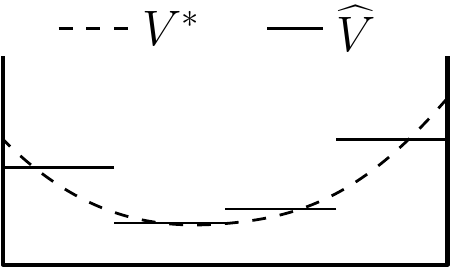}
    \caption{RTDP-Bel}
    \label{fig:v-grid}
  \end{subfigure}
  \begin{subfigure}{0.32\linewidth}
    \centering\includegraphics[width=0.92\linewidth]{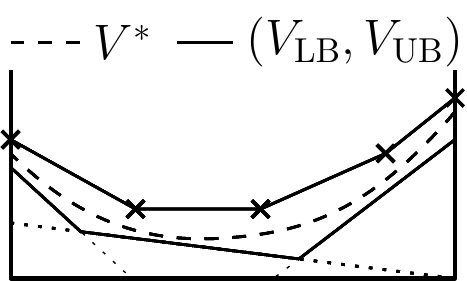}
    \caption{HSVI (PWLC)}
    \label{fig:v-pwlc}
  \end{subfigure}
  \begin{subfigure}{0.32\linewidth}
    \centering\includegraphics[width=0.92\linewidth]{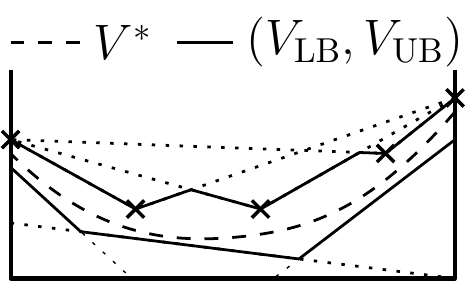}
    \caption{HSVI2 (sawtooth)}
    \label{fig:v-sawtooth}
  \end{subfigure}
  \caption{
    Comparison of value function approximation schemes
  }
\end{figure}

\paragraph{Heuristic search value iteration (HSVI)}
Heuristic search value iteration~\citep{smith2004-hsvi,smith2005-hsvi} is a representative of a class of point-based methods for solving POMDPs.
Unlike RTDP-Bel, it approximates $V^*$ using piecewise-linear functions.
We illustrate the difference between a grid-based approximation used in RTDP-Bel and a piecewise-linear approximation in Figures~\ref{fig:v-grid} and~\ref{fig:v-pwlc}.
Observe that unlike the grid-based approximation, a piecewise-linear approximation can yield a close approximation of $V^*$ even in regions with a rapid change of value.

In the original version of the \emph{heuristic-search value iteration} algorithm (HSVI)~\citep{smith2004-hsvi}, the algorithm keeps two piecewise-linear and convex (PWLC) functions $\uv$ and $\ov$ to approximate $V^*$ (see Figure~\ref{fig:v-pwlc}) and refines them over time.
The lower bound on the value is represented in the vector-set representation using a finite set of $\alpha$-vectors $\Gamma$, while the upper bound is formed as a lower convex hull of a set of points $\Upsilon=\lbrace (b_i,y_i) \,|\, i=1,\ldots,m \rbrace$ where $b_i \in \Delta(S)$ and $y_i \in \R$.
We then have
\begin{subequations}
\begin{align}
  \uv(b) &= \max_{\alpha \in \Gamma} \sum_{s \in S} b(s) \cdot \alpha(s) \\
  \ov(b) &= \min \lbrace \textstyle\sum_{i=1}^m \lambda_i y_i \mid \lambda \in \mathbb{R}_{\geq 0}^m: \textstyle\sum_{i=1}^m \lambda_i b_i = b \rbrace \ \text{.} \label{eq:hsvi-ub}
\end{align}
\end{subequations}

Computing $\ov(b)$ according to Equation~\eqref{eq:hsvi-ub} requires solving a linear program.
In the second version of the algorithm (HSVI2,~\citep{smith2005-hsvi}), the PWLC representation of upper bound has been replaced by a sawtooth-shaped approximation~\citep{hauskrecht2000-value-functions} (see Figure~\ref{fig:v-sawtooth}).
While the sawtooth approximation is less tight with the same set of points, the computation of $\ov(b)$ does not rely on the use of linear programming and can be done in linear time in the size of $\Upsilon$.

HSVI2 initializes the value function $\uv$ by considering policies `always play the action $a$' and construct one $\alpha$-vector for each action $a \in A$ corresponding to the expected cost for playing such policy.
For the initialization of the upper bound, the fast-informed bound is used~\citep{hauskrecht2000-value-functions}.

The refinement of $\uv$ and $\ov$ is done by adding new elements to the sets $\Gamma$ and $\Upsilon$.
Since the goal of each update is to improve the approximation quality in the selected belief $b$ as much as possible, we refer to them as \emph{point-based updates} (see Algorithm~\ref{alg:pb-update}).

\begin{algorithm}
  \SetKwFunction{Update}{$\mathtt{update}$}
  \SetKwProg{myproc}{procedure}{}{}
  \DontPrintSemicolon
  \small
  $\alpha^{a,o} \gets \argmax_{\alpha \in \Gamma} \sum_{s' \in S} \tau(b,a,o)(s') \cdot \alpha(s')$ for all $a \in A, o \in O$ \;
  $\alpha^a(s) \gets R(s,a) + \gamma \sum_{o,s'} T(o,s' \mid s, a) \cdot \alpha^{a,o}(s')$ for all $s \in S, a \in A$ \;
  $\Gamma \gets \Gamma \cup \lbrace \argmax_{\alpha^a} \sum_{s \in S} b(s) \cdot \alpha^a(s) \rbrace$ \;
  $\Upsilon \gets \Upsilon \cup \lbrace (b, \max_{a \in A} \left[ \sum_{s \in S} b(s) R(s,a) + \gamma\sum_{o \in O} \Pr_b[o \mid a] \cdot \ov(\tau(b,a,o)) \right])\rbrace$ 
  \caption{Point-based $\mathtt{update(}b\mathtt{)}$ procedure of $(\uv,\ov)$.}
  \label{alg:pb-update}
\end{algorithm}

\begin{algorithm}
  \caption{
    HSVI2 for discounted POMDPs.
    The pseudocode follows the ZMDP implementation and includes $\mathtt{update}$ on line~\ref{alg:hsvi-disc:update1}.
  }
  \label{alg:hsvi-disc}
  \SetKwFunction{Explore}{$\mathtt{explore}$}
  \SetKwProg{myproc}{procedure}{}{}

  \DontPrintSemicolon
  \small
  Initialize $\uv$ and $\ov$\;
  \lWhile{$\ov(b^{\mathrm{init}}) - \uv(b^{\mathrm{init}}) > \varepsilon$}{
    \Explore{$b^{\mathrm{init}},\varepsilon,0$}
  }

  \myproc{\Explore{$b,\varepsilon,t$}}{
    \lIf{$\ov(b) - \uv(b) \leq \varepsilon\gamma^{-t}$}{\Return\label{alg:hsvi-disc:termination}}
    $a^* \gets \argmax_{a \in A} \left[ \sum_{s} b(s) \cdot R(s,a) + \gamma\sum_{o \in O} \Pr_b[o \mid a] \ov(\tau(b,a,o)) \right]$\label{alg:hsvi-disc:action-choice}\;
    $\mathtt{update}(b)$ \label{alg:hsvi-disc:update1}\;
    $o^* \gets \argmax_{o \in O} \Pr_b[o \mid a] \cdot \mathrm{excess}_{t+1}(\tau(b,a^*,o))$\;
    \Explore{$\tau(b,a^*,o^*),\varepsilon,t+1$}\;
    $\mathtt{update}(b)$ \;
  }
\end{algorithm}

Similarly to RTDP-Bel, HSVI2 selects beliefs where the update should be performed based on the simulated play (selecting actions according to $\ov$).
Unlike RTDP-Bel, however, observations are not selected randomly.
Instead, HSVI2 selects an observation with the highest \emph{weighted excess gap}, i.e. the excess approximation error 
\begin{equation}
\mathrm{excess}_{t+1}(\tau(b,a^*,o)) = \ov(\tau(b,a^*,o))-\uv(\tau(b,a^*,o))-\varepsilon\gamma^{-(t+1)} \label{eq:hsvi:excess}
\end{equation}
in $\tau(b,a^*,o)$ weighted by the probability $\Pr_b[o \mid a^*]$.
This heuristic choice attempts to target beliefs where the update will have the most significant impact on $\ov(b^{\mathrm{init}}) - \uv(b^{\mathrm{init}})$.

The HSVI2 algorithm for discounted-sum POMDPs ($\gamma \in (0,1)$) is shown in Algorithm~\ref{alg:hsvi-disc}.
This algorithm provably converges to an $\varepsilon$-approximation of $V^*(b^{\mathrm{init}})$ using values $\uv(b^{\mathrm{init}})$ and $\ov(b^{\mathrm{init}})$, see~\citep{smith2004-hsvi}.

%% file: chapters/onesided.tex
\section{Game Model: One-Sided Partially Observable Stochastic Games (OS-POSGs)}
\label{sec:osposg:model}
We now define the model of one-sided POSGs and describe strategies for this class of games.

\begin{definition}[one-sided POSGs]
    A \emph{one-sided POSG} (or OS-POSG) is a tuple $G=(S,A_1,A_2,O,T,R,\gamma)$ where
    \begin{compactitem}
        \item $S$ is a finite set of of game \emph{states},
        \item $A_1$ and $A_2$ are finite sets of \emph{actions} of player~1 and player~2, respectively,
        \item $O$ is a finite set of \emph{observations}
        \item for every $(s,a_1,a_2) \in S \times A_1 \times A_2$, $T(\cdot \,|\, {s,a_1,a_2}) \in \Delta(O \times S)$ represents probabilistic transition function,
        \item $R: S \times A_1 \times A_2 \rightarrow \R$ is a reward function of player~1,
        \item $\gamma \in (0,1)$ is a discount factor.
    \end{compactitem}
\end{definition}

The game starts by sampling the initial state $s^{(1)} \sim b^{\mathrm{init}}$ from the \emph{initial belief} $b^{\mathrm{init}}$.
Then the game proceeds for an infinite number of \emph{stages} where the players choose their actions simultaneously and receive feedback from the environment.
At the beginning of $i$-th stage, the current state $s^{(i)}$ is revealed to player~2, but not to player~1.
Then player~1 selects action $a_1^{(i)} \in A_1$ and player~2 selects action $a_2^{(i)} \in A_2$.
Based on the current state of the game $s^{(i)}$ and the actions $(a_1^{(i)},a_2^{(i)})$ taken by the players, an unobservable reward $R(s^{(i)},a_1^{(i)},a_2^{(i)})$ is assigned\footnote{Note that we consider a zero-sum setting, hence the reward of player~2 is $-R(s^{(i)},a_1^{(i)},a_2^{(i)})$. We do however consider that player~2 focuses on minimizing the reward of player~1 instead of reasoning about the rewards of player~2 directly.} to player~1, and the game transitions to a state $s^{(i+1)}$ while generating observation $o^{(i)}$ with probability $T(o^{(i)},s^{(i+1)} \,|\, s^{(i)},a_1^{(i)},a_2^{(i)})$.
After committing to action $a_2^{(i)}$, player~2 observes the entire outcome of the current stage, including the action $a_1^{(i)}$ taken by player~1 and the observation $o^{(i)}$.
player~1, on the other hand, knows only his own action $a_1^{(i)}$ and the observation $o^{(i)}$, while the action $a_2^{(i)}$ of player~2 and both the past and new states of the system $s^{(i)}$ and $s^{(i+1)}$ remain unknown to him.

The information asymmetry in the game means that while player~2 can observe entire course of the game $(s^{(i)} a_1^{(i)} a_2^{(i)} o^{(i)})_{i=1}^t s^{(t+1)}  \in (S A_1 A_2 O)^*S$ up to the current decision point at time $t+1$, player~1 only knows his own actions and observations $(a_1^{(i)} o^{(i)}_{i=1})^t \in (A_1 O)^*$.\footnote{Recall that we use the standard notation where $X^* := $ all finite sequences over $X$ (and, if $Y$ is a set of sequences, $YZ$ denotes the set of sequences obtained by appending a single element of $Z$ at the end of some $y\in Y$).}
The players make decisions solely based on this information - formally, this is captured by the following definition:

\begin{definition}[Behavioral strategy]
\thlabel{def:osposg:behavioral}
    Let $G$ be a one-sided POSG.
    Mappings $\sigma_1: (A_1 O)^* \rightarrow \Delta(A_1)$ and $\sigma_2: (S A_1 A_2 O)^* S \rightarrow \Delta(A_2)$ are \emph{behavioral strategies} of imperfectly informed player~1 and perfectly informed player~2, respectively.
    The sets of all behavioral strategies of player~1 and player~2 are denoted $\Sigma_1$ and $\Sigma_2$, respectively.
\end{definition}

\paragraph{Plays in OS-POSGs}
Players use their behavioral strategies $(\sigma_1,\sigma_2)$ to play the game.
A \emph{play} is an infinite word $(s^{(i)} a_1^{(i)} a_2^{(i)} o^{(i)})_{i=1}^\infty$, while finite prefixes of plays $w=(s^{(i)} a_1^{(i)} a_2^{(i)} o^{(i)})_{i=1}^T s^{(T+1)}$ are called \emph{histories} of length~$T$, and plays having $w$ as a prefix are denoted $\mathsf{Cone}(w)$.
Formally, a \emph{cone} of $w$ is a set of all plays extending $w$,
\begin{align}
    \mathsf{Cone}(w) := \hspace{27em} \nonumber \\
    \left\lbrace (s^{(i)} a_1^{(i)} a_2^{(i)} o^{(i)})_{i=1}^\infty \in (S A_1 A_2 O)^* \mid (s^{(i)} a_1^{(i)} a_2^{(i)} o^{(i)})_{i=1}^\infty \text{ extends } w \right\rbrace \text{.}
\end{align}
At a decision point at time $t$, players extend a history $(s^{(i)} a_1^{(i)} a_2^{(i)} o^{(i)})_{i=1}^t s^{(t+1)}$ of length $t$ by sampling actions from their strategies $a_1^{(t+1)} \sim \sigma_1((a_1^{(i)} o^{(i)})_{i=1}^t)$ and $a_2^{(t+1)} \sim \sigma_2((s^{(i)} a_1^{(i)} a_2^{(i)} o^{(i)})_{i=1}^t s^{(t+1)})$.
We consider a discounted-sum objective with discount factor $\gamma \in (0,1)$.
The payoff associated with a play $(s^{(i)} a_1^{(i)} a_2^{(i)} o^{(i)})_{i=1}^\infty$ is thus $\Disc^\gamma := \sum_{i=1}^\infty \gamma^{i-1} R(s^{(i)},a_1^{(i)},a_2^{(i)})$.
Player~1 is aiming to maximize this quantity while player~2 is minimizing it.

\medskip

Apart from reasoning about decision rules of the players for the entire game (i.e., their behavioural strategies $\sigma_1$ and $\sigma_2$), we also consider the strategies they use for a single decision point---or stage---of the game only (i.e., assuming that the course of the previous stages $(s^{(i)} a_1^{(i)} a_2^{(i)} o^{(i)})_{i=1}^t$ is fixed and considered a parameter of the given stage).

\begin{definition}[Stage strategy]
\thlabel{def:osposg:stage-strategy}
    Let $G$ be a one-sided POSG.
    A \emph{stage strategy} of player~1 is a distribution $\pi_1 \in \Delta(A_1)$ over the actions player~1 can use at the current stage.
    A \emph{stage strategy} of player~2 is a mapping $\pi_2: S \rightarrow \Delta(A_2)$ from the possible current states of the game (player~2 observes the true state at the beginning of the current stage) to a distribution over actions of player~2.
    The sets of all stage strategies of player~1 and player~2 are denoted $\Pi_1$ and $\Pi_2$, respectively.
\end{definition}

Note that a stage strategy of player~2 is essentially a conditional probability distribution given the current state of the game.
For the reasons of notational convenience, we use notation $\pi_2(a_2 \,|\, s)$ instead of $\pi_2(s)(a_2)$ wherever applicable.

\subsection{Subgames}
Recall that both players know past actions of player~1 and all observations player~1 has received. The action-observation history is thus public knowledge.
This allows us to define a notion of \emph{subgames}.
A subgame induced by an action-observation history $\omega$ (or \emph{$\omega$-subgame}) is formed by histories $h$ such that the action-observation history $\omega(h)$ of player~1 in $h$ is a suffix of $\omega$, i.e., $\omega(h) \sqsupseteq \omega$.

Later in the text, we will specifically reason about subgames that follow directly after the first stage of the game---these correspond to $(a_1, o)$-subgames for some action $a_1$ and observation $o$.
Observe that, once $(a_1,o)$ is played and observed, both players know exactly which \emph{$(a_1,o)$-subgame} they are currently in.
Consequently, reasoning about $(a_1,o)$-subgame can be done without considering any other $(a_1',o')$-subgame.

\subsection{Probability measures}

We now proceed by defining a probability measure on the space of infinite plays in one-sided POSGs.
Assuming that $b \in \Delta(S)$ is the initial belief characterizing the distribution over possible initial states, and players use strategies $(\sigma_1,\sigma_2)$ to play the game from the current situation, we can define the probability distribution over histories (i.e., prefixes of plays) recursively as follows.
\begin{subequations}
\begin{align*}
    & \Pr_{b,\sigma_1,\sigma_2}[s^{(1)}] = b(s^{(1)}) \numberthis \\
    & \Pr_{b,\sigma_1,\sigma_2}[(s^{(i)} a_1^{(i)} a_2^{(i)} o^{(i)})_{i=1}^t s^{(t+1)}] = \Pr_{b,\sigma_1,\sigma_2}[(s^{(i)} a_1^{(i)} a_2^{(i)} o^{(i)})_{i=1}^{t-1} s^{(t)}] \ \cdot \numberthis \\
    & \qquad\qquad\qquad \cdot \sigma_1((a_1^{(i)} o^{(i)})_{i=1}^{t-1}, a_1^{(t)}) \cdot \sigma_2((s^{(i)} a_1^{(i)} a_2^{(i)} o^{(i)})_{i=1}^{t-1} s^{(t)}, a_2^{(t)}) \ \cdot \\
    & \qquad\qquad\qquad \cdot T(o^{(t)}, s^{(t+1)} \;|\; s^{(t)}, a_1^{(t)}, a_2^{(t)})
\end{align*}
\end{subequations}
This probability distribution also coincide with a measure $\mu$ defined over the cones, i.e. plays having $w$ as a prefix.
\begin{equation}
    \mu(\mathsf{Cone}(w)) = \Pr_{b,\sigma_1,\sigma_2}[w]
\end{equation}
The measure $\mu$ uniquely extends to the probability measure $\Pr_{b,\sigma_1,\sigma_2}[\cdot]$ over infinite plays of the game, which allows us to define the expected utility $\E_{b,\sigma_1,\sigma_2}[\Disc^\gamma]$ of the game when the initial belief of the game is $b$ and strategies $\sigma_1 \in \Sigma_1$ and $\sigma_2 \in \Sigma_2$ are played by player~1 and player~2, respectively.

In a similar manner, we can define a probability measure $\Pr_{b,\pi_1,\pi_2}[s, a_1, a_2, o, s']$ that predicts events only one step into the future (for \textit{stage} strategies $\pi_1 \in \Pi_1$, $\pi_2 \in \Pi_2$).
For belief $b$ and stage strategies $\pi_1$, $\pi_2$, we consider the probability that a stage starts in state $s \in S$ (sampled from $b$), players select actions $a_1 \sim \pi_1$ and $a_2 \sim \pi_2$, and that this results into a transition to a new state $s' \in S$ while generating an observation $o \in O$:
\begin{equation}
\Pr_{b,\pi_1,\pi_2}[s, a_1, a_2, o, s'] = b(s) \pi_1(a_1) \pi_2(a_2 \,|\, s) T(o, s' \,|\, s, a_1, a_2) \ \text{.} \label{eq:osposg:stage-prob}
\end{equation}
The probability distribution in Equation~\eqref{eq:osposg:stage-prob} can be marginalized to obtain, e.g., the probability that player~1 plays action $a_1 \in A_1$ and observes $o \in O$,
\begin{align}
\Pr_{b,\pi_1,\pi_2}[a_1, o] = &  \sum_{(s, a_2, s') \in S \times A_2 \times S} \Pr_{b,\pi_1,\pi_2}[s, a_1, a_2, o, s'] \nonumber \\
    = & \sum_{(s, a_2, s') \in S \times A_2 \times S} b(s) \pi_1(a_1) \pi_2(a_2 \,|\, s) T(o, s' \,|\, s, a_1, a_2) \ \text{.}
\end{align}

At the beginning of each stage, the imperfectly informed player~1 selects their action based on their belief about the current state of the game.
For a fixed current stage-strategy $\pi_2$ of player~2, player~1 can derive the distribution over possible states at the beginning of the next stage.
If player~1 starts with a belief $b$, takes an action $a_1 \in A_1$, and observes $o \in O$, his updated belief $b' = \tau(b,a_1,\pi_2,o)$ over states $s' \in S$ is going to be $\tau(b,a_1,\pi_2,o)(s') = $
\begin{subequations}\label{eq:osposg:tau}
\begin{align}
    &= \Pr_{b,\pi_1,\pi_2}[s' \,|\, a_1, o] = \sum_{(s,a_2) \in S \times A_2} \Pr_{b,\pi_1,\pi_2}[s, a_2, s' \,|\, a_1, o] \\
    &= \frac{1}{\Pr_{b,\pi_1,\pi_2}[a_1,o]} \sum_{(s,a_2) \in S \times A_2} \Pr_{b,\pi_1,\pi_2}[s, a_1, a_2, o, s'] \\
    &= \frac{1}{\Pr_{b,\pi_1,\pi_2}[a_1,o]} \sum_{(s,a_2) \in S \times A_2} b(s) \pi_1(a_1) \pi_2(a_2 \,|\, s) T(o, s' \,|\, s, a_1, a_2) \ \text{.}
\end{align}
\end{subequations}
In Section~\ref{sec:osposg:bellman}, this expression will prove useful for describing the Bellman equation in one-sided POSGs.

\section{Value of One-Sided POSGs}\label{sec:osposg:value}

We now proceed by establishing the value function of one-sided POSGs.
The value function represents the utility player~1 can achieve in each possible initial belief of the game.
First, we define the value of a strategy $\sigma_1 \in \Sigma_1$ of player~1, which assigns a payoff player~1 is guaranteed to get by playing $\sigma_1$ in the game (parameterized by the initial belief of the game).
Based on the value of strategies, we define the optimal value function of the game where player~1 chooses the best strategy for the given initial belief.

\begin{definition}[Value of strategy]\thlabel{def:osposg:strategy-value}
    Let $G$ be a one-sided POSG and $\sigma_1 \in \Sigma_1$ be a behavioral strategy of the imperfectly informed player~1.
    The \emph{value of strategy} $\sigma_1$, denoted $\val^{\sigma_1}$, is a function mapping each belief $b \in \Delta(S)$ to the expected utility that $\sigma_1$ guarantees against a best-responding player~2 given that the initial belief is $b$:
    \begin{equation}
        \val^{\sigma_1}(b) = \inf_{\sigma_2 \in \Sigma_2} \E_{b,\sigma_1,\sigma_2}[\Disc^\gamma] \ \text{.}
    \end{equation}
\end{definition}

When given an instance of a one-sided POSG with initial belief $b$, player~1 aims for a strategy that yields the best possible expected utility $\val^{\sigma_1}(b)$.
The value player~1 can guarantee in belief $b$ is characterized by the optimal value function $V^*$ of the game.

\begin{definition}[Optimal value function]
\thlabel{def:osposg:v-star}
    Let $G$ be a one-sided POSG.
    The \emph{optimal value function} $V^*: \Delta(S) \rightarrow \R$ of $G$ represents the supinf value of player~1 for each of the beliefs, i.e.
    \begin{equation}
        V^*(b) = \sup_{\sigma_1 \in \Sigma_1} \val^{\sigma_1}(b) \ \text{.}
    \end{equation}
\end{definition}

Note that according to von Neumann's minimax theorem~\cite{vonneumann1928-minimax} (resp. its generalization~\cite{sion1958-minimax}), every zero-sum POSG with discounted-sum objective $\Disc^\gamma$ is determined in the sense that the lower values (in the $\sup\inf$ sense) and the upper values (in the $\inf\sup$ sense) of the game coincide and represent the value of the game.
Therefore, $V^*(b)$ also represents the value of the game when the initial belief of the game is $b \in \Delta(S)$.

Since the $\Disc^\gamma$ objective is considered (for $0 < \gamma < 1$), the infinite discounted sum of rewards of player~1 converge.
As a result, the values of strategies $\val^{\sigma_1}(b)$ and the value of the game $V^*(b)$ can be bounded.

\begin{restatable}{proposition}{ValuesAreBounded}\label{thm:osposg:bounded}
    Let $G$ be a one-sided POSG.
    Then the payoff $\Disc^\gamma$ of an arbitrary play in $G$ is bounded by values
    \begin{equation}
        L = \min_{(s,a_1,a_2)} R(s,a_1,a_2) / (1-\gamma) \qquad U = \max_{(s,a_1,a_2)} R(s,a_1,a_2) / (1-\gamma) \ \text{.}
    \end{equation}
    It also follows that $L \leq V^*(b) \leq U$ and $L \leq \val^{\sigma_1}(b) \leq U$ holds for every belief $b \in \Delta(S)$ and strategy $\sigma_1 \in \Sigma_1$ of the imperfectly informed player~1.
\end{restatable}

Since the values $L$ and $U$ are uniquely determined by the given one-sided POSG, we will use these symbols in the remainder of the text.
We now focus on the discussion of structural properties of solutions of OS-POSGs.
First, we show that the value of an arbitrary strategy $\sigma_1 \in \Sigma_1$ of player~1 is linear in $b \in \Delta(S)$ --- that is, it can be represented as a convex combination of its values in the vertices of the simplex $\Delta (S)$.

In accordance with the notation used in the POMDP literature, we refer to linear functions defined over the $\Delta(S)$ simplex as \emph{$\alpha$-vectors}.
For $s \in S$, we overload the notation as $\alpha(s) := $ the value of $\alpha$ in the vertex corresponding to $s$. This allows us to write the following for every $b \in \Delta(S)$
\begin{equation}
    \alpha(b) = \sum_{s \in S} \alpha(s) \cdot b(s) \qquad \text{where } \alpha(s) = \alpha(\mathbbm{1}_s),\ \ \mathbbm{1}_s(s') = \begin{cases}
        1 & s = s' \\
        0 & \text{otherwise}
    \end{cases}
    \ \text{.}
\end{equation}

The following lemma shows the result we promised earlier:

\begin{lemma}\thlabel{thm:osposg:strategy-value-linear}
    Let $G$ be a one-sided POSG and $\sigma_1 \in \Sigma_1$ be an arbitrary behavioral strategy of player~1.
    Then the value $\val^{\sigma_1}$ of strategy $\sigma_1$ is a linear function in the belief space $\Delta(S)$.
\end{lemma}
\begin{proof}
    According to the \thref{def:osposg:strategy-value}, the value $\val^{\sigma_1}$ of strategy $\sigma_1$ is defined as the expected utility of $\sigma_1$ against the best-response strategy $\sigma_2$ of player~2.
    However, before having to act, player~2 observes the true initial state $s \sim b$.
    Therefore, he will play a best-response strategy $\sigma_2$ against $\sigma_1$ (with expected utility $\val^{\sigma_1}(s)$) given that the initial state is $s$.
    Since the probability that the initial state is $s$ is $b(s)$, we have
    \begin{equation}
        \val^{\sigma_1}(b) = \sum_{s \in S} b(s) \val^{\sigma_1}(s) \ \text{.}
    \end{equation}
    This shows that $\val^{\sigma_1}$ is a linear function in the belief $b \in \Delta(S)$.
\end{proof}

Since a point-wise supremum of a set of linear functions is convex, \thref{thm:osposg:strategy-value-linear} implies that the optimal value function $V^*$ is convex:

\begin{restatable}{lemma}{ValuesConvex}\label{thm:osposg:vs-convex}
    Optimal value function $V^*$ of a one-sided POSG is convex.
\end{restatable}

Unless otherwise specified, we endow any space $\Delta(X)$ over a finite set $X$ with the $\| \cdot \|_1$ metric.
To prepare the ground for the later proof of correctness of our main algorithm (presented in Section~\ref{sec:osposg:hsvi}), we now show that both the value of strategies and the optimal value function $V^*$ are Lipschitz continuous.
(Recall that for $k > 0$ a function $f: \Delta(X) \rightarrow \mathbb{R}$ is $k$-Lipschitz continuous if for every $p, q \in \Delta(X)$ it holds $| f(p) - f(q) | \leq k \cdot \| p - q \|_1$.)

\begin{restatable}{lemma}{LinearFsAreLipschitz}\label{thm:osposg:linbound-lipschitz}
    Let $X$ be a finite set and let $f: \Delta(X) \rightarrow [ y_{\mathrm{min}}, y_{\mathrm{max}} ]$ be a linear function.
    Then $f$ is $k$-Lipschitz continuous for $k=(y_{\mathrm{max}}-y_{\mathrm{min}})/2$.
\end{restatable}

Lemma~\ref{thm:osposg:linbound-lipschitz} directly implies that both values $\val^{\sigma_1}$ of strategies $\sigma_1$ of the imperfectly informed player~1, as well as the optimal value function $V^*$ are Lipschitz continuous.

\begin{lemma}\thlabel{thm:osposg:strategy-value-lipschitz}
    Let $\sigma_1 \in \Sigma_1$ be an arbitrary strategy of the imperfectly informed player~1.
    Then $\val^{\sigma_1}$ is $(U-L)/2$-Lipschitz continuous.\footnote{Recall that $L$ and $U$, introduced in Proposition~\ref{thm:osposg:bounded}, are the minimum and maximum possible utilities in the game.}
\end{lemma}
\begin{proof}
    Value $\val^{\sigma_1}$ of strategy $\sigma_1$ is linear (\thref{thm:osposg:strategy-value-linear}) and its values are bounded by $L$ and $U$ (Proposition~\ref{thm:osposg:bounded}).
    Therefore, according to Lemma~\ref{thm:osposg:linbound-lipschitz}, the function $\val^{\sigma_1}$ is $(U-L)/2$-Lipschitz.
\end{proof}

For notational convenience, we denote this constant as $\delta := (U-L)/2$ in the remainder of the text.

\begin{restatable}{proposition}{ValuesAreLipschitz}\label{thm:osposg:value-lipschitz}
    Value function $V^*$ of one-sided POSGs is $\delta$-Lipschitz continuous.
\end{restatable}

\begin{remark}
    In the remainder of the text, we will use term \emph{value function} to refer to an arbitrary function $V: \Delta(S) \rightarrow \R$ that assigns numbers $V(b)$ (estimates of the value achieved under optimal play) to beliefs $b \in \Delta(S)$ of player~1.
\end{remark}

\subsection{Elementary Properties of Convex Functions}
\label{sec:osposg:cvx}

In \thref{thm:osposg:vs-convex}, we have shown that the optimal value function $V^*$ of one-sided POSGs is convex.
In this section, we will explicitly state some of the important properties of convex functions that motivate our approach and are used throughout the rest of the text.

\begin{proposition}\thlabel{thm:osposg:sup-convex}
    Let $f: \Delta(S) \rightarrow \mathbb{R}$ be a point-wise supremum of linear functions, i.e.,
    \begin{equation}
        f(b) = \sup_{\alpha \in \Gamma} \alpha(b) \ , \qquad \Gamma \subseteq \left\lbrace \alpha: \Delta(S) \rightarrow \mathbb{R} \mid \alpha \text{ is linear} \right\rbrace \ \text{.}
    \end{equation}
    Then $f$ is convex and continuous.
    Furthermore, if every $\alpha \in \Gamma$ is $k$-Lipschitz continuous, $f$ is $k$-Lipschitz continuous as well.
\end{proposition}
\begin{proof}
    Let $b,b' \in \Delta(S)$ and $\lambda \in [0,1]$ be arbitrary.
    We have
    \begin{align*}
        \lambda f(b) + (1-\lambda) f(b') &= \lambda \sup_{\alpha \in \Gamma} \alpha(b) + (1-\lambda) \sup_{\alpha \in \Gamma} \alpha(b') \\
        &= \sup_{\alpha \in \Gamma} \lambda \alpha(b) + \sup_{\alpha \in \Gamma} (1-\lambda) \alpha(b') \\
        &\geq \sup_{\alpha \in \Gamma} \ \left[\lambda \alpha(b) + (1-\lambda) \alpha(b') \right] \\
        &= \sup_{\alpha \in \Gamma} \alpha(\lambda b + (1-\lambda) b') \\
        &= f(\lambda b + (1-\lambda)b') ,
    \end{align*}
    which shows that $f$ is convex.
    
    We now prove the continuity of $f$.
    Since every convex function is continuous on the interior of its domain, it remains to show that $f$ is continuous on the boundary of $\Delta(S)$.
    Assume to the contradiction that it is not continuous, i.e., there exists $b_0$ on the boundary such that for all $b$ from its neighborhood $f(b_0) > f(b) + C$ for some $C > 0$.
    Since $f$ is a pointwise supremum of linear functions, there exists $\alpha \in \Gamma$ such that $\alpha(b_0) > f(b_0) - C/2$.
    However, at the same time, we have $\alpha(b) \leq f(b_0) - C$.
    This is in contradiction with the fact that all $\alpha \in \Gamma$ are linear, and hence continuous.
    
    Furthermore, suppose that every $\alpha \in \Gamma$ is $k$-Lipschitz continuous and let $b,b'\in \Delta(S)$. We have
    \begin{align*}
        f(b) &= \sup_{\alpha \in \Gamma} \alpha(b) \\
        &\leq \sup_{\alpha \in \Gamma} \left[ \alpha(b') + k \| b - b' \|_1 \right] & \text{(since every $\alpha \in \Gamma$ is $k$-Lipschitz)} \\
        &= \left[ \sup_{\alpha \in \Gamma} \alpha(b') \right] + k \| b - b' \|_1 \\
        &= f(b') + k \| b - b' \|_1.
    \end{align*}
    Since the identical argument proves the inequality $f(b') \leq f(b) + k\|b-b'\|_1$, this shows that $f$ is $k$-Lipschitz continuous.
\end{proof}

Recall that we aim to emulate the HSVI algorithm from POMDPs, where the optimal value function $V^*$ is approximated by a series of piecewise linear and convex functions.
One of the common ways to represent these functions is as a point-wise maximum of a finite set of linear functions (typically called \textit{$\alpha$-vectors} in the POMDP context):

\begin{definition}[Piecewise linear and convex function on $\Delta(S)$]
\thlabel{def:osposg:pwlc}
    A function $f: \Delta(S) \rightarrow \mathbb{R}$ is said to be \emph{piecewise linear and convex} (PWLC) if it is of the form $f(b) = \max_{\alpha \in \Gamma} \alpha(b)$ (for each $b\in \Delta (S)$) for some finite set $\Gamma \subset \lbrace \alpha: \Delta(S) \rightarrow \mathbb{R} \mid \alpha \text{ is linear} \rbrace$.
\end{definition}

We immediately see that the preceding Proposition~\ref{thm:osposg:sup-convex} applies to any function of this type.
The next result shows that PWLC functions remain unchanged if we replace the set $\Gamma$ by its convex hull:

\begin{restatable}{proposition}{ConvexClosureDoesntIncreaseSupremum}
\label{thm:osposg:cvx-convexification}
    Let $\Gamma \subset \lbrace \alpha: \Delta(S) \rightarrow \mathbb{R} \mid \alpha \text{ is linear} \rbrace$ be a set of linear functions.
    Then for every $b \in \Delta(S)$ we have
    \begin{equation}
        \sup_{\alpha \in \Gamma} \alpha(b) = \sup_{\alpha \in \mathsf{Conv}(\Gamma)} \alpha(b) \ \text{.}
    \end{equation}
\end{restatable}

In the opposite direction, every convex function can be represented as a supremum over some set of linear functions.
The following proposition shows this using the largest possible set, i.e. $\{ \alpha \leq f \mid \alpha \textnormal{ linear} \}$:

\begin{restatable}{proposition}{ConvexFunctionsAsSupremaOfLinearFs}
\label{thm:osposg:cvx-sup-representable}
    Let $f: \Delta(S) \rightarrow \mathbb{R}$ be a convex continuous function.
    Then there exists a set $\Gamma$ of linear functions such that $\alpha \leq f$ for every $\alpha \in \Gamma$ and $f(b) = \sup_{\alpha \in \Gamma} \alpha(b)$ for every $b \in \Delta(S)$.
\end{restatable}

\section{Composing Strategies}
\label{sec:osposg:composing}

\begin{figure}
    \centering
    \includegraphics[width=\linewidth]{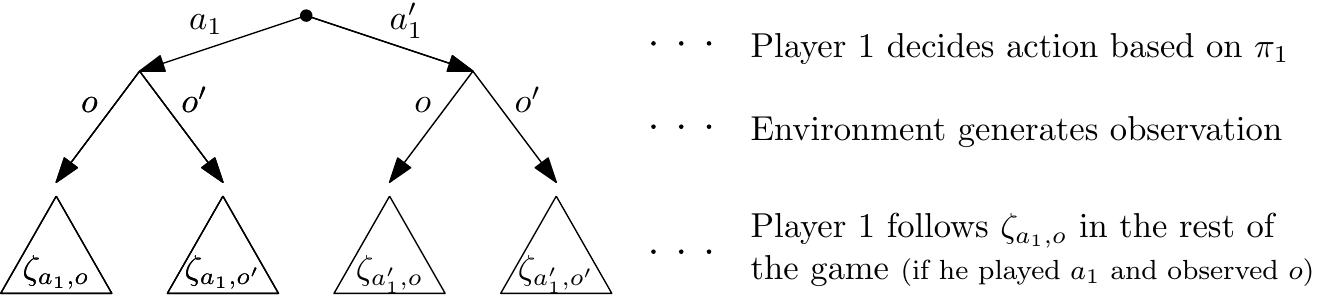}
    \caption{Composition of strategies $\zeta$ using a stage strategy $\pi_1$.}
    \label{fig:osposg:composition}
\end{figure}

Every behavioural strategy of the imperfectly informed player~1 can be split into the stage strategy $\pi_1$ player~1 uses in the first stage of the game, and behavioural strategies he uses in the rest of the game after he reaches an $(a_1,o)$-subgame.
We can also use the inverse principle, called \emph{strategy composition}, to form new strategies by choosing the stage strategy $\pi_1$ for the first stage and then selecting a separate behavioral strategy $\overline{\zeta}=(\zeta_{a_1,o})_{(a_1,o) \in A_1 \times O}$ for each subgame (see Figure~\ref{fig:osposg:composition} for illustration).

\begin{definition}[Strategy composition]
\thlabel{def:osposg:composite}
    Let $G$ be a one-sided POSG and $\pi_1 \in \Pi_1$ a stage strategy of player~1.
    Furthermore, let $\overline{\zeta} \in (\Sigma_1)^{A_1 \times O}$ be a vector representing behavioral strategies of player~1 for each $(a_1,o)$-subgame where $a_1 \in A_1$ and $o \in O$.
    The \emph{strategy composition} $\mathsf{comp}(\pi_1,\overline{\zeta})$ is a behavioral strategy of player~1 such that
    \begin{equation}
        \mathsf{comp}(\pi_1,\overline{\zeta})(\omega) = \begin{cases}
            \pi_1 & \omega = \emptyset \\
            \zeta_{a_1,o}(\omega') & \omega = a_1 o \omega'
        \end{cases}
        \qquad
        \text{ for each } \omega \in (A_1 O )^*
        \ \text{.}
        \label{eq:osposg:composite}
    \end{equation}
\end{definition}

By composing strategies $\overline{\zeta}$ using $\pi_1$, we obtain a new strategy where the probability of playing $a_1$ in the first stage of the game is $\pi_1(a_1)$, and strategy $\zeta_{a_1,o}$ is used after playing action $a_1$ and receiving observation $o$ in the first stage of the game.
Importantly, the newly formed strategy $\mathsf{comp}(\pi_1,\overline{\zeta}) \in \Sigma_1$ is also a behavioral strategy (of imperfectly informed player~1), and therefore the properties of strategies presented in Section~\ref{sec:osposg:value} apply also to $\mathsf{comp}(\pi_1,\overline{\zeta})$.
As the next result shows, the opposite property also holds --- for each strategy $\sigma_1 \in \Sigma_1$ of player~1, we can find the appropriate $\pi_1$ and $\overline{\zeta}$ such that $\sigma_1 = \mathsf{comp}(\pi_1,\overline{\zeta})$:

\begin{restatable}{proposition}{StrategyDecomposition}
\label{thm:osposg:decomposition}
    Every behavioral strategy $\sigma_1 \in \Sigma_1$ of player~1 can be represented as a strategy composition of some stage strategy $\pi_1 \in \Pi_1$ and player~1 behavioral strategies $\zeta_{a_1,o}$.
\end{restatable}

Importantly, we can obtain values $\val^{\mathsf{comp}(\pi_1,\overline{\zeta})}$ of composite strategies without considering the entire strategy $\mathsf{comp}(\pi_1,\overline{\zeta})$.
As the following lemma shows, it suffices to consider only the first stage of the game and the \emph{values} of the strategies $\overline{\zeta} \in (\Sigma_1)^{A_1 \times O}$.

\begin{restatable}{lemma}{StrategyCompositionValue}
\label{thm:osposg:composition}
    Let $G$ be a one-sided POSG and $\mathsf{comp}(\pi_1,\overline{\zeta})$ a composite strategy. Then the following holds:
    \begin{align}
        \val^{\mathsf{comp}(\pi_1,\overline{\zeta})}(s) = \min_{a_2 \in A_2} \E_{a_1 \sim \pi_1,\, (o,s') \sim T(\cdot \,|\, s,a_1,a_2)} \left[ R(s,a_1,a_2) + \gamma \val^{\zeta_{a_1,o}}(s') \right] \nonumber \\
                                             \!\!\!\!\!\!\!\!\!\! = \min_{a_2 \in A_2} \sum_{a_1 \in A_1} \!\! \pi_1(a_1) \left[ R(s,a_1,a_2) + \gamma \!\!\!\!\!\!\! \sum_{(o,s) \in O \times S} \!\!\!\!\!\!\! T(o,s' \,|\, s,a_1,a_2) \val^{\zeta_{a_1,o}}(s') \right] \text{.}
    \end{align}
\end{restatable}

The proof relies on the fact that when player~1 takes the action $a_1$, observes $o$, and ends up in $s'$, the strategy $\zeta_{a_1,o}$ guarantees the player gets at least $\val^{\zeta_{a_1,o}}(s')$ utility (in expectation), no matter what player~2 does.
Since the values in the rest of the game are known, it suffices to focus on the best-response strategy of player~2 in the first stage of the game.
\subsection{Generalized Composition}

Lemma~\ref{thm:osposg:composition} suggests that we can use composition of \emph{values} of strategies $\val^{\zeta_{a_1,o}}$ to form values of composite strategies $\val^{\mathsf{comp}(\pi_1,\overline{\zeta})}$.
In this section, still consider linear functions $\val^{\zeta_{a_1,o}}$, but we relax the assumption that these functions represent values of some specific behavioural strategy.
This allows us to derive a generalized principle of composition and approximate the value function $V^*$ by a supremum of arbitrary linear functions (as opposed to functions $\val^{\sigma_1}$).
Throughout the text, we will use $\linDS$ to denote the set of linear functions on $\Delta(S)$ (i.e., $\alpha$-vectors).
We will also use the term `linear' to refer to functions that satisfy $f( \lambda b + (1-\lambda)b') = \lambda f(b) + (1-\lambda)f(b')$ on $\Delta(S)$.

\begin{definition}[Value composition]
\thlabel{def:osposg:val-composition}
    Let $\pi_1 \in \Pi_1$ and $\overline{\alpha} \in (\linDS)^{A_1 \times O}$.
    \emph{Value composition} $\mathsf{valcomp}(\pi_1,\overline{\alpha}): \Delta(S) \rightarrow \R$ is a linear function defined by the values in vertices of the $\Delta(S)$ simplex as follows:
    \begin{align*}
        \mathsf{valcomp}(\pi_1,\overline{\alpha})(s) = & \min_{a_2 \in A_2} \sum_{a_1 \in A_1} \pi_1(a_1) \Big[ R(s,a_1,a_2) \ + \numberthis\label{eq:osposg:valcomp}\\
        & \qquad \gamma \sum_{(o,s') \in O \times S} T(o,s' \,|\, s,a_1,a_2) \alpha_{a_1,o}(s') \Big] \ \text{.}
    \end{align*}
\end{definition}

Observe that according to Lemma~\ref{thm:osposg:composition}, $\mathsf{valcomp}(\pi_1,\overline{\alpha}) = \val^{\mathsf{comp}(\pi_1,\overline{\zeta})}$ for $\alpha_{a_1,o} = \val^{\zeta_{a_1,o}}$.
The value composition $\mathsf{valcomp}(\pi_1,\overline{\alpha})$, however, admits arbitrary linear function $\alpha_{a_1,o}$ and not only the value $\val^{\zeta_{a_1,o}}$ of some strategy $\zeta_{a_1,o} \in \Sigma_1$.
Moreover, as long as linear functions $\alpha_{a_1,o}$ serve as lower bounds for values of some strategies, so will the corresponding value composition serve as a lower bound for the corresponding composite strategy:

\begin{restatable}{lemma}{GenCompositionValue}
\label{thm:osposg:gen-composition}
    Let $\pi_1 \in \Pi_1$ be a stage strategy of player~1 and $\overline{\alpha} \in (\linDS)^{A_1 \times O}$ a~vector of linear functions s.t. for each $\alpha_{a_1,o}$ there exists a strategy $\zeta_{a_1,o} \in \Sigma_1$ with $\val^{\zeta_{a_1,o}} \geq \alpha_{a_1,o}$.
    Then there exists a strategy $\sigma_1 \in \Sigma_1$ such that $\sigma_1(\emptyset) = \pi_1$ and $\val^{\sigma_1} \geq \mathsf{valcomp}(\pi_1,\overline{\alpha})$.
\end{restatable}

In case of value of composite strategies, we know that $\val^{\mathsf{comp}(\pi_1,\zeta)}$ is a $\delta$-Lipschitz continuous linear function (since $\mathsf{comp}(\pi_1,\zeta) \in \Sigma_1$ is a behavioral strategy of player~1 and \thref{thm:osposg:strategy-value-lipschitz} applies).
Additionally, we prove that as long as linear functions $\alpha_{a_1,o}$ are bounded by $L \leq \alpha_{a_1,o}(b) \leq U$ for every belief $b \in \Delta(S)$, and are therefore $\delta$-Lipschitz continuous, the value composition $\mathsf{valcomp}(\pi_1,\overline{\alpha})$ is also $\delta$-Lipschitz.

\begin{restatable}{lemma}{ValcompLipschitz}
\label{thm:osposg:valcomp-lipschitz}
    Let $\pi_1 \in \Pi_1$ and $\overline{\alpha} \in (\linDS)^{A_1 \times O}$ such that $L \leq \alpha^{a_1,o}(b) \leq U$ for every $b \in \Delta(S)$.
    Then $L \leq \mathsf{valcomp}(\pi_1,\overline{\alpha})(b) \leq U$ for every $b \in \Delta(S)$ and $\mathsf{valcomp}(\pi_1,\overline{\alpha})$ is a $\delta$-Lipschitz continuous function.
\end{restatable}

\section{Bellman Equation for One-Sided POSGs}
\label{sec:osposg:bellman}

In Section~\ref{sec:osposg:value}, we have defined the value function $V^*$ as the supremum over the strategies player~1 can achieve in each of the beliefs (see \thref{def:osposg:v-star}).
However, while this correctly defines the value function, it does not provide a straightforward recipe to obtaining value $V^*(b)$ for the given belief $b \in \Delta(S)$.
Obtaining the value for the given belief according to \thref{def:osposg:v-star} is as hard as solving the game itself.

In this section, we provide an alternative characterization of the optimal value function $V^*$ inspired by the value iteration methods, e.g., for Markov decision processes (MDPs) and their partially observable variant (POMDPs).
The high-level idea behind these approaches is to start with a coarse approximation  $V_0: \Delta(S) \rightarrow \mathbb{R}$ of the value function $V^*$, and then iteratively improve the approximation by applying the Bellman's operator $H$, i.e., generate a sequence such that $V_{i+1} = HV_i$.
In our case, the improvement is based on finding a new, previously unknown, strategy that achieves higher values for each of the beliefs by means of value composition principle (\thref{def:osposg:val-composition}).
Throughout this section, we will consider value functions that are represented as a point-wise supremum over a (possibly infinite) set $\Gamma$ of linear functions (called $\alpha$-vectors), i.e.,
\begin{equation}
    V(b) = \sup_{\alpha \in \Gamma} \alpha(b) \qquad \text{for } \Gamma \subset \left\lbrace \alpha: \Delta(S) \rightarrow \mathbb{R} \mid \alpha \text{ is linear} \right\rbrace \ \text{.}
\end{equation}
By Proposition~\ref{thm:osposg:cvx-convexification}, we can always assume that the set $\Gamma$ is convex (since this doesn't come at the loss of generality).
For more details on this representation of value functions see Section~\ref{sec:osposg:cvx}.

\begin{definition}[Max-composition]
\thlabel{def:osposg:H-valcomp}
    Let $V: \Delta(S) \rightarrow \R$ be a convex continuous function and let $\Gamma$ be a convex set of linear functions such that $V(b) = \sup_{\alpha \in \Gamma} \alpha(b)$.
    The \emph{max-composition} operator $H$ is defined as
    \begin{equation}
        [HV](b) = \max_{\pi_1 \in \Pi_1} \sup_{\overline{\alpha} \in \Gamma^{A_1 \times O}} \mathsf{valcomp}(\pi_1,\overline{\alpha})(b) \ \text{.} \label{eq:osposg:max-composition}
    \end{equation}
\end{definition}

We will now prove several fundamental properties of the max-composition operator $H$ from \thref{def:osposg:H-valcomp}.
First, we will show that this operator preserves continuity and convexity, allowing us to apply the operator iteratively.
Second, we introduce equivalent formulations of the operator $H$, which represent the solution of $[HV](b)$ in a more traditional form of finding a Nash equilibrium of a stage-game.
These formulations also allow us to show that the behaviour of $H$ is not sensitive to the choice of the set $\Gamma$ used to represent the value function $V$.
Finally, we conclude by showing that the operator $H$ can indeed be used to approximate the optimal value function $V^*$.
Namely, we show that $H$ is a contraction mapping (and thus iterated application converges to a unique fixpoint) and that its fixpoint is the optimal value function $V^*$.

\begin{restatable}{proposition}{HVLipschitzConvex}
Proposition~\label{thm:osposg:hv-convex}
    Let $V: \Delta(S) \rightarrow \R$ be a convex continuous function and let $\Gamma$ be a convex set of linear functions such that $V(b)=\sup_{\alpha \in \Gamma} \alpha(b)$.
    Then $HV$ is also convex and continuous.
    Furthermore, if $V$ is $\delta$-Lipschitz continuous, the function $HV$ is $\delta$-Lipschitz continuous as well.
\end{restatable}

The proof of this result goes by rewriting $HV$ as a supremum over all value-compositions and using our earlier observations about convexity and Lipschitz continuity of such suprema.

We will now prove that the max-composition operator $H$ can be alternatively characterized using max-min and min-max optimization.
Recall that $\tau(b,a_1,\pi_2,o)$ denotes the Bayesian update of belief $b$ given that player~1 played $a_1$ and observed $o$, and player~2 is assumed to follow stage strategy $\pi_2$ in the current round (see Equation~\eqref{eq:osposg:tau}).

\begin{restatable}{theorem}{Bellman}
\label{thm:osposg:bellman}
    Let $V: \Delta(S) \rightarrow \R$ be a convex continuous function and let $\Gamma$ be a convex set of linear functions on $\Delta(S)$ such that $V(b) = \sup_{\alpha \in \Gamma} \alpha(b)$ for every belief $b \in \Delta(S)$.
    Then the following definitions of operator $H$ are equivalent:
    \begin{subequations}
    \begin{align}
        & [HV](b) = \nonumber \\
        & = \max_{\pi_1 \in \Delta(S)} \sup_{\overline{\alpha} \in \Gamma^{A_1 \times O}} \mathsf{valcomp}(\pi_1,\overline{\alpha})(b) \label{eq:osposg:equiv-valcomp}\\
               &= \! \max_{\pi_1 \in \Pi_1} \!\! \min_{\pi_2 \in \Pi_2} \left[ \E_{b,\pi_1,\pi_2}[R(s,a_1,a_2)] + \gamma \sum_{a_1,o} \Pr_{b,\pi_1,\pi_2}[a_1,o] \cdot V(\tau(b,a_1,\pi_2,o)) \right] \label{eq:osposg:H-maxmin}\\
               &= \! \min_{\pi_2 \in \Pi_2} \!\! \max_{\pi_1 \in \Pi_2} \left[ \E_{b,\pi_1,\pi_2}[R(s,a_1,a_2)] + \gamma \sum_{a_1,o} \Pr_{b,\pi_1,\pi_2}[a_1,o] \cdot V(\tau(b,a_1,\pi_2,o)) \right] \ \text{.} \label{eq:osposg:H-minmax}
    \end{align}
    \end{subequations}
\end{restatable}

The proof consists of verifying the assumptions of von Neumann's minimax theorem, which shows the equivalence of \eqref{eq:osposg:H-maxmin} and \eqref{eq:osposg:H-minmax}.
The equivalence of \eqref{eq:osposg:H-maxmin} and \eqref{eq:osposg:equiv-valcomp} can be then shown by reformulating each stage game as a separate zero-sum game and verifying that it satisfies the assumptions of a Sion's generalization of the minimax theorem \cite{sion1958-minimax}.

\begin{corollary}\thlabel{thm:osposg:gamma-independent}
    Bellman's operator $H$ does not depend on the convex set $\Gamma$ of linear functions used to represent the convex value function $V$.
\end{corollary}

Since the maximin and minimax values of the game (from equations~\eqref{eq:osposg:H-maxmin} and \eqref{eq:osposg:H-minmax}) coincide, the value $[HV](b)$ corresponds to the Nash equilibrium in the stage game.
We define the stage game formally.

\begin{definition}[Stage game]
\thlabel{def:osposg:stage-game}
    A \emph{stage game} with respect to a convex continuous value function $V: \Delta(S) \rightarrow \R$ and belief $b \in \Delta(S)$ is a two-player zero sum game with strategy spaces $\Pi_1$ for the maximizing player~1 and $\Pi_2$ for the minimizing player~2, and payoff function
    \begin{equation}
        u^{V,b}(\pi_1,\pi_2) = \E_{b,\pi_1,\pi_2}[R(s,a_1,a_2)] + \gamma \sum_{a_1,o} \Pr_{b,\pi_1,\pi_2}[a_1,o] \cdot V(\tau(b,a_1,\pi_2,o)) \ \text{.}
    \end{equation}
    With a slight abuse of notation, we use $[HV](b)$ to refer both to the max-composition operator (\thref{def:osposg:H-valcomp}) as well as to this stage game.
\end{definition}

We will now show that the Bellman's operator $H$ is a contraction mapping.
Recall that the mapping $H$ is a contraction, if there exists $0 \leq k < 1$ such that $\| HV_1 - HV_2 \| \leq k \| V_1 - V_2 \|$.
We consider the metric $\| V_1 - V_2 \|_\infty = \max_{b \in \Delta(S)} | V_1(b) - V_2(b) |$  corresponding to the $l_\infty$.
First, we focus on a single belief point and identify a criterion which ensures that $| HV_1(b) - HV_2(b) | \leq k | V_1(b) - V_2(b) |$.
While somewhat technical, this criterion will enable us to demonstrate the contractivity of $H$.
Moreover, it will also be useful in Section~\ref{sec:osposg:hsvi:alg} to prove the correctness of the HSVI algorithm proposed therein.

\begin{restatable}{lemma}{ContractivityLemma}
\label{thm:osposg:point-contractivity}
    Let $V, W: \Delta(S) \rightarrow \R$ be two convex continuous value functions and $b \in \Delta(S)$ a belief such that $[HV](b) \leq [HW](b)$.
    Let $(\pi_1^V,\pi_2^V)$ and $(\pi_1^W,\pi_2^W)$ be Nash equilibrium strategy profiles in stage games $[HV](b)$ and $[HW](b)$, respectively, and $C \geq 0$.
    If $W(\tau(b,a_1,o,\pi_2^V)) - V(\tau(b,a_1,o,\pi_2^V)) \leq C$ for every action $a_1 \in \Supp(\pi_1^W)$ of player~1 and every observation $o \in O$ such that $\Pr_{b,\pi_1^W,\pi_2^V}[o \,|\, a_1] > 0$, then $[HW](b) - [HV](b) \leq \gamma C$.
\end{restatable}

\begin{lemma}\thlabel{thm:osposg:contraction}
    Operator $H$ is a contraction on the space of convex continuous functions $V: \Delta(S) \rightarrow \R$ (under the supremum norm), with contraction-factor $\gamma$.
\end{lemma}
\begin{proof}
    Let $V, W: \Delta(S) \rightarrow \R$ be convex functions such that $\| V - W \|_{\infty} = \max_{b \in \Delta(S)} | V(b) - W(b) | \leq C$.
    To prove the contractivity of $H$, it suffices to show that $\| HV - HW \|_{\infty} \leq \gamma C$, i.e., $| [HV](b) - [HW](b) | \leq \gamma C$ for every belief $b \in \Delta(S)$.
    Since $| V(b) - W(b) | \leq C$ holds for every belief $b$, Lemma~\ref{thm:osposg:point-contractivity} yields both $HV(b)-HW(b) \leq \gamma C$ and $HW(b)-HV(b) \leq \gamma C$.
\end{proof}

Next, we show that the optimal value function from \thref{def:osposg:v-star} is the fixpoint of the Bellman's operator $H$.
Intuitively, this holds because $V^*$ can be represented as a supremum over all possible value functions $\val^{\sigma_1}$, which remains unchanged as we apply the operator $H$ (resp. the value-compositions it consists of).

\begin{restatable}{lemma}{FixpointLemma}
Lemma~\label{thm:osposg:fixed-point}
    The optimal value function $V^*$ satisfies $V^*=HV^*$.
\end{restatable}

\noindent
Together, the two results ensure that $H$ can be applied iteratively to obtain $V^*$:

\begin{theorem}
\thlabel{thm:osposg:unique-fixpoint}
    $V^*$ is a unique fixpoint of $H$.
    Moreover, for any convex function $V_0$, the sequence $\lbrace V_i \rbrace_{i=0}^\infty$ such that $V_i = HV_{i-1}$ converges to $V^*$.
\end{theorem}
\begin{proof}
    By Lemma~\ref{thm:osposg:fixed-point}, $V^*$ is \textit{a} fixpoint of $H$.
    By \thref{thm:osposg:contraction}, $H$ is a contraction mapping on the space of convex value functions.
    Banach's fixed point theorem~\citep{ciesielski2007-banach} then implies the uniqueness and the ``moreover'' part.
\end{proof}

\section{Exact Value Iteration}
\label{sec:osposg:vi}

In Section~\ref{sec:osposg:bellman}, we have shown that the optimal value function can be approximated by means of composing strategies in the sense of max-composition introduced in \thref{def:osposg:H-valcomp}.
In this section, we provide a linear programming formulation to perform such optimal composition for value functions that are piecewise linear and convex, i.e., can be represented as a point-wise maximum of a finite set $\Gamma$ of linear functions.
Furthermore, we show that as long as the value function $V$ is piecewise linear and convex, $HV$ is also piecewise linear and convex.
This allows for using the same linear program (LP) iteratively to approximate the optimal value function $V^*$ by means of constructing a sequence of piecewise linear and convex value functions $\lbrace V_i \rbrace_{i=1}^\infty$ such that $V_i = HV_{i-1}$.

\subsection{Computing Max-Compositions}

In order to compute $HV$ given a piecewise linear and convex (PWLC) value function $V$, it is essential to solve Equation~\eqref{eq:osposg:max-composition}.
Every PWLC value function can be represented as a point-wise maximum over a finite set of linear functions $\lbrace \alpha_1, \ldots, \alpha_k \rbrace$ (see \thref{def:osposg:pwlc}).
Without loss of generality, we consider that the set $\Gamma$ used to represent the value function $V$ is the convex hull of the aforementioned set:
\begin{equation}
    \Gamma \coloneqq \mathsf{Conv}\left( \left\lbrace \alpha_1, \ldots, \alpha_k \right) \right\rbrace = \left\lbrace \sum_{i=1}^k \lambda_i \alpha_i \mid \lambda \in \mathbb{R}^k_{\geq 0}, \| \lambda \|_1 = 1 \right\rbrace \ \text{.} \label{eq:gamma-conv}
\end{equation}
Recall that forming a convex hull of the set of linear functions used to represent $V$ does not affect the values $V$ attains (by Proposition~\ref{thm:osposg:cvx-convexification}).
We will now show that when the set $\Gamma$ is represented as in Equation~\eqref{eq:gamma-conv}, linear programming can be used to compute $HV(b)$:
\begin{restatable}{lemma}{LP}
\label{thm:lp}
    Let $\Gamma = \mathsf{Conv}\left( \left\lbrace \alpha_1, \ldots, \alpha_k \right) \right\rbrace$ be a convex hull of a finite set of $\alpha$-vectors.
    Then $[HV](b)$ coincides with the solution of the following linear program:
    
    \begin{subequations}
    \label{eq:osposg:max-composition-lp}
    \begin{align*}
        \max_{\pi_1,\lambda,\overline{\alpha},V} \ & \sum_{s \in S} b(s) \cdot V(s) \numberthis\label{eq:osposg:max-composition-lp:objective}\\
        \text{s.t.} \ \ \ & V(s) \leq \sum_{a_1 \in A_1} \pi_1(a_1) R(s,a_1,a_2) \ + \gamma \!\!\!\!\!\!\!\!\!\!\!\! \sum_{(a_1,o,s') \in A_1 \times O \times S} \!\!\!\!\!\!\!\!\!\!\!\! T(o,s' \,|\, s,a_1,a_2) \hat{\alpha}^{a_1,o}(s') \!\!\!\!\!\!\!\!\!\!\!\!\!\!\!\!\!\!\!\!\!\!\!\!\!\!\!\!\!\!\!\!\!\! \\
        & \hspace{20em} \forall (s,a_2) \in S \times A_2 \numberthis\label{eq:osposg:max-composition-lp:br} \\
        & \hat{\alpha}^{a_1,o}(s') = \sum_{i=1}^k \hat{\lambda}_i^{a_1,o} \cdot \alpha_i(s') \hspace{4.5em}  \forall (a_1,o,s') \in A_1 \times O \times S \numberthis\label{eq:osposg:max-composition-lp:convexification}\\
        & \sum_{i=1}^k \hat{\lambda}_i^{a_1,o} = \pi_1(a_1) \hspace{11.5em} \forall (a_1,o) \in A_1 \times O \numberthis\label{eq:osposg:max-composition-lp:lambda-hat}\\
        & \sum_{a_1 \in A_1} \pi_1(a_1) = 1 \numberthis\label{eq:osposg:max-composition-lp:pi-sum}\\
        & \pi_1(a_1) \geq 0  \hspace{19em} \forall a_1 \in A_1 \numberthis\label{eq:osposg:max-composition-lp:pi-positive}\\
        & \hat{\lambda}_i^{a_1,o} \geq 0 \hspace{11em} \forall (a_1,o) \in A_1 \times O, 1 \leq i \leq k \numberthis\label{eq:osposg:max-composition-lp:last}\\
    \end{align*}
    \end{subequations}
\end{restatable}

In the latter text, we also use the following dual formulation of the linear program~\eqref{eq:osposg:max-composition-lp} (with some minor modifications to improve readability):
\begin{subequations}
\label{eq:osposg:max-composition-dual}
\begin{align}
    \min_{V,\pi_2,\hat{\tau}} \ & V \\
    \text{s.t.} \ & V \geq \!\!\!\!\!\!\!\! \sum_{(s,a_2) \in S \times A_2} \!\!\!\!\!\!\!\! \pi_2(s \wedge a_2) R(s,a_1,a_2) + \gamma \sum_{o \in O} \hat{V}(a_1,o) \!\!\!\!\!\!\!\! & \forall a_1 \label{eq:osposg:max-composition-dual:br} \\
    & \hat{V}(a_1,o) \geq \sum_{s' \in S} \hat{\tau}(b,a_1,o,\pi_2)(s') \cdot \alpha_i(s') & \forall (a_1,o), 1 \leq i \leq k \label{eq:osposg:max-composition-dual:subgame}\\
    & \hat{\tau}(b,a_1,\pi_2,o)(s') = \!\!\!\!\!\!\!\!\! \sum_{(s,a_2) \in S \times A_2} \!\!\!\!\!\!\!\!\! T(o,s' \,|\, s,a_1,a_2) \pi_2(s \wedge a_2) \!\!\!\!\!\!\!\!\!\! & \forall (a_1,o,s') \label{eq:osposg:max-composition-dual:tau}\\
    & \sum_{a_2 \in A_2} \pi_2(s \wedge a_2) = b(s) & \forall s \label{eq:osposg:max-composition-dual:pi-sum}\\
    & \pi_2(s \wedge a_2) \geq 0 & \forall (s, a_2) \label{eq:osposg:max-composition-dual:pi-positive}
\end{align}
\end{subequations}
Here, the stage strategy of player~2 is represented as a joint probability $\pi_2(s \wedge a_2)$ of playing action $a_2 \in A_2$ while being in state $s \in S$ (i.e., $\pi_2(a_2 \,|\, s) = \pi_2(s \wedge a_2) / b(s)$ where applicable).
Player~1 then seeks the best response $a_1 \in A_1$ (constraint~\eqref{eq:osposg:max-composition-dual:br}) that maximizes the sum of the expected immediate reward and the $\gamma$-discounted utility in the $(a_1,o)$-subgames.
The beliefs $\tau(b,a_1,\pi_2,o)$ in the subgames are multiplied by the probability of reaching the $(a_1,o)$-subgame (i.e., there is no division by $\Pr_{b,a_1,\pi_2}[a_1,o]$ in Equation~\eqref{eq:osposg:max-composition-dual:tau}), hence also the values of subgames $V(a_1,o)$ need not be multiplied by $\Pr_{b,a_1,\pi_2}[a_1,o]$.
The value of an $(a_1,o)$-subgame $V(a_1,o)$ is expressed as a maximum $\max_{\alpha \in \Gamma} \alpha(\tau(b,a_1,\pi_2,o))$ expressed by constraints~\eqref{eq:osposg:max-composition-dual:subgame}.

\subsection{Value Iteration}

To run a value iteration algorithm that would apply the linear program~\eqref{eq:osposg:max-composition-lp} repeatedly, we require that every $V_i$ in the sequence $\lbrace V_i \rbrace_{i=0}^\infty$, starting from an arbitrary PWLC value function $V_0$, is also piecewise linear and convex.
By \thref{thm:osposg:hv-pwlc} this is always the case.

\begin{lemma}
\label{thm:lp-vertices}
    Let $Q$ be the set of vertices of the polytope defined by constraints
    \eqref{eq:osposg:max-composition-lp:br}-\eqref{eq:osposg:max-composition-lp:last},
    and let $(\pi_1^q,\hat{\alpha}^q)$ be the assignment of the variables $\pi_1$ and $\hat{\alpha}$ corresponding to the vertex $q \in Q$.
    Then\footnote{Note that $\overline{\alpha}^q(a_1,o)$ for $a_1$ with $\pi_1^q(a_1)=0$ do not contribute to $\mathsf{valcomp}(\pi_1^q,\overline{\alpha}^q)$. In parts of the game that are not reached by player~1, we can thus define $\overline{\alpha}^q$ arbitrarily.}
    \begin{equation}
        [HV](b) = \max_{q \in Q} \mathsf{valcomp}(\pi_1^q, \overline{\alpha}^q) \qquad \text{ for } \overline{\alpha}^q(a_1,o)=\hat{\alpha}^q(a_1,o) / \pi_1^q(a_1) \ \text{.} \label{eq:thm:lp-vertices}
    \end{equation}
\end{lemma}
\begin{proof}
    Consider the LP~\eqref{eq:osposg:max-composition-lp} which computes the optimal value composition $\mathsf{valcomp}(\pi_1,\overline{\alpha})$ in $[HV](b)$ (see Lemma~\ref{thm:lp}).
    The polytope of feasible solutions of the LP defined by the constraints~\eqref{eq:osposg:max-composition-lp:br}--\eqref{eq:osposg:max-composition-lp:last} is independent of the belief $b$ (which only appears in the objective~\eqref{eq:osposg:max-composition-lp:objective}).
    Therefore, the set $Q$ of vertices of this polytope is also independent of belief $b \in \Delta(S)$.
    The optimal solution of a linear programming problem~\eqref{eq:osposg:max-composition-lp} representing $[HV](b)$ can be found within the vertices $Q$ of the polytope of feasible solutions~\citep{vanderbei2015-lp}.
    There is a finite number of vertices $q \in Q$, and each vertex $q \in Q$ corresponds to some assignment of variables defining the value composition $\mathsf{valcomp}(\pi_1^q,\overline{\alpha}^q)$.
    Since the set $Q$ of the vertices of the polytope is independent of the belief $b$, we get the desired result.
\end{proof}

\begin{lemma}
\thlabel{thm:osposg:hv-pwlc}
    If $V$ is a piecewise linear and convex function, then so is $HV$.
\end{lemma}
\begin{proof}
    This lemma is a direct consequence of \thref{thm:lp-vertices}.
    Since the number of vertices of the polytope of LP~\eqref{eq:osposg:max-composition-lp} is finite, the pointwise maximization in~\eqref{eq:thm:lp-vertices} defines a PWLC function.
\end{proof}

We can use the above-stated results to iteratively construct a sequence of value functions $\lbrace V_i \rbrace_{i=0}^\infty$ such that $V_0$ is an arbitrary PWLC function and $V_i = HV_{i-1}$.
Namely, we construct $V_i$ by enumerating the vertices of the polytope defined by the linear program~\eqref{eq:osposg:max-composition-lp} and constructing appropriate linear functions $\mathsf{valcomp}(\pi_1^q, \overline{\alpha}^q)$.
By \thref{thm:lp-vertices}, these linear functions form the set of $\alpha$-vectors needed to represent a PWLC (\thref{thm:osposg:hv-pwlc}) function $V_i$.
According to \thref{thm:osposg:unique-fixpoint} this sequence converges to $V^*$:

\begin{corollary}
    Starting from an arbitrary PWLC value function $V_0$, a repeated application of the LP \eqref{eq:osposg:max-composition-lp}, as described in \thref{thm:lp-vertices}, converges to $V^*$.
\end{corollary}

A more efficient algorithm can be devised based on, e.g., the linear support algorithm for POMDPs~\citep{cheng1988-pomdp-algorithms}.
Here, the set $\Gamma'$ of linear functions defining $HV$ is constructed incrementally, terminating once it is provably large enough to represent the value function $HV$.
Exact value iteration algorithms to solve POMDPs are, however, generally considered to only be capable of solving very small problems.
We cannot, therefore, expect a decent performance of such approaches when solving one-sided POSGs that are more general than POMDPs.
The next section remedies this issue by providing a point-based approach for solving one-sided POSGs

\section{Heuristic Search Value Iteration for OS-POSGs}
\label{sec:osposg:hsvi}

In this section, we provide a scalable algorithm for solving one-sided POSGs, inspired by the \emph{heuristic search value iteration} (HSVI) algorithm~\citep{smith2004-hsvi,smith2005-hsvi} for approximating value function of POMDPs (summarized in Section~\ref{sec:pomdps}).
Our algorithm approximates the convex optimal value function $V^*$ using a pair of piecewise linear and convex value functions $\uv$ (lower bound on $V^*$) and $\ov$ (upper bound on $V^*$).
These bounds are refined over time and, given the initial belief $b^{\mathrm{init}}$ and the desired precision $\varepsilon > 0$, the algorithm is guaranteed to approximate the value $V^*(b^{\mathrm{init}})$ within $\varepsilon$.
In Section~\ref{sec:osposg:playing}, we show that this process also generates value functions that allow us to extract $\varepsilon$-Nash equilibrium strategies of the game.

We first show the approximation schemes used to represent $\uv$ and $\ov$, and the methods to initialize these bounds (Section~\ref{sec:osposg:hsvi:vf}).
We then discuss the so-called ``point-based updates'' which are used to refine the bounds induced by $\uv$ and $\ov$ (Section~\ref{sec:osposg:hsvi:pb-updates}).
Finally, in Section~\ref{sec:osposg:hsvi:alg}, we describe the algorithm and prove its correctness.

\subsection{Value Function Representations}
\label{sec:osposg:hsvi:vf}
Following the results on POMDPs and the original HSVI algorithm~\citep{hauskrecht2000-value-functions,smith2004-hsvi}, we use two distinct methods to represent upper and lower PWLC bounds on $V^*$.

\paragraph{Lower bound $\uv$}
Similarly as in the previous sections, the lower bound $\uv: \Delta(S) \rightarrow \R$ is represented as a point-wise maximum over a finite set $\Gamma$ of linear functions called $\alpha$-vectors, i.e., $\uv(b) = \max_{\alpha \in \Gamma} \alpha(b)$.
Each $\alpha \in \Gamma$ is a linear function $\alpha: \Delta(S) \rightarrow \R$ represented by its values $\alpha(s)$ in the vertices of the $\Delta(S)$ simplex, i.e., $\alpha(b) = \sum_{s \in S} b(s) \cdot \alpha(s)$.

\paragraph{Upper bound $\ov$}
Upper bound $\ov: \Delta(S) \rightarrow \R$ is represented as a lower convex hull of a set of points $\Upsilon = \lbrace (b_i, y_i) \mid 1 \leq i \leq k \rbrace$.
Each point $(b_i,y_i) \in \Upsilon$ provides an upper bound $y_i$ on the value $V^*(b_i)$ in belief $b_i$, i.e., $y_i \geq V^*(b_i)$.
Since the value function $V^*$ is convex, it holds that
\begin{align}
    \left( \forall \lambda \in \R^k_{\geq 0} \ \textnormal{s.t. } \sum_{i=1}^k \lambda_i = 1 \right) \ : \ V^* \left( \sum_{i=1}^k \lambda_i b_i \right) \leq \sum_{i=1}^k \lambda_i \cdot V^*(b_i) \leq \sum_{i=1}^k \lambda_i \cdot y_i .
\end{align}
This fact is used in the first variant of the HSVI algorithm (HSVI1~\citep{smith2004-hsvi}) to obtain the value of the upper bound $V_{\mathrm{HSVI1}}^\Upsilon(b)$ for belief $b$:
A linear program can be used to find coefficients $\lambda \in \R^k_{\geq 0}$ such that $b = \sum_{i=1}^k \lambda_i \cdot y_i$ holds and $\sum_{i=1}^k \lambda_i \cdot y_i$ is minimal:
\begin{equation}
    V_{\mathrm{HSVI1}}^\Upsilon(b) = \min \left\lbrace \sum_{i=1}^k \lambda_i y_i \mid \lambda \in \R_{\geq 0}^k: \sum_{i=1}^k \lambda_i = 1 \ \wedge \ \sum_{i=1}^k \lambda_i b_i = b \right\rbrace \ \text{,}
    \label{eq:osposg:ub-hsvi}
\end{equation}
In the latter proof of the \thref{thm:osposg:correctness} showing the correctness of the algorithm, we require the bounds $\uv$ and $\ov$ to be $\delta$-Lipschitz continuous.
Since this needs not hold for $V_{\mathrm{HSVI1}}^\Upsilon$, we define $\ov$ as a lower $\delta$-Lipschitz envelope of $V_{\mathrm{HSVI1}}^\Upsilon$:
\begin{equation}
    \ov(b) = \min_{b' \in \Delta(S)} \left[ V_{\mathrm{HSVI1}}^\Upsilon(b') + \delta \| b - b' \|_1 \right] \ \text{.} \label{eq:osposg:projection}
\end{equation}
This computation can be expressed as a linear programming problem
\begin{subequations}
\label{lp:osposg:projection}
\begin{align}
    \ov(b) = \min_{\lambda,\Delta,b'} \ & \sum_{i=1}^k \lambda_i y_i + \delta \sum_{s \in S} \Delta_s  \\
      \text{s.t.} \ & \sum_{i=1}^k \lambda_i b_i(s) = b'(s) & \forall s \in S \\
                    & \Delta_s \geq b'(s) - b(s) & \forall s \in S \\
                    & \Delta_s \geq b(s) - b'(s) & \forall s \in S \\
                    & \sum_{i=1}^k \lambda_i = 1 \\
                    & \lambda_i \geq 0 & \forall 1 \leq i \leq k
\end{align}
\end{subequations}
Here, we have $\Delta_s=|b'(s) - b(s)|$ (and hence $\sum_{s \in S} \Delta_s = \| b - b' \|_1$).
Using the definitions of $\ov$ and $V_{\mathrm{HSVI1}}^\Upsilon$ together with the fact that $V^*$ is $\delta$-Lipschitz continuous and convex, we can prove that the function $\ov$ represents an upper bound on $V^*$:

\begin{restatable}{lemma}{UBIsUpperBound}
\label{thm:osposg:projection}
    Let $\Upsilon = \lbrace (b_i,y_i) \,|\, 1 \leq i \leq k \rbrace$ such that $y_i \geq V^*(b_i)$ for every $1 \leq i \leq k$.
    Then the value function $\ov$ is $\delta$-Lipschitz continuous and satisfies
    \begin{equation*}
    V^* \leq \ov \leq V_{\mathrm{HSVI1}}^\Upsilon.
    \end{equation*}
\end{restatable}

The dichotomy in representation of value functions $\uv$ and $\ov$ allows for easy refinement of the bounds.
By adding new elements to the set $\Gamma$, the value $\uv(b) = \max_{\alpha \in \Gamma} \alpha(b)$ can only increase---and hence the lower bound $\uv$ gets tighter.
Similarly, by adding new elements to the set of points $\Upsilon$, the solution of linear program~\eqref{lp:osposg:projection} can only decrease and hence the upper bound $\ov$ tightens.

\subsubsection{Initial Bounds}

We now describe our approach to obtaining the initial bounds $\uv$ and $\ov$ on the optimal value function $V^*$ of the game:

\paragraph{Lower bound $\uv$}
We initially set the lower bound to the value $\val^{\sigma_1^{\mathrm{unif}}}$ of the uniform strategy $\sigma_1^{\mathrm{unif}} \in \Sigma_1$ of player~1 (i.e., the strategy that plays every action with probability $1/|A_1|$ in all stages of the game).
Recall that the value $\val^{\sigma_1^{\mathrm{unif}}}$ of the strategy $\sigma_1^{\mathrm{unif}}$ is a linear function (see \thref{thm:osposg:strategy-value-linear}), and hence the initial lower bound $\uv$ is a piecewise linear and convex function represented as a pointwise maximum of the set $\Gamma = \lbrace \val^{\sigma_1^{\mathrm{unif}}} \rbrace$.

\paragraph{Upper bound $\ov$}
We use the solution of a perfect information variant of the game (i.e., where player~1 is assumed to know the entire history of the game, unlike in the original game).
We form a modified game $G'$ which is identical to the OS-POSG $G$ (i.e., has the same states $S$, actions $A_1$ and $A_2$, dynamics $T$ and rewards $R)$, except that all information is revealed to player~1 in each step.
$G'$ is a perfect information stochastic game, and we can apply the value iteration algorithm to solve $G'$~\cite{bowling2000-sg}.
The additional information player~1 in $G'$ (compared to $G$) can only increase the utility he can achieve.
Hence $V^*_s$ of the state $s$ of game $G'$ forms an upper bound on the utility player~1 can achieve in $G$ if he knew that the initial state of the game is $s$ (i.e., his belief is $b_s$ where $b_s(s)=1$).
We initially define $\Upsilon$ as the set that contains one point for each state $s \in S$ of the game (i.e., for each vertex of the $\Delta(S)$ simplex),
\begin{equation}
    \Upsilon = \lbrace (b_s, V^*_s) \;|\; s \in S \rbrace \qquad\qquad
    b_s(s') = \begin{cases}
        1 & s = s' \\
        0 & \text{otherwise} \ \text{.}
    \end{cases}
    \label{eq:osposg:upsilon-initial}
\end{equation}

\subsection{Point-based Updates}
\label{sec:osposg:hsvi:pb-updates}

Unlike the exact value iteration algorithm (Section~\ref{sec:osposg:vi}) which constructs all $\alpha$-vectors needed to represent $HV$ in each iteration, the HSVI algorithm focuses on a single belief at a time.
Performing a \emph{point-based} update in belief $b \in \Delta(S)$ corresponds to solving the stage-games $[H\uv](b)$ and $[H\ov](b)$ where the values of subsequent stages are represented using value functions $\uv$ and $\ov$, respectively.

\paragraph{Update of lower bound $\uv$}
First, the LP~\eqref{eq:osposg:max-composition-lp} is used to compute the optimal value composition $\mathsf{valcomp}(\pi_1^{\mathrm{LB}},\overline{\alpha}^{\mathrm{LB}})$ in $[H\uv](b)$, i.e.,
\begin{equation}
    (\pi_1^{\mathrm{LB}},\overline{\alpha}^{\mathrm{LB}}) \ \ \ \ = \argmax_{\substack{\pi_1 \in \Pi_1 \\ \overline{\alpha} \in \mathsf{Conv}(\Gamma)^{A_1 \times O}}} \mathsf{valcomp}(\pi_1,\overline{\alpha})(b) \ \text{.}
    \label{eq:osposg:pbupdate-lb}
\end{equation}
The $\mathsf{valcomp}(\pi_1^{\mathrm{LB}},\overline{\alpha}^{\mathrm{LB}})$ function is a linear function corresponding to a new $\alpha$-vector that forms a lower bound on $V^*$.
This new $\alpha$-vector is used to refine the bound by setting $\Gamma \coloneqq \Gamma \cup \lbrace \mathsf{valcomp}(\pi_1^{\mathrm{LB}},\overline{\alpha}^{\mathrm{LB}}) \rbrace$.
As the following lemma shows, refining the lower bound $\uv$ via a point-based update preserves its desirable properties:
\begin{restatable}{lemma}{LBUpdatesPreserveStuff}
\label{thm:osposg:lb-point}
    The lower bound $\uv$ initially satisfies the following conditions, which are subsequently preserved during point-based updates:
    \begin{compactenum}[(1)]
        \item $\uv$ is $\delta$-Lipschitz continuous.
        \item $\uv$ is lower bound on $V^*$.
    \end{compactenum}
\end{restatable}

\paragraph{Update of upper bound $\ov$}
Similarly to the case of the point-based update of the lower bound $\uv$, the update of upper bound is performed by solving the stage game $[H\ov](b)$.
Since $\ov$ is represented by a set of points $\Upsilon$, it is not necessary to compute the optimal value composition.
Instead, we form a refined upper bound $V_{\mathrm{UB}}^{\Upsilon'}$ (which corresponds to $\ov$ after the point-based update is made) by adding a new point $(b, [H\ov](b))$ to the set $\Upsilon'$ representing $V_{\mathrm{UB}}^{\Upsilon'}$, i.e., $\Upsilon' = \Upsilon \cup \lbrace (b, [H\ov](b)) \rbrace$.
We now show that the upper bound $\ov$ has the desired properties, and these properties are retained when applying the point-based update---and hence we can perform point-based updates of $\ov$ repeatedly.

\begin{restatable}{lemma}{UBPreservesStuff}
\label{thm:osposg:ub-point}
    The upper bound $\ov$ initially satisfies the following conditions, which are subsequently preserved during point-based updates:
    \begin{compactenum}[(1)]
        \item $\ov$ is $\delta$-Lipschitz continuous.
        \item $\ov$ is an upper bound on $V^*$.
    \end{compactenum}
\end{restatable}

The LPs~\eqref{eq:osposg:max-composition-lp} and~\eqref{eq:osposg:max-composition-dual} solve the stage game $[HV](b)$ when the value function $V$ is represented as a maximum over a set of linear functions (i.e., the way lower bound $\uv$ is).
It is, however, possible to adapt constraints in~\eqref{eq:osposg:max-composition-dual} to solve the $[H\ov](b)$ problem.
We replace constraint~\eqref{eq:osposg:max-composition-dual:subgame} by constraints inspired by the LP~\eqref{lp:osposg:projection} used to solve $\ov(b)$.
\begin{subequations}
\label{eq:osposg:dual-projection}
\begin{align}
    & \hat{V}(a_1,o) = \sum_{i=1}^{|\Upsilon|} \lambda^{a_1,o}_i y_i + \delta \sum_{s' \in S} \Delta^{s'}_{a_1,o} & \forall (a_1,o) \in A_1 \times O \\
    & \sum_{i=1}^{|\Upsilon|} \lambda^i_{a_1,o} b_i(s') = b'_{a_1,o}(s') & \forall (a_1,o,s') \in A_1 \times O \times S \\
    & \Delta^{s'}_{a_1,o} \geq b'_{a_1,o}(s') - \hat{\tau}(b,a_1,\pi_2,o)(s') & \forall (a_1,o,s') \in A_1 \times O \times S \\
    & \Delta^{s'}_{a_1,o} \geq \hat{\tau}(b,a_1,\pi_2,o)(s') - b'_{a_1,o}(s') & \forall (a_1,o,s') \in A_1 \times O \times S \\
    & \sum_{i=1}^{|\Upsilon|} \lambda^{a_1,o}_i = \sum_{s' \in S} \hat{\tau}(b,a_1,\pi_2,o)(s') & \forall (a_1,o) \in A_1 \times O \label{eq:osposg:dual-projection:lambda-sum}\\
    & \lambda^i_{a_1,o} \geq 0 & \forall (a_1,o) \in A_1 \times O, 1 \leq i \leq |\Upsilon|
\end{align}
\end{subequations}

\subsection{The Algorithm}
\label{sec:osposg:hsvi:alg}
We are now ready to present the heuristic search value iteration (HSVI) algorithm for one-sided POSGs (Algorithm~\ref{alg:osposg:hsvi}) and prove its correctness.
The algorithm is similar to the HSVI algorithm for POMDPs~\citep{smith2004-hsvi,smith2005-hsvi}.
First, the bounds $\uv$ and $\ov$ on the optimal value function $V^*$ are initialized (as described in Section~\ref{sec:osposg:hsvi:vf}) on line~\ref{alg:osposg:hsvi:initialization}.
Then, until the desired precision $\ov(b^{\mathrm{init}}) - \uv(b^{\mathrm{init}}) \leq \varepsilon$ is reached, the algorithm performs a sequence of trials using the $\mathtt{Explore}$ procedure, starting from the initial belief $b^{\mathrm{init}}$ (lines~\ref{alg:osposg:hsvi:termination}--\ref{alg:osposg:hsvi:exploration}).

\begin{algorithm}[t]
\caption{HSVI algorithm for one-sided POSGs}
\label{alg:osposg:hsvi}
\KwData{Game $G$, initial belief $b^{\mathrm{init}}$, discount factor $\gamma \in (0,1)$, desired precision $\varepsilon > 0$, neighborhood parameter $D$}
\KwResult{Approximate value functions $\uv$ and $\ov$ satisfying $\ov(b) - \uv(b) \leq \varepsilon$, sets $\Gamma$ and $\Upsilon$ constructed by point-based updates that represent $\uv$ and $\ov$}
\SetKwFunction{Explore}{Explore}
\SetKwProg{myproc}{procedure}{}{}
\DontPrintSemicolon
Initialize $\uv$ and $\ov$ (see Section~\ref{sec:osposg:hsvi:vf}) \label{alg:osposg:hsvi:initialization}\;
\While{$\excess_0(b^{\mathrm{init}}) > 0$}{\label{alg:osposg:hsvi:termination}
  \Explore{$b^{\mathrm{init}},0$} \label{alg:osposg:hsvi:exploration}\;
}
\Return{\normalfont $\uv$ and $\ov$, sets $\Gamma$ and $\Upsilon$ that represent $\uv$ and $\ov$}\;
\vspace{0.3em}
\myproc{\Explore{$b_t,t$}}{\label{alg:osposg:hsvi:explore}
  $(\pi_1^{\mathrm{LB}},\pi_2^{\mathrm{LB}}) \gets $ equilibrium strategy profile in $[H\uv](b_t)$ \label{alg:osposg:hsvi:follower-choice}\;
  $(\pi_1^{\mathrm{UB}},\pi_2^{\mathrm{UB}}) \gets $ equilibrium strategy profile in $[H\ov](b_t)$ \label{alg:osposg:hsvi:leader-choice}\;
  Perform point-based updates of $\uv$ and $\ov$ at belief $b_t$ (see Section~\ref{sec:osposg:hsvi:pb-updates}) \label{alg:osposg:hsvi:pb-update-1} \;
  $(a_1^*,o^*) \gets $ select according to forward exploration heuristic \label{alg:osposg:hsvi:fwd-exploration}\;
  \If{$\Pr_{b,\pi_1^{\mathrm{UB}},\pi_2^{\mathrm{LB}}}[a_1^*,o^*] \cdot \excess_{t+1}(\tau(b_t,a_1^*,\pi_2^{\mathrm{LB}},o^*)) > 0$ \label{alg:osposg:hsvi:recursion-termination}}{
      \Explore{$\tau(b_t,a_1^*,\pi_2^{\mathrm{LB}},o^*),t+1$} \label{alg:osposg:hsvi:recursion}\;
      Perform point-based updates of $\uv$ and $\ov$ at belief $b_t$ (see Section~\ref{sec:osposg:hsvi:pb-updates}) \label{alg:osposg:hsvi:pb-update-2}\;
  }
}
\end{algorithm}

The recursive procedure $\mathtt{Explore}$ generates a sequence of beliefs $\lbrace b_i \rbrace_{i=0}^k$ (for some $k \geq 0$) where $b_0 = b^{\mathrm{init}}$ and each belief $b_t$ reached at the recursion depth $t$ satisfied $\excess_t(b_t) > 0$ on line~\ref{alg:osposg:hsvi:termination} or~\ref{alg:osposg:hsvi:recursion-termination}.
The algorithm tries to ensure that values of beliefs $b_t$ reached at $t$-th level of recursion (i.e., $t$-th stage of the game) are approximated with sufficient accuracy and the gap between $\ov(b)$ and $\uv(b)$ is at most $\rho(t)$, where $\rho(t)$ is defined by
\begin{equation}
	\rho(0) = \varepsilon \qquad\qquad \rho(t+1) = [ \rho(t) - 2 \delta D ] / \gamma \ \text{.} \label{eq:osposg:rho}
\end{equation}
To ensure that the sequence $\rho$ is monotonically increasing and unbounded, we need to select the parameter $D$ from the interval $(0 , (1-\gamma)\varepsilon / 2\delta )$.
When the approximation quality $\ov(b_t) - \uv(b_t)$ of the value of a belief $b_t$ reached at the $t$-th recursion level of $\mathtt{Explore}$ (i.e., at the $(t+1)$-th stage of the game) exceeds the desired approximation quality $\rho(t)$, it is said to have a positive \emph{excess gap} $\excess_t(b_t)$,
\begin{equation}\label{eq:onesided:excess}
	\excess_t(b_t) = \ov(b_t) - \uv(b_t) - \rho(t) \ \text{.}
\end{equation}
Note that our definition of excess gap is more strict compared to the original HSVI algorithm for POMDPs, where the $-2 \delta D$ term from Equation~\eqref{eq:osposg:rho} is absent (see Equation~\eqref{eq:hsvi:excess}).
Unlike in POMDPs, which are single-agent, the belief transitions $\tau(b,a_1,\pi_2,o)$ in one-sided POSGs depend on player~2 as well (resp., on her strategy $\pi_2$).
The tighter bounds on the approximation quality allow us to prove the correctness of the proposed algorithm in \thref{thm:osposg:correctness}.

\paragraph{Forward exploration heuristic}
The algorithm uses a heuristic approach to select which belief $\tau(b,a_1,\pi_2^{\mathrm{LB}},o)$ will be considered in the next recursion level of the $\mathtt{Explore}$ procedure, i.e., what action-observation pair $(a_1,o) \in A_1 \times O$ will be chosen by player~1, on line~\ref{alg:osposg:hsvi:fwd-exploration}.
This selection is motivated by Lemma~\ref{thm:osposg:point-contractivity}---in order to ensure that $\excess_t(b_t) \leq 0$ (or more precisely $\excess_t(b_t) \leq -2 \delta D$) at the currently considered belief $b_t$ in $t$-th recursion level, all beliefs $\tau(b_t,a_1,\pi_2^{\mathrm{LB}},o)$ reached with positive probability when playing $\pi_1^{\mathrm{UB}}$ have to satisfy $\excess_{t+1}(\tau(b_t,a_1,\pi_2^{\mathrm{LB}},o)) \leq 0$.
Specifically, we focus on a belief that has the highest \emph{weighted excess gap}.
Inspired by the original HSVI algorithm for POMDPs~\citep{smith2004-hsvi,smith2005-hsvi}), we define the weighted excess gap as the excess gap $\excess_{t+1}(\tau(b_t,a_1,\pi_2^{\mathrm{LB}},o))$ multiplied by the probability that the action-observation pair $(a_1,o)$ that leads to the belief $\tau(b_t,a_1,\pi_2^{\mathrm{LB}},o)$ occurs.
As a result, the next action-observation pair $(a_1^*,o^*)$ for further exploration is selected according to the formula
\begin{equation}
	(a_1^*,o^*) = \argmax_{(a_1,o) \in A_1 \times O} \Pr_{b,\pi_1^{\mathrm{UB}},\pi_2^{\mathrm{LB}}}[a_1,o] \cdot \excess_{t+1}(\tau(b_t,a_1,\pi_2^{\mathrm{LB}},o)) \ \text{.}
\end{equation}
We now show formally that if the weighted excess gap of the optimal $(a_1^*,o^*)$ satisfies $\Pr_{b,\pi_1^{\mathrm{UB}},\pi_2^{\mathrm{LB}}}[a_1^*,o^*] \cdot \excess_{t+1}(\tau(b_t,a_1^*,\pi_2^{\mathrm{LB}},o^*)) \leq 0$, performing the point based update at $b_t$ ensures that $\excess_t(b_t) \leq -2 \delta D$.

\begin{restatable}{lemma}{ExcessContractivity}
\label{thm:osposg:hsvi:excess-contractivity}
	Let $b_t$ be a belief encountered at $t$-th recursion level of $\mathtt{Explore}$ procedure and assume that the corresponding action-observation pair $(a_1^*,o^*)$ (from line~\ref{alg:osposg:hsvi:fwd-exploration} of Algorithm~\ref{alg:osposg:hsvi}) satisfies
	\begin{equation}
		\Pr_{b,\pi_1^{\mathrm{UB}},\pi_2^{\mathrm{LB}}}[a_1^*,o^*] \cdot \excess_{t+1}(\tau(b_t,a_1^*,\pi_2^{\mathrm{LB}},o^*)) \leq 0 \ \text{.}
	\end{equation}
	Then $\excess_t(b_t) \leq -2 \delta D$ after performing a point-based update at $b_t$.
	Furthermore, all beliefs $b_t' \in \Delta(S)$ such that $\| b_t - b_t' \|_1 \leq D$ satisfy $\excess_t(b_t') \leq 0$.
\end{restatable}

The proof goes by verifying the assumptions of Lemma~\ref{thm:osposg:point-contractivity} (``a criterion for contractivity''), which allows us to bound the difference between $\ov$ and $\uv$ by $\rho(t+1)$. The ``furthermore'' part then follows from $\delta$-Lipschitz continuity of the bounds.

We now use Lemma~\ref{thm:osposg:hsvi:excess-contractivity} (especially its second part) to prove that Algorithm~\ref{alg:osposg:hsvi} terminates with $\ov(b^{\mathrm{init}}) - \uv(b^{\mathrm{init}}) \leq \varepsilon$.
As we mentioned earlier, we can also use value functions $\uv$ and $\ov$ to play the game and obtain $\varepsilon$-Nash equilibrium of the game (see Section~\ref{sec:osposg:playing}).

\begin{theorem}\thlabel{thm:osposg:correctness}
	For any $\varepsilon > 0$ and $0 < D < (1-\gamma)\varepsilon / 2\delta$, Algorithm~\ref{alg:osposg:hsvi} terminates with $\ov(b^{\mathrm{init}}) - \uv(b^{\mathrm{init}}) \leq \varepsilon$.
\end{theorem}
\begin{proof}
	By the choice of parameter $D$, the sequence $\rho(t)$ (for $\rho(0)=\varepsilon$) is monotonically increasing and unbounded, and the difference between value functions $\uv$ and $\ov$ is bounded by $U-L$ (since $L \leq \uv(b) \leq \ov(b) \leq U$ for every belief $b \in \Delta(S)$).
	Therefore, there exists $T_{\mathrm{max}}$ such that $\rho(T_{\mathrm{max}}) \geq U - L \geq \ov(b) - \uv(b)$ for every $b \in \Delta(S)$, so the recursive procedure $\mathtt{Explore}$ always terminates.
	
	To prove that the whole algorithm terminates, we reason about sets $\Psi_t \subset \Delta(S)$ of belief points where the trials performed by the $\mathtt{Explore}$ terminated.
	Initially, $\Psi_t = \emptyset$ for every $0 \leq t < T_{\mathrm{max}}$.
	Whenever the $\mathtt{Explore}$ recursion terminates at recursion level $t$ (i.e., the condition on line~\ref{alg:osposg:hsvi:recursion-termination} does not hold), the belief $b_t$ (which was the last belief considered during the trial) is added into set $\Psi_t$ ($\Psi_t \coloneqq \Psi_t \cup \lbrace b_t \rbrace$).
	Recall that since $\Delta(S)$ is compact, it is, in particular, totally bounded (that is, if any two distinct elements $b,b'$ of a set $\Psi_t \subset \Delta(S)$ satisfy $\| b - b' \|_1 > D$, the set $\Psi_t$ must be finite).
	Since the number of possible termination depths is finite ($0 \leq t \leq T_{\mathrm{max}}$), the algorithm has to terminate unless some $\Psi_t$ is infinite.
	To show that the algorithm terminates, it thus remains that every two distinct points $b, b' \in \Psi_t$ are at least $D$ apart.
	
	Assume to the contradiction that two trials terminated at recursion level $t$ with the last beliefs considered $b_t^{(1)}$ (for the earlier trial) and $b_t^{(2)}$ (for the trial that occurred at a later time), and that these beliefs satisfy $\| b_t^{(1)} - b_t^{(2)} \|_1 \leq D$.
	When the former trial has been terminated in belief $b_t^{(1)}$, all reachable beliefs from $b_t^{(1)}$ had a negative excess gap (otherwise the trial would have continued as the condition on line~\ref{alg:osposg:hsvi:recursion-termination} would have been satisfied).
	According to Lemma~\ref{thm:osposg:hsvi:excess-contractivity}, after the point-based update is performed in $b_t^{(1)}$, the excess gap of all beliefs $b_t'$ with $\| b_t^{(1)} - b_t' \|_1 \leq D$ have negative excess gap $\excess_t(b_t') \leq 0$.
	When $b_t^{(2)}$ has been selected for exploration in $(t-1)$-th level of recursion, the condition on line~\eqref{alg:osposg:hsvi:recursion-termination} was met and $b_t^{(2)}$ must have had positive excess gap $\excess_t(b_t^{(2)}) > 0$.
	This, however, contradicts the assumption that all beliefs $b_t'$ with $\| b_t^{(1)} - b_t' \|_1 \leq D$ (i.e., including $b_t^{(2)}$) already have negative excess gap.
	
	Now that we know that Algorithm~\ref{alg:osposg:hsvi} always terminates,
	note that at least one trial must have terminated in the first level of recursion (unless the Algorithm~\ref{alg:osposg:hsvi} has terminated on line~\ref{alg:osposg:hsvi:termination} with $\excess_0(b^{\mathrm{init}}) \leq 0$ beforehand).
	By Lemma~\ref{thm:osposg:hsvi:excess-contractivity}, the update in $b^{\mathrm{init}}$ then renders $\excess_0(b^{\mathrm{init}}) \leq -2\delta D \leq 0$.
	We then have that $\ov(b^{\mathrm{init}}) - \uv(b^{\mathrm{init}}) \leq \rho(0) = \varepsilon$ which completes the proof.
\end{proof}

\section{Using Value Function to Play}
\label{sec:osposg:playing}

In the previous section, we have presented an algorithm that can approximate the value $V^*(b^{\mathrm{init}})$ of the game within an arbitrary given precision $\varepsilon > 0$ starting from an arbitrary initial belief $b^{\mathrm{init}}$.
However, in many games, knowing only the game's value is not enough. Indeed, to solve the game, we also need access to strategies that achieve the desired near-optimal performance.
In this section, we show that using the value functions $\uv$ and $\ov$ computed by the HSVI algorithm (Algorithm~\ref{alg:osposg:hsvi}) enables us to obtain $\varepsilon$-Nash equilibrium strategies for both players.

The Bellman's equation from Theorem~\ref{thm:osposg:bellman} may suggest that the near-optimal strategies can be extracted by employing the lookahead decision rule (similarly to POMDPs) and obtaining strategies to play in the current stage by computing the Nash equilibrium of stage games $[H\uv](b)$ and $[H\ov](b)$, respectively.
However, unlike in POMDPs and Markov games of imperfect information, this approach does \emph{not} work in one-sided POSGs because the belief of player~1 does not constitute a sufficient statistic for playing the game.
The reasons for this are similar to the usage of unsafe resolving~\citep{burch2018-thesis,seitz2019-value} in the realm of extensive-form games.
We use the following example to demonstrate the insufficiency of the belief to play the game.

\begin{example}
Consider a \emph{matching pennies} game shown in Figure~\ref{fig:belief-insufficient:nfg}.
This game can be formalized as a one-sided POSG that is shown in Figure~\ref{fig:belief-insufficient:osposg}.
The game starts in the state $s_0$ (i.e., the initial belief is $b^{\mathrm{init}}(s_0)=1$) and player~2 chooses her action $H$ or $T$.
Next, after transitioning to $s_H$ or $s_T$ (based on the decision of player~2), player~1 is \emph{unaware} of the true state of the game (i.e., the past decision of player~1) and chooses his action $H$ or $T$.
Based on the combination of decisions taken by the players, player~1 gets either $1/\gamma$ or $-1/\gamma$ and the game transitions to the state $s_\infty$ where it stays forever with zero future rewards.
\begin{figure}
    \centering
    \begin{subfigure}{0.35\linewidth}
        \begin{tabular}{c||r|r}
              &  H &  T \\
            \hline\hline
            H &  1 & -1 \\
            \hline
            T & -1 &  1
        \end{tabular}
        \caption{Normal form}
        \label{fig:belief-insufficient:nfg}
    \end{subfigure}
    \begin{subfigure}{0.5\linewidth}
        \includegraphics[scale=1]{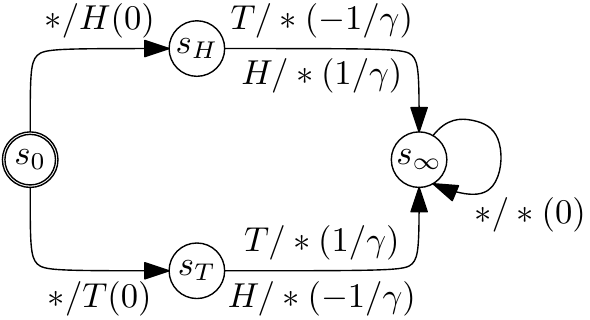}
        \caption{OS-POSG representation}
        \label{fig:belief-insufficient:osposg}
    \end{subfigure}
    \\
    \begin{subfigure}{0.45\linewidth}
        \includegraphics[scale=1]{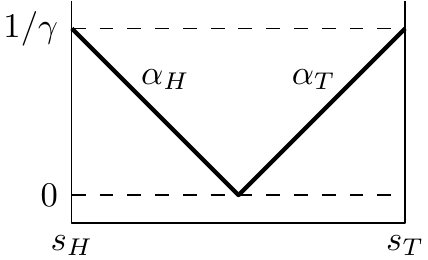}
        \caption{Value function $V^*$}
        \label{fig:belief-insufficient:value}
    \end{subfigure}
    \caption{A game where belief is not a sufficient statistic for the imperfectly informed player.}
    \label{fig:belief-insufficient}
\end{figure}

To understand the caveats of using belief $b \in \Delta(S)$ to derive the stage strategy to play, let us consider the optimal value function $V^*$ of the OS-POSG representation (Figure~\ref{fig:belief-insufficient:osposg}) of the matching pennies game.
Figure~\ref{fig:belief-insufficient:value} shows the values of $V^*$ over simplex $\Delta(\lbrace s_H, s_T \rbrace)$.
If it is more likely that the player~2 played $H$ in the first stage of the game (i.e., the current state is $s_H$), it is optimal for player~1 to play strategy prescribing him to play $H$ in the current stage (with value $\alpha_H$).
Conversely, if it is more likely that the current state is $s_T$, player~1 is better off with playing $T$ (with value $\alpha_T$).
The value function $V^*$ is then a point-wise maximum over these two linear functions.

Now, since the uniform mixture between $H$ and $T$ is the Nash equilibrium strategy for both players in the matching pennies game, player~1 will find himself in a situation when he assumes that the current belief is $\lbrace s_H: 0.5, s_T: 0.5 \rbrace$.
In this belief, any decision of player~1 yields expected reward 0---hence based purely on the belief, player~1 may opt to play, e.g., ``always $T$''.
However, such strategy is not in equilibrium and player~2 is able to exploit it by playing ``always $H$''.
This example illustrates that the belief alone does not provide sufficient information to choose the right strategy $\pi_1$ for the current stage based on the Equation~\eqref{eq:osposg:H-maxmin}.
\end{example}

\subsection{Justified Value Functions}

First of all, we define conditions under which it makes sense to use value function to play a one-sided POSG.
The conditions are similar to \emph{uniform improvability} in, e.g., POMDPs.
Our definitions, however, reflect the fact that we deal with a two-player problem (and we thus introduce the condition for each player separately).
Moreover, we use a stricter condition for player~1 who does \emph{not} have perfect information about the belief---and thus defining the condition based solely on the beliefs is not sufficient.

\begin{definition}[Min-justified value function]
\thlabel{def:osposg:min-justified}
    Convex continuous value function $V$ is said to be \emph{min-justified} (or, justified from the perspective of the minimizing player~2) if for every belief $b \in \Delta(S)$ it holds that $[HV](b) \leq V(b)$.
\end{definition}

\begin{definition}[Max-justified value function]
\thlabel{def:osposg:max-justified}
    Let $\Gamma$ be a compact set of linear functions, and $V$ be a value function such that $V(b) = \sup_{\alpha \in \Gamma} \alpha(b)$ for every $b$.
    $V$ is said to be \emph{max-justified} by $\Gamma$ (or, justified from the perspective of the maximizing player~1) if for every $\alpha \in \Gamma$ there exists $\pi_1 \in \Pi_1$ and $\overline{\alpha} \in \Gamma^{A_1 \times O}$ such that $\mathsf{valcomp}(\pi_1,\overline{\alpha}) \geq \alpha$.
\end{definition}

While the reason for the terminology is not apparent just yet, we will show in Sections~\ref{sec:osposg:p1-strategy} that the ``max-justifying'' set $\Gamma$ can be used to construct a strategy $\sigma_1$ of player~1 such that $\val^{\sigma_1}(b) \geq V(b)$ for every $b$.
Similarly, we will show in Section~\ref{sec:osposg:p2-strategy} that if the value function $V$ is min-justified, we can construct a strategy $\sigma_2$ of player~2 that \emph{justifies} the value $V(b)$ for every belief $b \in \Delta(S)$, i.e., we have $\E_{b,\sigma_1,\sigma_2}[\Disc^\gamma] \leq V(b)$ against every strategy $\sigma_1$ of player~1.

As preparation for more substantial proofs that follow, the remainder of this subsection presents several basic properties of min- and max-justified functions.

Recall that no matter how well things go for the maximizing player, the corresponding utility will never get above $U$. Similarly, the minimizing player cannot push the utility below $L$.
Lemma~\ref{thm:osposg:max-justified-bounded} and Lemma~\ref{thm:osposg:min-justified-bounded} prove that max- and min-justified functions obey the same restrictions.
This is in agreement with our intuition that max-justification should guarantee utility of \textit{at least} some value (which therefore cannot be higher than $U$) and min-justification should guarantee utility of \textit{no more than} some value (which therefore cannot be lower than $L$).

\begin{restatable}{lemma}{MinJustified}
\label{thm:osposg:min-justified-bounded}
    Let $V$ be a value function that is min-justified.
    Then $V(b) \geq L$.
\end{restatable}

\begin{restatable}{lemma}{MaxJustified}
\label{thm:osposg:max-justified-bounded}
    Let $V$ be a value function that is max-justified by a set of $\alpha$-vectors $\Gamma$.
    Then for every $\alpha \in \Gamma$ we have $\alpha \leq U$.
\end{restatable}

To prepare for showing that the value function $\uv$ resulting produced by Algorithm~\ref{alg:osposg:hsvi} is max-justified by $\mathsf{Conv}(\Gamma)$, we state the following technical lemma:

\begin{restatable}{lemma}{ConvMaxJustified}
\label{thm:osposg:conv-max-justified}
  Let $\Gamma$ be a set of linear functions, and $V$ a value function that is max-justified by $\Gamma$.
  Then $V$ is also max-justified by $\mathsf{Conv}(\Gamma)$.
\end{restatable}

\subsection{Strategy of Player~1}
\label{sec:osposg:p1-strategy}
In this section, we will show that when the value function $V$ is max-justified by a set of $\alpha$-vectors $\Gamma$, we can implicitly form a strategy $\sigma_1$ of player~1 that achieves utility of at least $V(b^{\mathrm{init}})$ for any given initial belief $b^{\mathrm{init}}$.
We provide an online game-playing algorithm (Algorithm~\ref{alg:osposg:cr}) which implicitly constructs the desired strategy. This algorithm is inspired by the ideas of continual resolving for extensive-form games~\citep{moravcik2017-deepstack}.

While playing the game, Algorithm~\ref{alg:osposg:cr} maintains a lower bound $\rho$ on the values the reconstructed strategy has to achieve.
Inspired by the terminology of continual resolving for extensive-form games, we call this lower-bounding linear function a \emph{gadget}.
The goal of the $\mathtt{Act}(b,\rho)$ method is to reconstruct a strategy $\sigma_1$ of player such that its value satisfies $\val^{\sigma_1} \geq \rho$.
We will now show that the $\mathtt{Act}$ method achieves precisely this.
The reasoning about the current gadget allows us to obtain guarantees on the quality of the reconstructed strategy, even when player~1 does not have an accurate belief because he does not have access to the stage strategies used by the adversary.

\begin{algorithm}
    \caption{Continual resolving algorithm for one-sided POSGs}
    \label{alg:osposg:cr}
    \DontPrintSemicolon
    \SetKwBlock{Repeat}{repeat}{}
    \SetKwInOut{Input}{input}
    
    \Input{one-sided POSG $G$ \\ a finite set $\Gamma$ of linear functions representing convex value function $V$}
    \SetKwFunction{Act}{Act}
    \SetKwProg{myproc}{procedure}{}{}
    
    $b \gets b^{\mathrm{init}}$ \label{alg:osposg:cr:binit}\;
    $\rho^{\mathrm{init}} \gets \argmax_{\alpha \in \Gamma} \alpha(b_{\mathrm{init}})$ \label{alg:osposg:cr:rho-zero}\;
    \Act{$b^{\mathrm{init}}, \rho^{\mathrm{init}}$}
    
    \myproc{\Act{$b, \rho$}}{
      $(\pi_1^*,\overline{\alpha}^*) \gets \argmax_{\pi_1,\overline \alpha} \lbrace \mathsf{valcomp}(\pi_1,\overline{\alpha})(b) \mid \pi_1 \in \Pi_1, \overline{\alpha} \in \textsf{Conv}(\Gamma)^{A_1 \times O} \textnormal{ s.t. } \mathsf{valcomp}(\pi_1,\overline{\alpha}) \geq \rho \rbrace$ \label{alg:osposg:cr:resolve}\;
      $\pi_2 \gets$ solve $[HV](b)$ to obtain assumed stage strategy of the adversary \;
      sample and play $a_1 \sim \pi_1^*$ \;
      $o \gets$ observed observation \;
      $b' \gets \tau(b, a_1, \pi_2, o)$ \;
      \Act{$b', \alpha^*_{a_1, o}$} \;
    }
\end{algorithm}

\begin{restatable}{proposition}{PlOneStrategy}
\label{thm:osposg:p1-strategy}
    Let $V$ be a value function that is max-justified by a set of $\alpha$-vectors $\Gamma$.
    Let $b^{\mathrm{init}} \in \Delta(S)$ and $\rho^{\mathrm{init}} \in \Gamma$.
    By playing according to $\mathtt{Act}(b^{\mathrm{init}},\rho^{\mathrm{init}})$, player~1 implicitly forms a strategy $\sigma_1$ for which $\val^{\sigma_1} \geq \rho^{\mathrm{init}}$.
\end{restatable}
This proposition is proven by constructing a sequence of strategies under which player~1 follows Algorithm~\ref{alg:osposg:cr} for $K$ steps (for $K=0,1,\ldots$).
We provide a lower bound on the value each of these strategies, and show that the limit of these lower bounds coincides with $\rho^{\mathrm{init}}$, as well as with the lower bound on the value guaranteed by following Algorithm~\ref{alg:osposg:cr} for \emph{infinite} period of time.

\begin{corollary}
\thlabel{thm:osposg:p1-strategy-guarantee}
    Let $V$ be a value function that is max-justified by a compact set $\Gamma$ and let $b^{\mathrm{init}}$ be the initial belief of the game.
    The Algorithm~\ref{alg:osposg:cr} implicitly constructs a strategy $\sigma_1$ which guarantees that the utility to player~1 will be at least $V(b^{\mathrm{init}})$.
\end{corollary}
\begin{proof}
    $\rho^{\mathrm{init}}$ from line~\ref{alg:osposg:cr:rho-zero} of Algorithm~\ref{alg:osposg:cr} has value $\rho^{\mathrm{init}}(b^{\mathrm{init}}) = V(b^{\mathrm{init}})$ in the initial belief $b^{\mathrm{init}}$.
    By Proposition~\ref{thm:osposg:p1-strategy}, we can construct a strategy $\sigma_1$ with value $\val^{\sigma_1} \geq \rho^{\mathrm{init}}$.
    Hence $\val^{\sigma_1}(b^{\mathrm{init}}) \geq \rho^{\mathrm{init}}(b^{\mathrm{init}}) =V(b^{\mathrm{init}})$.
\end{proof}

\subsection{Strategy of Player~2}
\label{sec:osposg:p2-strategy}
We will now present an analogous algorithm to obtain a strategy for player~2 when the value function $V$ is min-justified.
Recall that the stage strategies $\pi_2$ of player~2 influence the belief of player~1 (Equation~\ref{eq:osposg:tau}).
Unlike player~1, player~2 knows which stage strategies $\pi_2$ have been used in the past, and he is thus able to infer the current belief of player~1.
As a result, the $\mathtt{Act}$ method of Algorithm~\ref{alg:osposg:cr2} depends on the current belief of player~1, but not on the gadget $\rho$ as it did in Algorithm~\ref{alg:osposg:cr}.
\begin{algorithm}
    \caption{Strategy of player~2}
    \label{alg:osposg:cr2}
    \DontPrintSemicolon
    \SetKwBlock{Repeat}{repeat}{}
    \SetKwInOut{Input}{input}
    
    \Input{one-sided POSG $G$ \\ convex value function $V$}
    
    \SetKwFunction{Act}{Act}
    \SetKwProg{myproc}{procedure}{}{}
    
    \Act{$b^{\mathrm{init}}$}
    
    \myproc{\Act{$b$}}{
      $\pi_2^* \gets $ optimal strategy of player~2 in the stage game $[HV](b)$ \label{alg:osposg:cr2:resolve}\;
      $s \gets$ currently observed state \;
      sample and play $a_2 \sim \pi_2^*(\cdot \,|\, s)$ \;
      $(a_1,o) \gets$ action of the adversary and the corresponding observation\;
      \Act{$\tau(b, a_1, \pi_2^*, o)$} \;
    }
\end{algorithm}

We will now show that if the value function $V$ is min-justified, playing according to Algorithm~\ref{alg:osposg:cr2} guarantees that the utility will be at most\footnote{In other words, this is a performance guarantee for the (minimizing) player 2.} $V(b^{\mathrm{init}})$.

\begin{restatable}{proposition}{PlTwoStrategy}
\label{thm:osposg:p2-strategy}
    Let $V$ be a min-justified value function and let $b^{\mathrm{init}}$ be the initial belief of the game.
    The Algorithm~\ref{alg:osposg:cr2} implicitly constructs a strategy $\sigma_2$ which guarantees that the utility to player~1 will be at most $V(b^{\mathrm{init}})$.
\end{restatable}

The proof of Proposition~\ref{thm:osposg:p2-strategy} is similar to the proof of Proposition~\ref{thm:osposg:p1-strategy}.
We derive an upper bound on the utility player~1 can achieve against player~2 who follows Algorithm~\ref{alg:osposg:cr2} for $K$ steps (for $K=0,1,\ldots$).
We show that the limit of these upper bounds coincides with $V(b^{\mathrm{init}})$ and with the upper bound on the utility player~1 can achieve when player~2 follows \ref{alg:osposg:cr2} for an infinite number of iterations.

\subsection{Using Value Functions \texorpdfstring{$\uv$}{VLB} and \texorpdfstring{$\ov$}{VUB} to Play the Game}
\label{sec:osposg:playing-hsvi}

In Sections~\ref{sec:osposg:p1-strategy} and~\ref{sec:osposg:p2-strategy}, we have shown that we can obtain strategies to play the game when the value functions are max-justified or min-justified, respectively.
In this section, we will show that the heuristic search value iteration algorithm for solving one-sided POSGs (Section~\ref{sec:osposg:hsvi}) generates value functions with these properties.
Namely, at any time, the lower bound $\uv$ is max-justified value function by the set of $\alpha$-vectors $\mathsf{Conv}(\Gamma)$, and the upper bound $\ov$ is min-justified.

This allows us to derive two important properties of the algorithm.
First, since \thref{thm:osposg:correctness} guarantees that the algorithm terminates with $\ov(b^{\mathrm{init}}) - \uv(b^{\mathrm{init}}) \leq \varepsilon$, we can use the resulting value functions $\uv$ (represented by $\Gamma$) and $\ov$ to obtain $\varepsilon$-Nash equilibrium strategies for both players.
Next, we can also run the algorithm in anytime fashion and, since the bounds $\uv$ and $\ov$ satisfy the properties at any point of time, use these bounds to extract strategies with performance guarantees.

We will first prove that at any point of time in the execution of Algorithm~\ref{alg:osposg:hsvi}, the lower bound $\uv$ is max-justified by the set $\mathsf{Conv}(\Gamma)$, and the upper bound $\ov$ is a min-justified value function.
To prove this, it suffices to show initial lower-bound value function $\uv$ is max-justified by $\mathsf{Conv}(\Gamma)$ and the initial upper-bound value function $\ov$ is min-justified, and that this property is preserved after any sequence of point-based updates performed on $\uv$ and $\ov$.
With the help of Lemma~\ref{thm:osposg:conv-max-justified}, we can prove that this is true for $\uv$:

\begin{restatable}{lemma}{LBmaxJustByConv}
\label{thm:osposg:max-justified}
    Let $\Gamma$ be the set of $\alpha$-vectors that have been generated at any time during the execution of the HSVI algorithm for one-sided POSGs (Algorithm~\ref{alg:osposg:hsvi}).
    Then the lower bound $\uv$ is max-justified by the set $\mathsf{Conv}(\Gamma)$.
\end{restatable}

Even though the proof is more complicated, the analogous result holds for $\ov$ as well:

\begin{lemma}
\thlabel{thm:osposg:min-justified}
    Let $\ov$ be the upper bound considered at any time of the execution of the HSVI algorithm for one-sided POSGs (Algorithm~\ref{alg:osposg:hsvi}).
    Then $\ov$ is min-justified.
\end{lemma}
\begin{proof}
    Upper bound $\ov$ is only modified by means of point-based update on lines~\ref{alg:osposg:hsvi:pb-update-1} and~\ref{alg:osposg:hsvi:pb-update-2} of Algorithm~\ref{alg:osposg:hsvi}.
    Therefore, it suffices to show that (1) the initial upper bound is min-justified and that (2) the upper bound $V_{\mathrm{UB}}^{\Upsilon'}$ resulting from applying a point-based update on a min-justified upper bound $\ov$ is min-justified as well.
    
    First, let us prove that the initial value function $\ov$ is min-justified.
    Initially, $\ov(b)$ is set to the value of a \emph{perfect information} version of the game, where the imperfectly informed player~1 gets to know the initial state of the game.
    By removing this information from player~1, the utility player~1 can achieve can only decrease.
    It follows that $[H\ov](b) \leq \ov(b)$, so the initial value function $\ov(b)$ is min-justified.
    
    Now, let us consider an upper bound $\ov$ represented by a set $\Upsilon = \lbrace (b_i, y_i) \mid 1 \leq i \leq k \rbrace$ that is considered by the Algorithm~\ref{alg:osposg:hsvi} and let us assume that $\ov$ is min-justified.
    Consider that a point-based update in $b_{k+1}$ is to be performed.
    We show that the function $V_{\mathrm{UB}}^{\Upsilon'}$ resulting from the point-based update in $b_{k+1}$ is min-justified as well.
    Recall that $\Upsilon' = \Upsilon \cup \lbrace (b_{k+1}, y_{k+1}) \rbrace$ and $y_{k+1} = [H\ov](b_{k+1})$.
    Clearly, since $\Upsilon \subset \Upsilon'$, it holds $V_{\mathrm{UB}}^{\Upsilon'}(b) \leq \ov(b)$ and $[HV_{\mathrm{UB}}^{\Upsilon'}](b) \leq [H\ov](b)$ for every $b \in \Delta(S)$.
    Due to this and since $\ov$ is assumed to be min-justified, we have $y_i \geq [HV_{\mathrm{UB}}^{\Upsilon'}](b)$ for every $1 \leq i \leq k+1$.
    We will now prove that $V_{\mathrm{UB}}^{\Upsilon'}$ is min-justified by showing that $[HV_{\mathrm{UB}}^{\Upsilon'}](b) \leq V_{\mathrm{UB}}^{\Upsilon'}(b)$ holds for arbitrary belief $b \in \Delta$.
    Let $\lambda_i$ and $b'$ correspond to the optimal solution of the linear program~\eqref{lp:osposg:projection} for solving $V_{\mathrm{UB}}^{\Upsilon'}(b)$.
    We have
    \begin{align*}
        V_{\mathrm{UB}}^{\Upsilon'}(b) &= \sum_{i=1}^{k+1} \lambda_i y_i + \delta \| b - b' \|_1 \hspace{-10em} \\
        && \textit{ $\lambda_i$ and $b'$ represent an optimal solution of $V_{\mathrm{UB}}^{\Upsilon'}(b)$} \\
               &\geq \sum_{i=1}^{|\Upsilon|} \lambda_i \cdot [HV_{\mathrm{UB}}^{\Upsilon'}](b_i) + \delta \| b - b' \|_1 \hspace{-10em} \\
               &\geq [HV_{\mathrm{UB}}^{\Upsilon'}](b') + \delta \| b - b' \|_1 \hspace{-10em} \\
               && \textit{ $HV_{\mathrm{UB}}^{\Upsilon'}$ is convex, see Proposition~\ref{thm:osposg:hv-convex}} \\
               &\geq [HV_{\mathrm{UB}}^{\Upsilon'}](b) \hspace{-10em} & \textit{ $V_{\mathrm{UB}}^{\Upsilon'}$ is $\delta$-Lipschitz continuous, and hence,} \\
               && \textit{by Proposition~\ref{thm:osposg:hv-convex}, $HV_{\mathrm{UB}}^{\Upsilon'}$ is as well} \ \text{.}
    \end{align*}
    This shows that any point-based update results in a min-justified value function $V_{\mathrm{UB}}^{\Upsilon'}$.
    As a result, Algorithm~\ref{alg:osposg:hsvi} only considers upper bounds $\ov$ that are min-justified.
\end{proof}

We are now in a position to show that Algorithm~\ref{alg:osposg:hsvi} produces $\varepsilon$-Nash equilibrium strategies.

\begin{theorem}\thlabel{thm:equilibrium}
    In any OS-POSG, applying Algorithms~\ref{alg:osposg:cr} and~\ref{alg:osposg:cr2} to the output of Algorithm~\ref{alg:osposg:hsvi} yields an $\varepsilon$-Nash equilibrium.
\end{theorem}
\begin{proof}
    According to \thref{thm:osposg:correctness}, Algorithm~\ref{alg:osposg:hsvi} terminates and the value functions $\uv$ and $\ov$ that result from the execution of the algorithm satisfy $\ov(b^{\mathrm{init}}) - \uv(b^{\mathrm{init}}) \leq \varepsilon$.
    Furthermore, we know that lower bound $\uv$ is max-justified by the set $\Gamma$ resulting from the execution of Algorithm~\ref{alg:osposg:hsvi} (\thref{thm:osposg:max-justified}), and the upper bound $\ov$ is min-justified (\thref{thm:osposg:min-justified}).
    We can therefore use Algorithm~\ref{alg:osposg:cr} to obtain a strategy for player~1 that achieves utility of at least $\uv(b^{\mathrm{init}})$ for player~1 (\thref{thm:osposg:p1-strategy-guarantee}).
    Similarly, we can use Algorithm~\ref{alg:osposg:cr2} to obtain a strategy for player~2 that ensures that the utility of player~1 will be at most $\ov(b^{\mathrm{init}})$ (Proposition~\ref{thm:osposg:p2-strategy}).
    It follows that if either player were to deviate from the strategy prescribed by the algorithm, they would not be able to improve their utility by more than $\ov(b^{\mathrm{init}}) - \uv(b^{\mathrm{init}})$.
    Since $\ov(b^{\mathrm{init}}) - \uv(b^{\mathrm{init}}) \leq \varepsilon$, these strategies must form a $\varepsilon$-Nash equilibrium of the game.
\end{proof}

\section{Experimental evaluation}
\label{sec:osposg:experiments}

In this section, we focus on the experimental evaluation of the heuristic search value iteration algorithm for solving one-sided partially observable stochastic games from Section~\ref{sec:osposg:hsvi}.
We demonstrate the scalability of the algorithm in three security domains.
Rewards in all of the domains have been scaled to the interval $[0,100]$ or $[-100,0]$, respectively, and we report the runtime required to reach $\ov(b^{\mathrm{init}}) - \uv(b^{\mathrm{init}}) \leq 1$.
We first outline the details of our experimental setup.

\subsection{Algorithm Settings}\label{sec:alg:settings}

Compared to the version of the HSVI algorithm presented in Section~\ref{sec:osposg:hsvi}, we adopt several modifications to improve the scalability of the algorithm.
In this section, we describe these modifications and show that the theoretical guarantees of the algorithm still hold.

\paragraph{Pruning the Sets $\Gamma$ and $\Upsilon$}
Each time a point-based update is performed, the size of the sets $\Gamma$ and $\Upsilon$ used to represent value functions $\uv$ and $\ov$ increases.
As new elements are generated, some of the elements in these sets may become unnecessary to accurately represent the bounds $\uv$ and $\ov$.
Since the sizes of sets $\uv$ and $\ov$ have a direct impact on the sizes of linear programs used throughout the algorithm, removing unnecessary elements from $\uv$ and $\ov$ improves the performance.
Whenever a new $\alpha$-vector $\mathsf{valcomp}(\pi_1^{\mathrm{LB}}, \overline{\alpha}^{\mathrm{LB}})$ is generated according to Equation~\eqref{eq:osposg:pbupdate-lb}, all dominated elements in the set $\Gamma$ get removed and only those elements of $\alpha \in \Gamma$ that dominate $\mathsf{valcomp}(\pi_1^{\mathrm{LB}}, \overline{\alpha}^{\mathrm{LB}})$ in at least one state remain, i.e.,
\begin{align}
  \Gamma \ \ \ & \coloneqq & \left\lbrace \alpha' \in \Gamma \mid \exists s \in S: \alpha'(s) > \mathsf{valcomp}(\pi_1^{\mathrm{LB}}, \overline{\alpha}^{\mathrm{LB}})(s) \right\rbrace \hspace{5em} \nonumber \\
  && \cup \ \ \ \left\lbrace \mathsf{valcomp}(\pi_1^{\mathrm{LB}}, \overline{\alpha}^{\mathrm{LB}}) \right\rbrace \ \text{.} \label{eq:osposg:uv-pruning}
\end{align}
For the set $\Upsilon$ used to represent the upper bound $\ov$, we use a batch approach instead of removing dominated elements immediately.
We remove dominated elements every time the size of the set $\Upsilon$ increases by 10\% compared to the size after the last pruning was performed (this is analogous to the pruning technique proposed in~\citep{smith2004-hsvi}).
Algorithm~\ref{alg:osposg:ov-pruning} inspects each point $(b_i,y_i) \in \Upsilon$ and checks whether it is needed to represent value function $\ov$---and if it is not needed, the point gets removed.
\begin{algorithm}
    \caption{Pruning set $\Upsilon$ representing the upper bound $\ov$}
    \label{alg:osposg:ov-pruning}
    \DontPrintSemicolon
    \SetKwInOut{Input}{input}
    
    \Input{Set $\Upsilon$ used to represent $\ov$}
    \For{$(b_i,y_i) \in \Upsilon$}{
        \lIf{$y_i > \ov(b_i)$}{$\Upsilon \coloneqq \Upsilon \setminus \lbrace (b_i,y_i) \rbrace$}
    }
\end{algorithm}

Removing elements from sets $\Gamma$ and $\Upsilon$ does not violate the theoretical properties of the algorithm.
First of all, only elements that are not necessary to represent currently considered bounds are removed---hence the values of value functions $\uv$ and $\ov$ considered at each step of the algorithm remain unchanged, and the convergence property is hence retained.
Furthermore, we can still use pruned value functions to extract strategies with guaranteed performance.
Since the resulting upper bound value function $\ov$ is identical to the one obtained without pruning, it is still min-justified. It can thus be used to obtain a strategy of the minimizing player~2 with guaranteed utility at most $\ov(b^{\mathrm{init}})$ (Section~\ref{sec:osposg:p2-strategy}).
Similarly, $\uv$ can be used to obtain a strategy of player~1 (Section~\ref{sec:osposg:p1-strategy}).
Despite the fact that the resulting set $\Gamma$ of $\alpha$-vectors is different from the set constructed by Algorithm~\ref{alg:osposg:hsvi} when no pruning is used, we can see that for every missing element $\alpha'$ there has to exist an element $\alpha$ such that $\alpha \geq \alpha'$ (see Equation~\eqref{eq:osposg:uv-pruning}).
Therefore, we can always replace missing $\alpha$-vectors in value compositions (i.e., linear functions $\alpha^{a_1,o}$) without decreasing the values of the resulting value composition---and hence $\uv$ remains max-justified by the set of $\alpha$-vectors $\mathsf{Conv}(\Gamma)$.

\paragraph{Partitioning States and Value Functions}
In many games, even the imperfectly informed player~1 has access to some information about the game.
For example, in the pursuit-evasion games we discuss below, the pursuer \emph{knows} his position---and representing his uncertainty about his position within the belief is unnecessary.
To reduce the dimension of the beliefs, we allow for partitioning states into disjoint sets such that the imperfectly informed player~1 always \emph{knows} in which set he is currently.
Formally, let $S = \bigcup_{i=1}^K S_i$ such that $S_i \cap S_j = \emptyset$ for every $i \neq j$.
Player~1 has to know the initial partition, i.e., $\Supp(b^{\mathrm{init}}) \subseteq S_i$ for some $1 \leq i \leq K$.
Furthermore, he has to be able to infer which partition he is in at any time, i.e., for every belief $b$ over a partition $S_i$ (i.e., $\Supp(b) \subseteq S_i$), every achievable action-observation pair $(a_1,o)$ and every stage strategy $\pi_2 \in \Pi_2$ of player~2, we have $\Supp(\tau(b,a_1,\pi_2,o)) \subseteq S_j$ for some $1 \leq j \leq K$.
We use $T(S_i, a_1, o)$ to denote such $S_j$.

This partitioning allows for reducing the size of LP~\eqref{eq:osposg:max-composition-lp} used to compute stage game solutions.
Namely, the quantification over $s \in S$ can be replaced by $s \in S_i$, where $S_i$ is the current partition.
Furthermore, since also the partition of the next stage has to be known, we can also replace $(a_1,o,s') \in A_1 \times O \times S$ by $(a_1,o,s') \in A_1 \times O \times T(S_i,a_1,o)$.

\paragraph{Parameters and Hardware}
We use value iteration for stochastic games, or MDPs, respectively, to initialize the upper and lower bounds.
The upper bound is initialized by solving a perfect-information variant of the game (see Section~\ref{sec:osposg:hsvi:vf}).
The lower bound is computed by fixing the uniform strategy $\sigma_1^{\mathrm{unif}}$ for player~1 and solving the resulting Markov decision process from the perspective of player~2.
We terminate the algorithms when either change in valuations between iterations of value iteration is lower than $0.025$, or $20$ minutes time limit has expired. 
The initialization time is included in the computation times of the HSVI algorithm. 

We use $\varepsilon=1$.
However, similarly to~\cite{smith2004-hsvi}, we adjust $\varepsilon$ in each iteration, and we get $\varepsilon_{\mathrm{imm}}$ that is about to be used in the current iteration using formula $\varepsilon_{\mathrm{imm}} = 0.25 + \eta(\ov(b^{\mathrm{init}}) - \uv(b^{\mathrm{init}}) - 0.25)$ with $\eta=0.9$.
We set the parameter $D$ to the largest value such that $\rho(t) \geq 0.25^{-t}$ holds for every $t \geq 0$.

Each experiment has been run on a single core of Intel Xeon Platinum 8160. We have used CPLEX 12.9 to solve the linear programs.

\subsection{Experimental Results}
We now turn our attention to the discussion of experimental results.
We introduce the domains used in our experiments and comment on the scalability of the proposed algorithm.

\begin{figure}
\begin{subfigure}{0.2\textwidth}
\centering\includegraphics[width=0.9\textwidth]{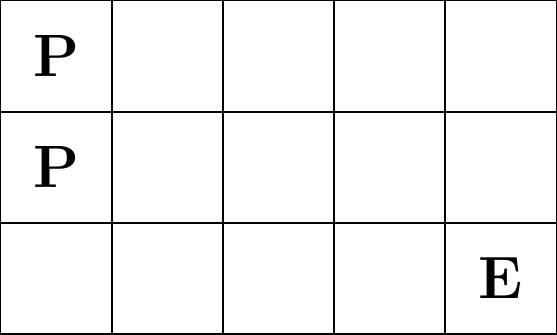}
\caption{\ }\label{fig:peg-game}
\end{subfigure}
\begin{subfigure}{0.78\textwidth}
\centering\includegraphics[width=0.9\textwidth]{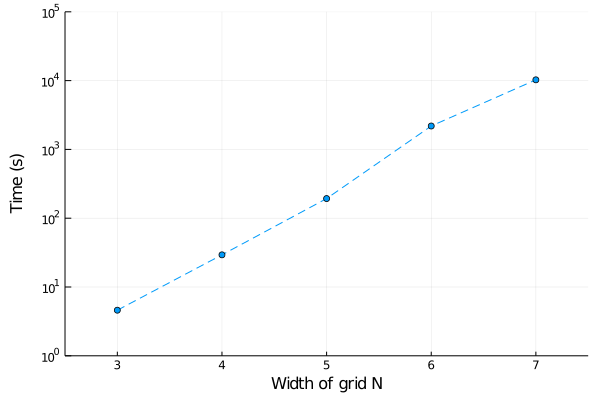}
\caption{\ }\label{fig:peg-results}
\end{subfigure}
\caption{Pursuit evasion games: (a) Pursuit evasion game $5 \times N$. The {\bf P}s denote the initial positions of the pursuers, the {\bf E} denotes the initial position of the evader. (b) Time to reach gap 1 for different grid-widths ($N$).}
\end{figure}
\paragraph{Pursuit-Evasion Games (inspired by~\citep{chung2011search,isler2008-os-peg})}
In pursuit-evasion games, a team of $K$ centrally controlled pursuers (we consider a team of $K=2$) is trying to locate and capture the evader---who is trying to avoid getting captured.
The game is played on a grid (dimensions $3 \times N$), with the pursuers starting in the top-left corner and the evader in the bottom-right corner -- see Figure~\ref{fig:peg-game}.
In each step, the units move to one of their adjacent locations (i.e., the actions of the evader are $A_2 = \lbrace \mathrm{left}, \mathrm{right}, \mathrm{up}, \mathrm{down} \rbrace$, while the actions available to the team of pursuers are joint actions for all units in the team, $A_1 = (A_2)^K$).
The game ends when one of the units from the team of pursuers enters the same cell as the evader---and the team of pursuers (player~1) then receives a reward of $+100$.
The reward for all other transitions in the game is zero.
The pursuer knows the location of their units, but the current location of the evader is not known.

The game with $N=3$ was solved in 4.5\,s on average, while the game with $N=7$ took 10\,267\,s to be solved to the gap $\varepsilon=1$ -- full results can be found in Figure~\ref{fig:peg-results}.
The game $8 \times N$ has not been solved successfully within 10 hours time limit, and the gap of $\ov(b_{\mathrm{init}})-\uv(b_{\mathrm{init}}) = 1.245$ has been reached after 10 hours.
Sizes of the games range from 143 states and 2\,671 transitions (for $3 \times N$ game) to 3\,171 states and 92\,531 transitions (for $8 \times N$ game).

\begin{figure}
\begin{subfigure}[t]{0.23\linewidth}
\caption{}
\label{fig:bpg}
\centering
\vspace{5.5em}
\includegraphics[width=0.9\linewidth]{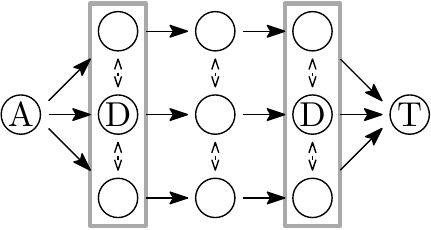}
\end{subfigure}
\begin{subfigure}[t]{0.75\linewidth}
\caption{}
\label{fig:bpg-results}
\centering\includegraphics[width=0.9\linewidth]{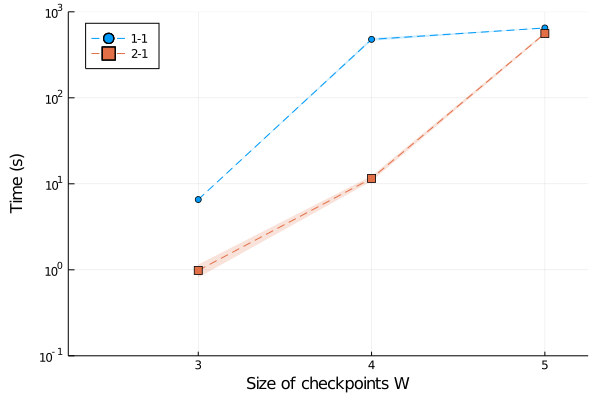}\hspace{1em}
\end{subfigure}
\caption{Intrusion search games: (a) Intrusion-search game with width $W=3$ in configuration 1-1: A denotes the initial position of the attacker and D the positions of the defender's units. T is the attacker's target. (b) Time to reach $\ov(b_{\mathrm{init}}) - \uv(b_{\mathrm{init}}) \leq 1$.} 
\end{figure}

\paragraph{Search Games (inspired by~\citep{bosansky2014-jair})}
In search games that model intrusion, the defender patrols checkpoint zones (see Figure~\ref{fig:bpg}, the zones are marked with box).
The attacker aims to cross the graph while not being captured by the defender.
She can either wait for one move to conceal her presence (and clean up the trace), or move further.
Each unit of the defender can move to adjacent nodes within its assigned zone.
The goal of the attacker is to cross the graph to reach node marked by $T$ without encountering any unit of the defender.
If she manages to do so, the defender receives a reward of $-100$.

We consider games with two checkpoint zones with a varying number of nodes in a zone $W$ (i.e. the width of the graph). We use two configurations of the defending forces: (1) one defender in each checkpoint and (2) two defenders in the first checkpoint and one defender in the second checkpoint. We denote these settings as 1-1 and 2-1.

The results are shown in Figure~\ref{fig:bpg-results} (with five runs for each parameterization, the confidence intervals mark the standard error in our graphs).
The largest game ($W=5$ and two defenders in the first zone) has 4\,656 states and 121\,239 transitions and can be solved within 560\,s.
This case highlights that our algorithm can solve even large games.
However, a much smaller game with  $W=5$ and configuration 1-1 (964 states and 9\,633 transitions) is more challenging, since the coordination problem with just one defender in the first zone is harder, and despite its smaller size it is solved within 640\,s.

\paragraph{Patrolling Games (inspired by~\citep{Basilico2009,vorobeychik2014-icaps})}
In a patrolling game, a patroller (player~1) aims to protect a set of targets $V$.
The targets are represented by vertices of a graph, and the possible movements of the patroller are represented by the edges of the graph.
The attacker observes the movement of the patroller and decides which target $v \in V$ he will attack, or whether he will postpone the decision.
Once the attacker decides to attack a target $v$, the defender has $t_\times$ steps to reach the attacked vertex.
If he fails to do so, he receives a negative reward $-C(v)$ associated to the target $v$---otherwise, he successfully protects the target, and the reward is zero.
The patroller does not know whether and where the attack has already started.
The costs $C(v)$ are scaled so the $\max_{v \in V} C(v) = 100/\gamma^{t_{\times}}$, i.e., the minimum possible payoff for the defender is $-100$.

\begin{figure}
    \centering
    \includegraphics[width=0.7\linewidth]{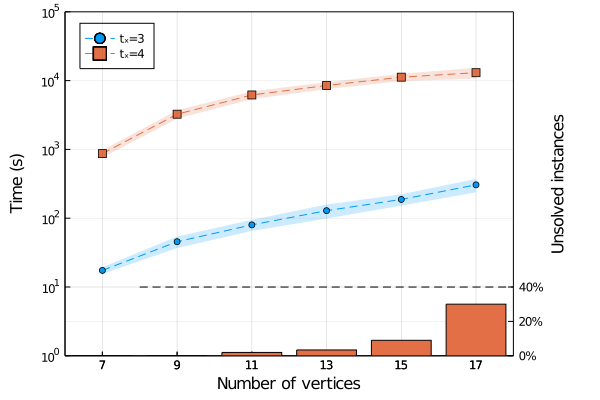}
    \caption{Time to reach $\ov(b_{\mathrm{init}})-\uv(b_{\mathrm{init}})\leq1$ for patrolling games with attack times $t_\times=3$ and $t_\times=4$. Bars indicate percentage of unsolved instances for $t_\times=4$.}
    \label{fig:patrolling-results}
\end{figure}

Following the setting in~\citep{vorobeychik2014-icaps}, we focus on graphs generated from Erdos-Renyi model~\citep{newman2010networks} with parameter $p=0.25$ (denoted $ER(0.25)$) with attack times $t_\times \in \lbrace 3, 4 \rbrace$ and number of vertices $|\mathcal{V}|$ ranging from 7 to 15.
The time to solve even the largest instances ($V=17$) with $t_\times=3$ was $305.5$\,s.
For attack time $t_\times=4$, however, some number of instances failed to reach the precision $\ov(b^{\mathrm{init}}) - \uv(b^{\mathrm{init}}) \leq 1$ within the time limit of 10 hours.
For the most difficult setting, $|\mathcal{V}|=17$ and $t_\times=4$, the algorithm reached desired precision in $70\%$ of instances.
For unsolved instances in this setting, mean $\ov(b^{\mathrm{init}}) - \uv(b^{\mathrm{init}})$ after the cutoff after 10 hours is however reasonably small at 3.77$\pm$0.54.
The results include games with up to 856 states and 6\,409 transitions.
See Figure~\ref{fig:patrolling-results} for more details.

\begin{figure}
    \centering
    \includegraphics[width=0.7\linewidth]{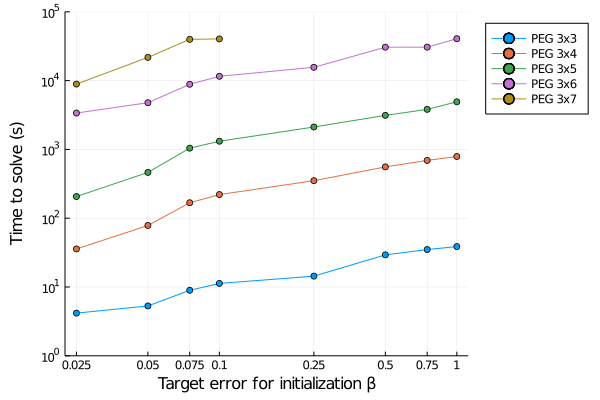}
    \caption{Effect of initialization on runtime. The target error is measured as Bellman residual $\| TV - V \|_\infty$ of the value iteration algorithms used to obtain initial bounds.}
    \label{fig:residual}
\end{figure}
\subsection{Impact of Initialization on Solution Time}
Recall that we use value iteration algorithms for solving perfect information stochastic games and Markov decision processes, respectively, to initialize upper and lower bounds on $V^*$.
In our experiments, we terminate the algorithms whenever the change in valuation between iterations of value iteration is smaller than $\beta=0.025$.
In Figure~\ref{fig:residual}, we analyze the impact of the choice of $\beta$ on the running time of the algorithm when applied to pursuit evasion games.
Observe that the tighter initial bounds are used, the faster the convergence of the algorithm.
In fact, the difference between $\beta=1$ and $\beta=0.025$ is approximately an order of magnitude in run time.
Recall that the bounds $\ov$ and $\uv$ not only serve as bounds on $V^*$, but they are also used to obtain strategies that are considered during the forward exploration phase of the algorithm (see lines~\ref{alg:osposg:hsvi:follower-choice} and~\ref{alg:osposg:hsvi:leader-choice} of Algorithm~\ref{alg:osposg:hsvi}).
We believe that these results indicate that the use of, e.g., domain-dependent initialization of the bounds can greatly improve the run time of the algorithm in complex domains.

\subsection{Performance Analysis}

Based on the the algorithm's runtime data, we observed that most of the computation time is split between solving the linear programs used to compute $H\ov$ and $H\uv$ and pruning the representations of these bounds.
Together, these three tasks typically took around 85\% of the total runtime (and always at least 70\%), with the remaining time being spent on computation of initial bounds, construction of the linear programs, and other smaller tasks.
More specifically, solving $H\ov$ took 30-50\% of the runtime in typical games while reaching as far as 60\% in large pursuit evasion games (e.g., 60.5\% in the $3 \times 7$ pursuit evasion game).
Solving $H\uv$ was faster --- in most games, it took between 10 and 20\% of the total runtime.
Finally, time required to perform pruning of the bounds $\ov$ and $\uv$ also took 10-20\% of the runtime, with the exception of the patrolling games with attack time $t_\times=4$, where it required over 40\% of the total runtime.

%% file: chapters/conclusion.tex
\section{Conclusions}

We cover two-player zero-sum partially observable stochastic games (POSGs) with discounted rewards and one-sided observability --- that is, those where the second player has perfect information about the game.
We describe the theoretical properties of the value function in these games and show that algorithms based on value-iteration converge to an optimal value of the game.
We also propose the first approximate algorithm that generalizes the ideas behind point-based algorithms designed for partially observable Markov decision processes (POMDPs) and transfers these techniques to POSGs. 

The presented work shows that it is possible to translate selected results from the single-agent setting to zero-sum games. 
Moreover, in future work, this work could be further extended in several ways:
First, as already demonstrated by existing follow-up works~\cite{horak2019-cose}, the scalability of the algorithm can be substantially improved for specific security games.
Second, many heuristics and methods proven useful in the POMDP setting can be translated and evaluated in the game-theoretic setting, further improving the scalability.
Third, generalization beyond the strictly adversarial setting (e.g., by computing a Stackelberg equilibrium) is another key direction supporting the applicability of these game-theoretic models to security.

\section*{Acknowledgements}
This research was supported by the Czech Science Foundation (no. 19-24384Y), by the OP VVV MEYS funded project CZ.02.1.01/0.0/0.0/16 019/0000765 ``Research Center for Informatics'', and by the U.S. Army Combat Capabilities Development Command Army Research Laboratory and was accomplished under Cooperative Agreement Number W911NF-13-2-0045 (ARL Cyber Security CRA). The views and conclusions contained in this document are those of the authors and should not be interpreted as representing the official policies, either expressed or implied, of the Combat Capabilities Development Command Army Research Laboratory or the U.S. Government. The U.S. Government is authorized to reproduce and distribute reprints for Government purposes notwithstanding any copyright notation here on.

%% file: chapters/appendix.tex
\section{Proofs}\label{sec:app:proofs}

\ValuesAreBounded*
\begin{proof}
    The smallest payoff player~1 can hypothetically achieve in any play consists of getting $\underline{r} = \min_{(s,a_1,a_2)} R(s,a_1,a_2)$ in every timestep.
    The infinite discounted sum $\sum_{t=1}^\infty \gamma^{t-1} \underline{r}$ converges to $\underline{r}/(1-\gamma)=L$.
    Conversely, the maximum payoff can be achieved if player~1 obtains $\overline{r} = \max_{(s,a_1,a_2)} R(s,a_1,a_2)$ in every timestep.
    Expected values of strategies (and therefore also the value of the game) are expectation over the payoffs of individual plays---hence are bounded by $L$ and $U$ as well.
\end{proof}

\ValuesConvex*
\begin{proof}
    Definition~\ref{def:osposg:v-star} defines $V^*$ as the point-wise supremum over linear functions $\val^{\sigma_1}$ (over all strategies $\sigma_1 \in \Sigma_1$ of player~1).
    This implies the convexity of $V^*$~\citep[p.81]{boyd2004-convex}.
\end{proof}

\LinearFsAreLipschitz*
\begin{proof}
    Let $p,q \in \Delta(X)$ be arbitrary two points in the probability simplex over the finite set $X$.
    Since $f$ is a linear function, it can be represented as a convex combination of values $\alpha(x)$ in the vertices of the simplex corresponding to the elements $u \in X$,
    \begin{subequations}
    \begin{equation}
        f(p) = \sum_{u \in X} \alpha(u) \cdot p(u) \qquad \text{where} \qquad \alpha(u) = f(\mathbbm{1}_u), \ \ \mathbbm{1}_u(v) = \begin{cases}
            1 & v = u \\
            0 & \text{otherwise}
        \end{cases} \ \text{.}
    \end{equation}
    Without loss of generality, let us assume $f(p) \geq f(q)$.
    Now, the difference $| f(p) - f(q) |$ satisfies
    \begin{equation}
        | f(p) - f(q) | = f(p) - f(q) = \sum_{u \in X} \alpha(u) \cdot [ p(u) - q(u) ] \ \text{.} \label{thm:osposg:linbound-lipschitz:step1}
    \end{equation}
    Denote $X^+ = \lbrace u \in X \ |\ p(u) - q(u) \geq 0 \rbrace$ and $X^- = \lbrace u \in X \ |\ p(u) - q(u) < 0 \rbrace$.
    We can now bound the difference from Equation~\eqref{thm:osposg:linbound-lipschitz:step1} by $| f(p) - f(q) | = $
    \begin{align}
        &= \sum_{u \in X^+} \alpha(u) \cdot [ p(u) - q(u) ] + \sum_{u \in X^-} \alpha(u) \cdot [ p(u) - q(u) ] \\
        &\leq \sum_{u \in X^+} y_{\mathrm{max}} \cdot [ p(u) - q(u) ] + \sum_{u \in X^-} y_{\mathrm{min}} \cdot [ p(u) - q(u) ] \\
        &= y_{\mathrm{max}} \sum_{u \in X^+} [ p(u) - q(u) ] + y_{\mathrm{min}} \sum_{u \in X^-} [ p(u) - q(u) ] \ \text{.}
    \end{align}
    Since both $p$ and $q$ belong to $\Delta(X)$, we have $\| p \|_1 = \| q \|_1 = 1$. Since $p(u),q(u) \geq 0$ are non-negative, we have
    \begin{equation*}
    \| p \|_1 = \| q \|_1 + \sum_{u \in X^+} [ p(u) - q(u) ] - \sum_{u \in X^-} [ q(u) - p(u) ] . \end{equation*}
    It follows that
    \begin{equation}\label{eq:blabla}
        \sum_{u \in X^+} [ p(u) - q(u) ] = -\sum_{u \in X^-} [ p(u) - q(u) ] \text{.}
    \end{equation}
    From equation \eqref{eq:blabla}, we further see that both terms in \eqref{eq:blabla} are equal to $\| p - q \|_1 / 2 $.
    This implies that
    \begin{align*}
        | f(p) - f(q) | \leq \ & \ y_{\mathrm{max}} \| p - q \|_1/2 + y_{\mathrm{min}} (- \| p - q \|_1/2) \\
        = \ & \ (y_{\mathrm{max}} - y_{\mathrm{min}})/2 \cdot \| p - q \|_1
    \end{align*}
    and completes the proof.
    \end{subequations}
\end{proof}

\ValuesAreLipschitz*
\begin{proof}
    $V^*$ is defined as a supremum over $\delta$-Lipschitz continuous values $\val^{\sigma_1}$ of strategies $\sigma_1 \in \Sigma_1$ of the imperfectly informed player~1.
    Therefore for arbitrary $b, b' \in \Delta(S)$, we have the following
    \begin{equation}
        V^*(b) = \sup_{\sigma_1 \in \Sigma_1} \val^{\sigma_1}(b) \leq \sup_{\sigma_1 \in \Sigma_1} [ \val^{\sigma_1}(b') + \delta \| b - b' \|_1 ] = V^*(b') + \delta \| b - b' \|_1 \ \text{.}
    \end{equation}
\end{proof}

\ConvexClosureDoesntIncreaseSupremum*
\begin{proof}
    Clearly, it suffices to prove the inequality $\geq$.
    Let $b \in \Delta(S)$ and let $\sum_{i=1}^k \lambda_i \alpha_i$ be an arbitrary\footnote{Recall that according to the Carath\'eodory's theorem, it suffices to consider finite convex combinations.} convex combination of linear functions from $\Gamma$ (i.e., we have $\alpha_i \in \Gamma$).
    We need to show that $\alpha(b) \geq \sum_{i=1}^k \lambda_i \alpha_i(b)$ holds for some $\alpha \in \Gamma$.
    This is straightforward, as can be witnessed by the function $\alpha_{i^*} \in \Gamma$, $i^* := \argmax_i \alpha_i(b)$:
    \begin{align*}
        \sum_{i=1}^k \lambda_i \alpha_i(b) \leq \sum_{i=1}^k \lambda_i \max_{1 \leq i \leq k} \alpha_i(b) = \max_{1 \leq i \leq k} \alpha_i(b) = \alpha_{i^*}(b) \ \text{.}
    \end{align*}
\end{proof}

\ConvexFunctionsAsSupremaOfLinearFs*
\begin{proof}
    Let $\Gamma := \{ \alpha : \Delta(S) \rightarrow \mathbb{R} \textnormal{ linear} \mid \alpha \leq f \}$.
    Clearly, the pointwise supremum of $\Gamma$ is no greater than $f$.
    It remains to show that $\sup_{\alpha \in \Gamma} \alpha(b_0) \geq f(b_0)$ for each $b_0$.
    Let $b_0$ be an interior point of $\Delta(S)$.
    By the standard convex-analysis result, there exists a subdifferential of $f$ at $b_0$, that is, a vector $v$ such that $f(b) \geq f(b_0) + v\cdot (b-b_0)$ holds for each $b \in \Delta(S)$.
    The function $\alpha (b) := f(b_0) + v\cdot (b-b_0)$ therefore belongs to $\Gamma$ and witnesses that $\sup_{\alpha \in \Gamma} \alpha(b_0) \geq f(b_0)$.
    
    Suppose that $b_0$ lies at the boundary of $\Delta(S)$ and let $\eta$, $\|\eta \|_1=1$, be a direction in which every nearby point $b_\delta := b_0 - \delta \eta$, $\delta \in (0,\Delta]$, lies in the interior of $\Delta(S)$ (for some $\Delta>0$).
    Since $f$ is convex, the directional derivatives $f'_\eta(b_\delta) = \lim_{g\to 0_+} \frac{f(b_\delta+g\eta)-f(b_\delta)}{g}$ are non-decreasing as the points $b_\delta$ get closer to $b_0$.
    In particular, the linear functions $\alpha_\delta$ found for $b_\delta$ in the previous step satisfy
    \[
    \alpha_\delta(b_0) \geq f(b_\delta) + f'_\eta(b_\delta)\delta \geq f(b_\delta) + f'_\eta(b_\Delta)\delta .
    \]
    The right-hand side converges to $f(b_0) + f'_\eta(b_\Delta)\cdot 0 = f(b_0)$, which shows that the supremum of $\alpha_\delta(b_0)$ is at least $f(b_0)$. Since $\alpha_\delta \in \Gamma$, this proves the remaining part of the proposition.
\end{proof}

\StrategyDecomposition*
\begin{proof}
    Let $\sigma_1 \in \Sigma_1$ be an arbitrary behavioral strategy of player~1, and let $\pi_1 = \sigma_1(\emptyset)$ and $\zeta_{a_1,o}(\omega')=\sigma_1(\omega')$ for every $(a_1,o) \in A_1 \times O$ and $\omega' \in (A_1 O)^*$.
    It can be easily verified that $\mathsf{comp}(\pi_1,\overline{\zeta})$ defined in Definition~\ref{def:osposg:composite} satisfies $\mathsf{comp}(\pi_1,\overline{\zeta})=\sigma_1$.
\end{proof}

\StrategyCompositionValue*
\begin{proof}
    Let us evaluate the payoff if player~2 uses $a_2$ in the first stage of the game given that the initial state of the game is $s$.
    The expected reward of playing action $a_2$ against $\mathsf{comp}(\pi_1,\overline{\zeta})$ in the first stage is $\sum_{a_1 \in A_1} \pi_1(a_1) R(s,a_1,a_2)$, i.e., the expectation over the actions player~1 can take.
    Now, at the beginning of the next stage, player~2 knows everything about the past stage---including action $a_1$ taken by player~1, observation $o$ he received, and the new state of the game $s'$.
    Therefore, player~2 knows the strategy $\zeta_{a_1,o}$ player~1 is about to use in the rest of the game.
    By definition of $\val^{\zeta_{a_1,o}}$ (Definition~\ref{def:osposg:strategy-value}), the best payoff player~2 can achieve in $(a_1,o)$-subgame is $\val^{\zeta_{a_1,o}}(s')$.
    After reaching the subgame, however, one stage has already passed and the rewards originally received at time $t$ are now received at time $t+1$.
    As a result, the reward $\val^{\zeta_{a_1,o}}(s')$ gets discounted by $\gamma$.
    The probability that the $(a_1,o)$-subgame is reached is $\sum_{(a_1,o,s') \in A_1 \times O \times S} \pi_1(a_1) T(o,s' \,|\, s,a_1,a_2)$, and the expectation over $\gamma \val^{\zeta_{a_1,o}}(s')$ is thus computed.
    Player~2 chooses an action which achieves the minimum payoff which completes the proof.
\end{proof}

\GenCompositionValue*
\begin{proof}
    Let $\overline{\zeta} \in (\Sigma_1)^{A_1 \times O}$ be as in the lemma, and let $\overline{\alpha}^\zeta$ be such that $\alpha_{a_1,o}^\zeta = \val^{\zeta_{a_1,o}}$.
    According to the assumption we have $\alpha_{a_1,o}^\zeta \geq \alpha_{a_1,o}$.
    Replacing $\alpha_{a_1,o}$ by $\alpha_{a_1,o}^\zeta$ in Equation~\eqref{eq:osposg:valcomp} can only increase the objective value, hence
    \begin{equation}
        \mathsf{valcomp}(\pi_1,\overline{\alpha})(s) \leq \mathsf{valcomp}(\pi_1,\overline{\alpha}^\zeta)(s) = \val^{\mathsf{comp}(\pi_1,\overline{\zeta})}(s) \ \text{.}
    \end{equation}
    Composite strategies are behavioral strategies of player~1, hence $\sigma_1 = \mathsf{comp}(\pi_1,\overline{\zeta})$.
\end{proof}

\ValcompLipschitz*
\begin{proof}
    Since $\mathsf{valcomp}(\pi_1,\overline{\alpha})(b)$ is calculated as a convex combination of the values $\mathsf{valcomp}(\pi_1,\overline{\alpha})(s)$ in the vertices of the $\Delta(S)$ simplex, it suffices to show that
    \begin{equation*}
    \left( \forall s \in S \right) : L \leq \mathsf{valcomp}(\pi_1,\overline{\alpha})(s) \leq U .
    \end{equation*}
    Let $a_2^* \in A_2$ be the minimizing action of player~2 in Equation~\eqref{eq:osposg:valcomp}.
    It holds $\underline{r} \leq R(s,a_1,a_2^*) \leq \overline{r}$, where $\underline{r}$ and $\overline{r}$ are minimum and maximum rewards in the game.
    Hence $\underline{r} \leq \sum_{a_1 \in A_1} \pi_1(a_1) R(s,a_1,a_2^*) \leq \overline{r}$.
    Similarly, from the assumption of the lemma, we have $L \leq \alpha_{a_1,o}(s') \leq U$ and hence $L \leq \sum_{(a_1,o,s') \in A_1 \times O \times S} \pi_1(a_1) T(o, s' \,|\, s, a_1, a_2^*) \alpha_{a_1,o}(s') \leq U$.
    We will now prove that $\mathsf{valcomp}(\pi_1,\overline{\alpha})(s) \leq U$ (the proof of $\mathsf{valcomp}(\pi_1,\overline{\alpha})(s) \geq L$ is analogous):
    \begin{align*}
        & \mathsf{valcomp}(\pi_1,\overline{\alpha})(s) = \\
        & = \sum_{a_1 \in A_1} \pi_1(a_1) R(s,a_1,a_2^*) + \gamma \!\!\!\!\!\!\!\!\!\!\!\!\!\!\! \sum_{(a_1,o,s') \in A_1 \times O \times S} \!\!\!\!\!\!\!\!\!\!\!\!\!\!\! \pi_1(a_1) T(o, s' \,|\, s, a_1, a_2^*) \alpha_{a_1,o}(s') \\
        & \leq \ \overline{r} + \gamma U \ = \ \overline{r} + \gamma \frac{\overline{r}}{1-\gamma} \ = \ U \ \text{.}
    \end{align*}
    The $\delta$-Lipschitz continuity of $\mathsf{valcomp}(\pi_1,\overline{\alpha})$ then follows directly from Lemma~\ref{thm:osposg:linbound-lipschitz}.
\end{proof}

\HVLipschitzConvex*
\begin{proof}
    According to Definition~\ref{def:osposg:H-valcomp}, operator $H$ can be rewritten as a supremum over all possible value compositions:
    \begin{subequations}
    \begin{align}
        & [HV](b) = \max_{\pi_1 \in \Pi_1} \sup_{\overline{\alpha} \in \Gamma^{A_1 \times O}} \mathsf{valcomp}(\pi_1, \overline{\alpha})(b) = \!\!\!\!\!\!\!\!\! \sup_{(\pi_1,\overline{\alpha}) \in \Pi_1 \times \Gamma^{A_1 \times O}} \!\!\!\!\!\!\!\!\! \mathsf{valcomp}(\pi_1, \overline{\alpha})(b) \ \text{, and} \\
        & [HV](b) = \sup_{\alpha \in \Gamma'} \alpha(b) \qquad \Gamma' = \left\lbrace \mathsf{valcomp}(\pi_1,\overline{\alpha}) \;|\; \pi_1 \in \Pi_1, \overline{\alpha} \in \Gamma^{A_1 \times O} \right\rbrace \ \text{.}
        \label{eq:osposg:hv-pointwise}
    \end{align}
    \end{subequations}
    In Equation~\eqref{eq:osposg:hv-pointwise}, $HV$ is represented as a point-wise supremum from a set $\Gamma'$ of linear functions $\mathsf{valcomp}(\pi_1,\overline{\alpha})$, which is a convex continuous function (see \thref{thm:osposg:sup-convex}).
    
    Moreover, in case $V$ is $\delta$-Lipschitz continuous, the set $\Gamma$ representing $V$ can be assumed to contain only $\delta$-Lipschitz continuous linear functions.
    According to Lemma~\ref{thm:osposg:valcomp-lipschitz}, $\mathsf{valcomp}(\pi_1,\overline{\alpha})$ is $\delta$-Lipschitz continuous for every $\pi_1 \in \Pi_1$ and $\alpha^{a_1,o} \in \Gamma$.
    Hence, $\Gamma'$ contains $\delta$-Lipschitz continuous linear functions only and the point-wise maximum $HV$ over $\Gamma'$ is $\delta$-Lipschitz continuous.
\end{proof}

\Bellman*
\begin{proof}
    We first prove the equality of \eqref{eq:osposg:H-maxmin} and \eqref{eq:osposg:H-minmax}.
    Let us define a payoff function $u: \Pi_1 \times \Pi_2 \rightarrow \mathbb{R}$ to be the objective of the maximin and minimax optimization in \eqref{eq:osposg:H-maxmin} and \eqref{eq:osposg:H-minmax}.
    \begin{subequations}
    \begin{equation}
        u(\pi_1,\pi_2) = \E_{b,\pi_1,\pi_2}[R(s,a_1,a_2)] + \gamma \sum_{a_1,o} \Pr_{b,\pi_1,\pi_2}[a_1,o] \cdot V(\tau(b,a_1,\pi_2,o))
    \end{equation}
    After expanding the expectation $\E_{b,\pi_1,\pi_2}[R(s,a_1,a_2)]$ and expressing $V$ as a supremum over linear functions $\alpha \in \Gamma$, we get
    \begin{align*}
        u(\pi_1,\pi_2) &= \sum_{s,a_1,a_2} b(s) \pi_1(a_1) \pi_2(a_2 | s) R(s,a_1,a_2) \ + \\
                       & \qquad + \gamma \sum_{a_1,o} \Pr_{b,\pi_1,\pi_2}[a_1,o] \cdot \sup_{\alpha \in \Gamma} \sum_{s'} \tau(b,a_1,\pi_2,o)(s') \cdot \alpha(s') \numberthis\\
                       &= \sum_{s,a_1,a_2} b(s) \pi_1(a_1) \pi_2(a_2 | s) R(s,a_1,a_2) \ + \numberthis\label{eq:osposg:stage-utility}\\
                       & \qquad + \gamma \sum_{a_1,o} \pi_1(a_1) \cdot \sup_{\alpha \in \Gamma} \sum_{s,a_2,s'} b(s)\pi_2(a_2|s)T(o,s' \,|\, s,a_1,a_2)\alpha(s') \ \text{.}
    \end{align*}
    \end{subequations}
    Note that the term $\Pr_{b,\pi_1,\pi_2}[a_1,o]$ cancels out after expanding $\tau(b,a_1,\pi_2,o)$ in Equation~\eqref{eq:osposg:stage-utility}.
    
    We now show that the von Neumann's minimax theorem~\citep{vonneumann1928-minimax,nikaido1953-minimax} applies to the game with utility function $u$ and strategy spaces $\Pi_1$ and $\Pi_2$ for player~1 and player~2, respectively.
    The von Neumann's minimax theorem requires that the strategy spaces $\Pi_1$ and $\Pi_2$ are convex compact sets (which is clearly the case), and that the utility function $u$ (as characterized by Equation~\eqref{eq:osposg:stage-utility}) is continuous, convex in $\Pi_2$ and concave in $\Pi_1$.
    We will now prove the latter and show that $u$ is a convex-concave utility function.
    Clearly, for every $\pi_2 \in \Pi_2$, the function $u(\cdot, \pi_2): \Pi_1 \rightarrow \R$ (where $\pi_2$ is considered constant) is linear in $\pi_1$, and hence also concave.
    The convexity of $u(\pi_1,\cdot): \Pi_2 \rightarrow \R$ (after fixing arbitrary $\pi_1 \in \Pi_1)$ is more involved.
    As weighted sum of convex functions with positive coefficients $\pi_1(a_1) \geq 0$ is also convex, it is sufficient to show that $f(\pi_2) = \sup_{\alpha \in \Gamma} \sum_{s,a_2,s'} b(s) \pi_2(a_2|s) T(o,s' \,|\, s,a_1,a_2) \alpha(s')$ is convex.
    Observe that for every $\alpha \in \Gamma$, the expression $\sum_{s,a_2,s'} b(s) \pi_2(a_2|s) T(o,s' \,|\, s,a_1,a_2) \alpha(s')$ is linear in $\pi_2$ and, as a result,  the supremum over such linear expressions in $\pi_2$ is convex in $\pi_2$ (see \thref{thm:osposg:sup-convex}).
    According to von Neumann's minimax theorem $\max_{\pi_1 \in \Pi_1} \min_{\pi_2 \in \Pi_2} u(\pi_1,\pi_2) = \min_{\pi_2 \in \Pi_2} \max_{\pi_1 \in \Pi_1} u(\pi_1,\pi_2)$ which concludes the proof of equality of~\eqref{eq:osposg:H-maxmin} and~\eqref{eq:osposg:H-minmax}.
    
    We now proceed by showing the equality of \eqref{eq:osposg:equiv-valcomp} and \eqref{eq:osposg:H-maxmin}.
    By further rearranging Equation~\eqref{eq:osposg:stage-utility}, we get
    \begin{align*}
        u(\pi_1,\pi_2) &= \sup_{\overline{\alpha} \in \Gamma^{A_1 \times O}} \Big[ \sum_{s,a_1,a_2} b(s) \pi_1(a_1) \pi_2(a_2 | s) R(s,a_1,a_2) \ + \numberthis\label{eq:osposg:stage-utility-supoutside}\\
        & \qquad\qquad + \gamma \sum_{a_1,o} \pi_1(a_1) \sum_{s,a_2,s'} b(s)\pi_2(a_2|s)T(o,s' \,|\, s,a_1,a_2)\alpha_{a_1,o}(s') \Big] \ \text{.}
    \end{align*}
    Let us define a game with strategy spaces $\Gamma$ and $\Pi_2$ and payoff function $u'_{\pi_1}: \Gamma \times \Pi_2 \rightarrow \R$ where $u'_{\pi_1}$ is the objective of the supremum in Equation~\eqref{eq:osposg:stage-utility-supoutside} (Equation~\eqref{eq:osposg:uprime-2} is an algebraic simplification of Equation~\eqref{eq:osposg:uprime-1}).
    \begin{subequations}
    \begin{align*}
        u'_{\pi_1}(\overline{\alpha},\pi_2) &= \sum_{s,a_1,a_2} b(s) \pi_1(a_1) \pi_2(a_2 | s) R(s,a_1,a_2) \ + \numberthis\label{eq:osposg:uprime-1}\\
        & \qquad\qquad + \gamma \sum_{a_1,o} \pi_1(a_1) \sum_{s,a_2,s'} b(s)\pi_2(a_2|s)T_{s,a_1,a_2}(o,s')\alpha_{a_1,o}(s') \\
        &= \sum_s b(s) \sum_{a_2} \pi_2(a_2|s) \sum_{a_1} \pi_1(a_1) \Big[ R(s,a_1,a_2) \ + \numberthis\label{eq:osposg:uprime-2}\\
        & \qquad\qquad + \gamma \sum_{o,s'} T(o,s' \,|\, s,a_1,a_2)\alpha_{a_1,o}(s') \Big] \ \text{.}
    \end{align*}
    \end{subequations}
    Plugging~\eqref{eq:osposg:uprime-2} into~\eqref{eq:osposg:stage-utility-supoutside}, we can write
    \begin{equation}
        \max_{\pi_1 \in \Pi_1} \min_{\pi_2 \in \Pi_2} u(\pi_1,\pi_2) = \max_{\pi_1 \in \Pi_1} \min_{\pi_2 \in \Pi_2} \sup_{\overline{\alpha} \in \Gamma^{A_1 \times O}} u'_{\pi_1}(\pi_2,\overline{\alpha}) \ \text{.}
    \end{equation}
    To prove the equivalence of \eqref{eq:osposg:equiv-valcomp} and \eqref{eq:osposg:H-maxmin}, we need to show that the minimum and supremum can be swapped.
    Since $u'_{\pi_1}$ is linear in both $\pi_2$ and $\overline{\alpha}$, $\Pi_2$ is a compact convex set and $\Gamma$ (and therefore also the set of mappings $\overline{\alpha} \in \Gamma^{A_1 \times O}$) is convex, it is possible to apply Sion's minimax theorem~\citep{sion1958-minimax} to get
    \begin{equation}
        \max_{\pi_1 \in \Pi_1} \min_{\pi_2 \in \Pi_2} \sup_{\overline{\alpha} \in \Gamma^{A_1 \times O}} u'_{\pi_1}(\pi_2,\overline{\alpha})
        = \max_{\pi_1 \in \Pi_1} \sup_{\overline{\alpha} \in \Gamma^{A_1 \times O}} \min_{\pi_2 \in \Pi_2} u'_{\pi_1}(\pi_2,\overline{\alpha}) \ \text{.}
    \end{equation}
    As $u'_{\pi_1}$ is linear in $\pi_2$ (for fixed $\pi_1$ and $\overline{\alpha}$), the minimum over $\pi_2$ is attained in pure strategies. Denote $\hat{\pi}_2: S \rightarrow A_2$ a pure strategy of player~2 assigning action $\hat{\pi}_2(s)$ to be played in state $s$, and $\hat{\Pi}_2$ the set of all pure strategies of player~2.
    We now rewrite $u'_{\pi_1}$ to use pure strategies $\hat{\Pi}_2$ instead of randomized stage strategies $\Pi_2$.
    First, in Equation~\eqref{eq:osposg:bellman-1}, we replace the maximization over $\Pi_2$ by maximization over the pure strategies $\hat{\Pi}_2$ and replace expectation over actions of player~2 by using the deterministic action $\hat{\pi}_2(s)$ where appropriate.
    Then, in Equation~\eqref{eq:osposg:bellman-2}, we leverage the fact that, unlike player~1, player~2 knows the state before having to act, and hence he can optimize his actions $\hat{\pi}_2(s)$ independently.
    And, finally, in Equation~\eqref{eq:osposg:bellman-3}, we use Definition~\ref{def:osposg:val-composition}.
    \begin{subequations}
    \begin{align*}
        & \max_{\pi_1 \in \Pi_1} \min_{\pi_2 \in \Pi_2} u(\pi_1,\pi_2) = \max_{\pi_1 \in \Pi_1} \sup_{\overline{\alpha} \in \Gamma^{A_1 \times O}} \min_{\pi_2 \in \Pi_2} u'_{\pi_1}(\pi_2,\overline{\alpha}) = \\
        & \qquad = \max_{\pi_1 \in \Pi_1} \sup_{\overline{\alpha} \in \Gamma^{A_1 \times O}} \min_{\hat{\pi}_2 \in \hat{\Pi}_2} \sum_s b(s) \sum_{a_1} \pi_1(a_1) \Big[ R(s,a_1,\hat{\pi_2}(s)) \ + \numberthis\label{eq:osposg:bellman-1}\\
        & \qquad\qquad\qquad\qquad\qquad\qquad\qquad\qquad\qquad + \gamma \sum_{o,s'} T(o,s' \mid s,a_1,\hat{\pi}_2(s))\alpha_{a_1,o}(s') \Big] \\
        & \qquad = \max_{\pi_1 \in \Pi_1} \sup_{\overline{\alpha} \in \Gamma^{A_1 \times O}} \sum_s b(s) \min_{\hat{\pi}_2(s) \in A_2} \sum_{a_1} \pi_1(a_1) \Big[ R(s,a_1,\hat{\pi_2}(s)) \ + \numberthis\label{eq:osposg:bellman-2}\\
        & \qquad\qquad\qquad\qquad\qquad\qquad\qquad\qquad\qquad + \gamma \sum_{o,s'} T(o,s' \mid s,a_1,\hat{\pi}_2(s))\alpha_{a_1,o}(s') \Big] \\
        & \qquad = \max_{\pi_1 \in \Pi_1} \sup_{\overline{\alpha} \in \Gamma^{A_1 \times O}} \sum_s b(s) \cdot \mathsf{valcomp}(\pi_1,\overline{\alpha})(s) \\
        & \qquad = \max_{\pi_1 \in \Pi_1} \sup_{\overline{\alpha} \in \Gamma^{A_1 \times O}} \mathsf{valcomp}(\pi_1,\overline{\alpha})(b) \ \text{.} \numberthis\label{eq:osposg:bellman-3}
    \end{align*}
    \end{subequations}
    This concludes the proof of the equality of Equations~\eqref{eq:osposg:equiv-valcomp} and~\eqref{eq:osposg:H-maxmin}.
\end{proof}

\ContractivityLemma*
\begin{proof}
    By deviating from the equilibrium strategy profiles in stage games $[HV](b)$ and $[HW](b)$, the players can only worsen their payoffs.
    Therefore, we have
    \begin{align*}
        & u^{V,b}(\pi_1^W,\pi_2^V) \leq u^{V,b}(\pi_1^V,\pi_2^V) = [HV](b) \leq \numberthis\\
        & \qquad\qquad\qquad \leq [HW](b) = u^{W,b}(\pi_1^W,\pi_2^W) \leq  u^{W,b}(\pi_1^W,\pi_2^V) \ \text{.}
    \end{align*}
    We can thus bound the difference $[HW](b) - [HV](b)$ by $u^{W,b}(\pi_1^W,\pi_2^V) - u^{V,b}(\pi_1^W,\pi_2^V)$ where, according to Definition~\ref{def:osposg:stage-game},
    \begin{align*}
        &u^{W,b}(\pi_1^W,\pi_2^V) - u^{V,b}(\pi_1^W,\pi_2^V) = \numberthis\label{eq:thm:osposg:point-contractivity}\\
        &\qquad\qquad = \gamma \sum_{a_1,o} \Pr_{b,\pi_1^W,\pi_2^V}[a_1,o] \cdot [ W(\tau(b,a_1,\pi_2^V,o)) - V(\tau(b,a_1,\pi_2^V,o)) ] \ \text{.}
    \end{align*}
    Since every $W(\tau(b,a_1,o,\pi_2^V)) - V(\tau(b,a_1,o,\pi_2^V)$ considered in Equation~\eqref{eq:thm:osposg:point-contractivity} with non-zero probability $\Pr_{b,\pi_1^W,\pi_2^V}[a_1,o]$ is assumed to be bounded by $C$, the expectation over such $W(\tau(b,a_1,o,\pi_2^V)) - V(\tau(b,a_1,o,\pi_2^V)$ is likewise bounded by $C$.
    It follows that $u^{W,b}(\pi_1^W,\pi_2^V) - u^{V,b}(\pi_1^W,\pi_2^V) \leq \gamma C$, and hence we also have $[HW](b) - [HV](b) \leq \gamma C$.
\end{proof}

\FixpointLemma*
\begin{proof}
    \begin{subequations}
    According to \thref{thm:osposg:gamma-independent}, the Bellman's operator does not depend on the set $\Gamma$ used to represent the value function $V^*$.
    To this end, we will assume that the set $\Gamma$ used to represent $V^*$ is
    \begin{equation}
        \Gamma = \mathsf{Conv} \lbrace \val^{\sigma_1} \mid \sigma_1 \in \Sigma_1 \rbrace \ \text{.}
    \end{equation}
    
    To prove the equivalence of value functions $V^*$ and $HV^*$ we consider that these functions are represented as follows:
    \begin{align}
        V^*(b) &= \sup_{\alpha \in \Gamma_{V^*}} \alpha(b) & \Gamma_{V^*} &= \left\lbrace \val^{\sigma_1} \mid \sigma_1 \in \Sigma_1 \right\rbrace \\
        [HV^*](b) &= \sup_{\alpha \in \Gamma_{HV^*})} & \Gamma_{HV^*} &= \left\lbrace \mathsf{valcomp}(\pi_1, \overline{\alpha}) \mid \pi_1 \in \Pi_1, \overline{\alpha} \in \Gamma^{A_1 \times O} \right\rbrace \ \text{.} \label{thm:osposg:fixpoint:gamma-hv}
    \end{align}
    To prove the equivalence of $V^*$ and $HV^*$, it suffices to show that for every $\alpha \in \Gamma_{V^*}$ there exists $\alpha' \in \Gamma_{HV^*}$ such that $\alpha' \geq \alpha$, and vice versa.
    
    First, from Proposition~\ref{thm:osposg:decomposition}, Lemma~\ref{thm:osposg:composition} and Definition~\ref{def:osposg:val-composition}, it follows that every strategy $\sigma_1 \in \Sigma_1$ can be represented as a value composition $\mathsf{valcomp}(\pi_1, \overline{\zeta})$, and we have
    \begin{equation}
        \val^{\sigma_1} = \val^{\mathsf{comp}(\pi_1,\overline{\zeta})} = \mathsf{valcomp}(\pi_1, \overline{\alpha}^{\overline{\zeta}})
    \end{equation}
    where $\alpha^{\overline{\zeta}}_{a_1,o} = \val^{\overline{\zeta}_{a_1,o}} \in \Gamma$.
    Hence $\val^{\sigma_1} = \mathsf{valcomp}(\pi_1, \overline{\zeta}) \in \Gamma_{HV^*}$.
    
    The opposite direction of the proof, i.e., that for every $\alpha \in \Gamma_{HV^*}$ there exists $\alpha' \in \Gamma_{V^*}$ such that $\alpha' \geq \alpha$, is more involved.
    Let $\alpha = \mathsf{valcomp}(\pi_1, \overline{\alpha}) \in \Gamma_{HV^*}$ be arbitrary.
    From~\eqref{thm:osposg:fixpoint:gamma-hv}, each $\alpha_{a_1,o}$ can be written as a convex combination of finitely many elements of $\lbrace \val^{\sigma_1} \mid \sigma_1 \in \Sigma_1 \rbrace$.
    \begin{equation}
        \alpha_{a_1,o} = \sum_{i=1}^K \lambda_i^{a_1,o} \val^{\sigma_1^{a_1,o,i}} \label{thm:osposg:fixpoint:cvx-combination}
    \end{equation}
    Let us form a vector of strategies $\overline{\zeta} \in (\Sigma_1)^{A_1 \times O}$ such that each $\zeta_{a_1,o}$ is a convex combination of strategies $\sigma_1^{a_1,o,i}$ using coefficients from Equation~\eqref{thm:osposg:fixpoint:cvx-combination},
    \begin{equation}
        \zeta_{a_1,o} = \sum_{i=1}^K \lambda_i^{a_1,o} \sigma_1^{a_1,o,i} \ \text{.}
    \end{equation}
    We can interpret strategy $\zeta_{a_1,o}$ as player~1 first randomly choosing among strategies $\sigma_1^{a_1,o,i}$, and then following the chosen strategy in the rest of the game.
    If the player~2 knew which strategy $\sigma_1^{a_1,o,i}$ has been chosen, he is able to achieve utility $\val^{\sigma_1^{a_1,o,i}}$.
    However, he has no access to this information, and hence $\val^{\zeta^{a_1,o}} \geq \sum_{i=1}^K \lambda_i^{a_1,o} \val^{\sigma_1^{a_1,o,i}} = \alpha^{a_1,o}$.
    Now, we have
    \begin{equation}
        \alpha' = \val^{\mathsf{comp}(\pi_1,\overline{\zeta})} \geq \mathsf{valcomp}(\pi_1,\overline{\alpha}) = \alpha
    \end{equation}
    \end{subequations}
    which concludes the proof.
\end{proof}

\LP*
\begin{proof}
Since the set $\Gamma$ is convex and compact, the dynamic programming operator $H$ can be used:
\begin{subequations}
\begin{align*}
    [HV](b) &= \max_{\pi_1 \in \Pi_1} \sup_{\overline{\alpha} \in \Gamma^{A_1 \times O}} \mathsf{valcomp}(\pi_1,\overline{\alpha})(b) \numberthis\\
            &= \max_{\pi_1 \in \Pi_1} \max_{\overline{\alpha} \in \Gamma^{A_1 \times O}} \mathsf{valcomp}(\pi_1,\overline{\alpha})(b) \numberthis\label{eq:osposg:max-composition-tolp-1}\\
            &= \max_{\pi_1 \in \Pi_1} \max_{\overline{\alpha} \in \Gamma^{A_1 \times O}} \sum_{s \in S} b(s) \cdot \mathsf{valcomp}(\pi_1,\overline{\alpha})(s) \numberthis\label{eq:osposg:max-composition-tolp-2}\\
            &= \max_{\pi_1 \in \Pi_1} \max_{\overline{\alpha} \in \Gamma^{A_1 \times O}} \sum_{s \in S} b(s) \cdot \min_{a_2} \Bigg[ \sum_{a_1} \pi_1(a_1) R(s,a_1,a_2) \ + \label{eq:osposg:max-composition-tolp}\numberthis\\
            & \qquad\qquad\qquad\qquad\qquad + \gamma \!\!\!\!\!\!\!\!\!\!\!\! \sum_{(a_1,o,s') \in A_1 \times O \times S} \!\!\!\!\!\!\!\!\!\!\!\! T(o,s' \,|\, s,a_1,a_2)\pi_1(a_1)\alpha^{a_1,o}(s') \Bigg] \ \text{.}
\end{align*}
\end{subequations}

Equation~\eqref{eq:osposg:max-composition-tolp-1} follows from the fact that $\mathsf{valcomp}(\pi_1,\overline{\alpha})$ is continuous in $\overline{\alpha}$, and $\Gamma$ is a compact set (and hence also $\Gamma^{A_1 \times O}$ is).
The Equation~\eqref{eq:osposg:max-composition-tolp-2} represents value of the linear function $\mathsf{valcomp}(\pi_1,\overline{\alpha})$ as the convex combination of its values in the vertices of the $\Delta(S)$ simplex, and, finally, Equation~\eqref{eq:osposg:max-composition-tolp} rewrites $\mathsf{valcomp}(\pi_1,\overline{\alpha})(s)$ using Definition~\ref{def:osposg:val-composition}.

Equation~\eqref{eq:osposg:max-composition-tolp} can be directly formalized as a mathematical program \eqref{eq:osposg:max-composition-nlp} whose solution is $[HV](b)$.
Indeed, the minimization over $a_2 \in A_2$ can be rewritten as a set of constraints for each value of state $V(s)$ (one for each action $a_2 \in A_2$ of player~2) in Equation~\eqref{eq:osposg:max-composition-nlp:br}.
The convex hull of set $\lbrace \alpha_1, \ldots, \alpha_k \rbrace$ is represented by~\eqref{eq:osposg:max-composition-nlp:convexification} where variables $\lambda_i^{a_1,o}$ represent coefficients of the convex combination.
The stage strategy $\pi_1$ is characterized by~\eqref{eq:osposg:max-composition-nlp:pi-sum} and~\eqref{eq:osposg:max-composition-nlp:pi-positive}.
\begin{subequations}
\label{eq:osposg:max-composition-nlp}
\begin{align*}
    \max_{\pi_1,\lambda,\overline{\alpha},V} \ & \sum_{s \in S} b(s) \cdot V(s) \numberthis\\
    \text{s.t.} \ \ \ & V(s) \leq \sum_{a_1 \in A_1} \pi_1(a_1) R(s,a_1,a_2) \ + & \forall (s,a_2) \in S \times A_2 \numberthis\label{eq:osposg:max-composition-nlp:br}\\[-0.5em]
    & \qquad\qquad\qquad + \gamma \!\!\!\!\!\!\!\!\!\!\!\! \sum_{(a_1,o,s') \in A_1 \times O \times S} \!\!\!\!\!\!\!\!\!\!\!\! T(o,s' \,|\, s,a_1,a_2) \pi_1(a_1) \alpha^{a_1,o}(s') \!\!\!\!\!\!\!\!\!\!\!\!\!\!\!\!\!\!\!\!\!\!\!\!\!\!\!\!\!\!\!\!\!\!\!\!\!\!\!\!\!\!\!\!\! \\
    & \alpha^{a_1,o}(s') = \sum_{i=1}^k \lambda_i^{a_1,o} \cdot \alpha_i(s') & \forall (a_1,o,s') \in A_1 \times O \times S \numberthis\label{eq:osposg:max-composition-nlp:convexification}\\[-0.5em]
    & \sum_{i=1}^k \lambda_i^{a_1,o} = 1 & \forall (a_1,o) \in A_1 \times O \numberthis\\
    & \sum_{a_1 \in A_1} \pi_1(a_1) = 1 \numberthis\label{eq:osposg:max-composition-nlp:pi-sum}\\
    & \pi_1(a_1) \geq 0  & \forall a_1 \in A_1 \numberthis\label{eq:osposg:max-composition-nlp:pi-positive}\\
    & \lambda_i^{a_1,o} \geq 0 & \forall (a_1,o) \in A_1 \times O, 1 \leq i \leq k \numberthis\label{eq:osposg:max-composition-nlp:last}\\
\end{align*}
\end{subequations}

This mathematical program is not linear since it contains a product of variables $\pi_1(a) \cdot \alpha^{a_1,o}(s')$.
It can, however, be linearized by introducing substitution $\hat{\alpha}^{a_1,o}(s') = \pi_1(a_1) \alpha^{a_1,o}(s')$ and $\hat{\lambda}_i^{a_1,o} = \pi_1(a_1) \lambda_i^{a_1,o}$ to obtain~\eqref{eq:osposg:max-composition-lp}.
\end{proof}

\UBIsUpperBound*
\begin{proof}
    The inequality $\ov \leq V_{\mathrm{HSVI1}}^\Upsilon$ follows trivially from eq. \eqref{eq:osposg:projection} (with $b' := b$).
    Proving $V^*(b) \leq \ov(b)$ is more involved.
    Suppose that $b'$ is the minimizer from the definition of $\ov$, i.e., that $\ov(b) = V_{\mathrm{HSVI1}}^\Upsilon(b') + \delta \| b - b' \|_1$.
    By definition of $V_{\mathrm{HSVI1}}^\Upsilon$, this $b'$ can be represented as a convex combination $\sum_i \lambda_i b_i = b'$ for which $\sum_i \lambda_i y_i = V_{\mathrm{HSVI1}}^\Upsilon(b')$.
    We thus have 
    \begin{equation}
        \ov(b) = \sum_{i=1}^k \lambda_i y_i + \delta \left\| b - b' \right\|_1 .
    \end{equation}
    Our assumptions imply that every pair $(b_i,y_i)$ satisfies $V^*(b_i) \leq y_i$.
    Combining this observations with the fact that $V^*$ is convex and $\delta$-Lipschitz continuous (Lemma~\ref{thm:osposg:vs-convex} and Proposition~\ref{thm:osposg:value-lipschitz}), we have
    \begin{align*}
        & V^*(b) \ \leq \ V^* \left( b' \right) + \delta \left\| b - b' \right\|_1 \ =  \ V^* \left( \sum_i \lambda_i b_i \right) + \delta \left\| b - b' \right\|_1 \leq \\
        & \leq \sum_{i=1}^k \lambda_i V^*(b_i) + \delta \left\| b - b' \right\|_1 \leq \sum_{i=1}^k \lambda_i y_i + \delta \left\| b - b' \right\|_1 = \ov(b) \ \text{.}
    \end{align*}
    
    Finally, let us prove that $\ov$ is $\delta$-Lipschitz continuous.
    Let us consider beliefs $b_1, b_2 \in \Delta(S)$.
    Without loss of generality, assume that $\ov(b_1) \geq \ov(b_2)$.
    Let $b_{\argmin}$ be the minimizer of $\ov(b_2)$, i.e.,
    \begin{equation}
        b_{\argmin} = \argmin_{b'} [ V_{\mathrm{HSVI1}}^\Upsilon(b') + \delta \| b_2 - b' \|_1 ] \text{.}
    \end{equation}
    By triangle inequality, we have
    \begin{align*}
        & \ov(b_1) = \\
        & = \min_{b' \in \Delta(S)} [ V_{\mathrm{HSVI1}}^\Upsilon(b') + \delta \| b_1 - b' \|_1 ] \leq V_{\mathrm{HSVI1}}^\Upsilon(b_{\argmin}) + \delta \| b_1 - b_{\argmin} \|_1 \leq \\
        & \leq [ V_{\mathrm{HSVI1}}^\Upsilon(b_{\argmin}) + \delta \| b_2 - b_{\argmin} \|_1 ] + \delta \| b_1 -  b_2 \|_1 = \ov(b_2) + \delta \| b_1 - b_2 \|_1
    \end{align*}
    which completes the proof.
\end{proof}

\LBUpdatesPreserveStuff*
\begin{proof}
    Initially, value function $\uv$ satisfies both conditions.
    Indeed, the set $\Gamma$ contains only the value $\val^{\sigma_1^{\mathrm{unif}}}$ of the uniform strategy $\sigma_1^{\mathrm{unif}}$, i.e., $\uv(b) = \val^{\sigma_1^{\mathrm{unif}}}(b)$ for every belief $b \in \Delta(S)$.
    Value $\val^{\sigma_1^{\mathrm{unif}}}$ is the value for a valid strategy $\sigma_1^{\mathrm{unif}}$ of player~1---hence it is $\delta$-Lipschitz continuous (Lemma~\ref{thm:osposg:strategy-value-lipschitz}) and lower bounds $V^*$.

    Assume that every $\alpha$-vector in the set $\Gamma$ is $\delta$-Lipschitz continuous, and that for each $\alpha \in \Gamma$ there exists strategy $\sigma_1 \in \Sigma_1$ with $\val^{\sigma_1} \geq \alpha$ (which holds also for the initial $\uv$).
    Let $\mathsf{valcomp}(\pi_1^{\mathrm{LB}},\overline{\alpha}^{\mathrm{LB}})$ be the value composition from Equation~\eqref{eq:osposg:pbupdate-lb} obtained when performing the point-based update of $\uv$ by solving $[H\uv](b)$.
    We will now show that the refined function $V_{\mathrm{LB}}^{\Gamma'}$ represented by the set $\Gamma' = \Gamma \cup \lbrace \mathsf{valcomp}(\pi_1^{\mathrm{LB}},\overline{\alpha}^{\mathrm{LB}}) \rbrace$ satisfies both properties, and hence any sequence of application of the point-based updates of $\uv$ preserves the aforementioned properties.
    \begin{compactenum}[(1)]
        \item By Lemma~\ref{thm:osposg:valcomp-lipschitz}, $\mathsf{valcomp}(\pi_1^{\mathrm{LB}},\overline{\alpha}^{\mathrm{LB}})$ is $\delta$-Lipschitz continuous (and thus so is the value function $V_{\mathrm{LB}}^{\Gamma'}$ represented by the set $\Gamma' = \Gamma \cup \lbrace \mathsf{valcomp}(\pi_1,\overline{\alpha}) \rbrace$).
        \item Each $\alpha$-vector in $\Gamma$ forms lower bound on the value of some strategy of player~1.
              Since $\overline{\alpha}^{\mathrm{LB}} \in \Gamma^{A_1 \times O}$, we have that every $\alpha_{a_1,o}$ lower bounds the value of some strategy of player~1.
              The fact that $\mathsf{valcomp}(\pi_1^{\mathrm{LB}},\overline{\alpha}^{\mathrm{LB}})$ is also a lower bound follows from Lemma~\ref{thm:osposg:gen-composition}---and hence every $\alpha$-vector from the set $\Gamma' = \Gamma \cup \lbrace \mathsf{valcomp}(\pi_1^{\mathrm{LB}},\overline{\alpha}^{\mathrm{LB}}) \rbrace$ is a lower bound on $V^*$.
              Hence also $V_{\mathrm{LB}}^{\Gamma'}(b)=\sup_{\alpha \in \Gamma'} \alpha(b) \leq V^*(b)$.
    \end{compactenum} 
\end{proof}

\UBPreservesStuff*
\begin{proof}
    $\ov$ has been defined as a lower $\delta$-Lipschitz envelope of $V_{\mathrm{HSVI1}}^\Upsilon$, hence it is $\delta$-Lipschitz continuous (Lemma~\ref{thm:osposg:projection}).
    We will therefore focus only on the property (2).
    Since the upper bound is initialized by a solution of a perfect information variant of the game, we have that $y_i \geq V^*(b_i)$ for every $(b_i,y_i)$ from the initial set $\Upsilon$ (Equation~\eqref{eq:osposg:upsilon-initial}).
    Hence, applying Lemma~\ref{thm:osposg:projection}, $\ov$ is an upper bound on $V^*$.
    
    We will now show that if $y_i \geq V^*(b_i)$ holds for $(b_i,y_i) \in \Upsilon$ (and $\ov$ is thus an upper bound on $V^*$), the application of a point-based update in any belief yields set $\Upsilon'$ such that $y_i \geq V^*(b_i)$ also holds for every $(b_i,y_i) \in \Upsilon'$---and the resulting value function $V_{\mathrm{UB}}^{\Upsilon'}$ is therefore upper bound on $V^*$ as well.
    Since $\ov \geq V^*$, the utility function of any stage game satisfies $u^{\ov,b}(\pi_1,\pi_2) \geq u^{V^*,b}(\pi_1,\pi_2)$ for every $b \in \Delta(S)$, $\pi_1 \in \Pi_1$ and $\pi_2 \in \Pi_2$.
    This implies that $[H\ov](b) \geq [HV^*](b) = V^*(b)$.
    We already know that $y_i \geq V^*(b_i)$ holds for $(b_i,y_i) \in \Upsilon$, and now we have $[H\ov](b) \geq V^*(b)$.
    Therefore, for every $(b_i,y_i) \in \Upsilon \cup \lbrace (b, [H\ov](b)) \rbrace$, we have $y_i \geq V^*(b_i)$, and applying the Lemma~\ref{thm:osposg:projection}, we have that the value function $V_{\mathrm{UB}}^{\Upsilon'}$ is an upper bound on $V^*$.
\end{proof}

\ExcessContractivity*
\begin{proof}
	Since $\uv \leq V^* \leq \ov$, it holds that $[H\uv](b_t) \leq [H\ov](b_t)$.
	Applying Lemma~\ref{thm:osposg:point-contractivity} with $C=\rho(t+1)$ implies that when the beliefs $\tau(b_t,a_1,\pi_2^{\mathrm{LB}},o)$ satisfy
	\begin{equation*}
	    \ov(\tau(b_t,a_1,\pi_2^{\mathrm{LB}},o)) - \uv(\tau(b_t,a_1,\pi_2^{\mathrm{LB}},o)) \leq \rho(t+1),
    \end{equation*}
	we have $[H\ov](b_t) - [H\uv](b_t) \leq \gamma \rho(t+1)$.
	Luckily, this assumption is satisfied in the considered situation --- indeed, otherwise there would be some $(a_1,o) \in A_1 \times O$ with 
	\begin{align*}
	\ov(\tau(b_t,a_1,\pi_2^{\mathrm{LB}},o)) - \uv(\tau(b_t,a_1,\pi_2^{\mathrm{LB}},o)) > \rho(t+1)    ,
	\end{align*}
	i.e., one satisfying $\excess_{t+1}(\tau(b_t,a_1,\pi_2^{\mathrm{LB}},o)) > 0$, for which $\Pr_{b,\pi_1^{\mathrm{UB}},\pi_2^{\mathrm{LB}}}[a_1,o] > 0$.
	This would contradict the assumption \begin{align*}
	\Pr_{b,\pi_1^{\mathrm{UB}},\pi_2^{\mathrm{LB}}}[a_1^*,o^*] \, \cdot \,  \excess_{t+1}(\tau(b_t,a_1^*,\pi_2^{\mathrm{LB}},o^*)) \leq 0 .
	\end{align*}

	Now, according to Equation~\eqref{eq:osposg:rho}, we have $[H\ov](b_t)-[H\uv](b_t) \leq \gamma \rho(t+1) = \rho(t) - 2 \delta D$.
	It follows that the excess gap after performing the point-based update in $b_t$ satisfies 
	\begin{align}
		\excess_t(b_t) & = \ov(b_t) - \uv(b_t) - \rho(t) \leq \gamma \rho(t+1) - \rho(t) \nonumber \\
		& = [\rho(t) - \rho(t)] - 2 \delta D = -2 \delta D ,
	\end{align}
	which completes the proof of the first part of the lemma.

	Now since the value functions $\uv$ and $\ov$ are $\delta$-Lipschitz continuous (Lemma~\ref{thm:osposg:lb-point} and Lemma~\ref{thm:osposg:ub-point}), the difference $\ov - \uv$ is $2\delta$-Lipschitz continuous.
	Thus for every belief $b_t' \in \Delta(S)$ satisfying $\| b_t - b_t' \|_1 \leq D$, we have
	\begin{equation}
		\ov(b_t') - \uv(b_t') \leq \ov(b_t) - \uv(b_t) + 2\delta \| b_t - b_t' \|_1 \leq \ov(b_t) - \uv(b_t) + 2\delta D \ \text{.}
	\end{equation}
	Now since $\excess_t(b_t) \leq -2\delta D$, we have $\excess_t(b_t') \leq 0$ which proves the second part of the lemma.
\end{proof}

\MinJustified*
\begin{proof}
    Assume for the contradiction that $V(b) < L$ for some belief $b \in \Delta(S)$.
    We pick $b = \argmin_{b' \in \Delta(S)} V(b')$ and denote $\varepsilon=L-V(b)$.
    Now, using the utility $u^{V,b}$ from Definition~\ref{def:osposg:stage-game} and using our choice of $b$, we have
    \begin{align*}
        u^{V,b}(\pi_1,\pi_2) &= \E_{b,\pi_1,\pi_2}[R(s,a_1,a_2)] + \gamma \sum_{a_1,o} \Pr_{b,\pi_1,\pi_2}[a_1,o] V(\tau(b,a_1,\pi_2,o)) \\
        &\geq \underline{r} + \gamma \sum_{a_1,o} \Pr_{b,\pi_1,\pi_2}[a_1,o] V(b) = \underline{r} + \gamma V(b) = \underline{r} + \gamma (L - \varepsilon)
    \end{align*}
    where $\underline{r}$ is the minimum reward in the game.
    Since $L = \sum_{t=1}^\infty \gamma^{t-1} \underline{r} = \underline{r} + \sum_{t=2}^\infty \gamma^{t-1} \underline{r} = \underline{r} + \gamma L$, we also have that $u^{V,b}(\pi_1,\pi_2) \geq L - \gamma \varepsilon$.
    Therefore it would have to also hold that
    $[HV](b) = \max_{\pi_1 \in \Pi_1} \min_{\pi_2 \in \Pi_2} u^{V,b}(\pi_1,\pi_2) \geq L - \gamma \varepsilon > L - \varepsilon = V(b)$
    which contradicts that $V$ is min-justified.
\end{proof}

\MaxJustified*
\begin{proof}
    Let $V$ be max-justified by $\Gamma$ and let us assume for contradiction that there exists $\alpha \in \Gamma$ and $s \in S$ such that $\alpha(s) > U$.
    We pick $\alpha$ and $s$ such that $(\alpha,s) = \argmax_{\alpha \in \Gamma, s \in S} \alpha(s)$ and denote $\varepsilon = \alpha(s) - U$.
    Using Definition~\ref{def:osposg:val-composition} and our choice of $(\alpha,s)$, we get the following for every $\pi_1 \in \Pi_1$ and $\overline{\alpha} \in \Gamma^{A_1 \times O}$:
    \begin{align*}
        & \mathsf{valcomp}(\pi_1,\overline{\alpha})(s) =\\
        & = \min_{a_2 \in A_2} \sum_{a_1 \in A_1} \pi_1(a_1) \Big[ R(s,a_1,a_2) + \gamma \!\!\! \sum_{o,s' \in O \times S} \!\!\! T(o,s' \mid s,a_1,a_2) \alpha_{a_1,o}(s') \Big] \\
        &\leq \min_{a_2 \in A_2} \sum_{a_1 \in A_1} \pi_1(a_1) \Big[ \overline{r} + \gamma \!\!\! \sum_{o,s' \in O \times S} \!\!\! T(o,s' \mid s,a_1,a_2) \alpha(s) \Big] \\
        &= \min_{a_2 \in A_2} \left[ \overline{r} + \gamma \alpha(s) \right]
    \end{align*}
    where $\overline{r} = \max_{(s,a_1,a_2)} R(s,a_1,a_2)$ is the maximum reward in the game.
    Since $U = \sum_{t=1}^\infty \gamma^{t-1}\overline{r} = \overline{r} + \sum_{t=2}^\infty \gamma^{t-1}\overline{r} = \overline{r} + \gamma U$, we have the following inequality for every $\pi_1 \in \Pi_1$ and $\overline{\alpha} \in \Gamma^{A_1 \times O}$
    \begin{equation}\label{eq:thm:osposg:max-justified-bounded}
        \mathsf{valcomp}(\pi_1,\overline{\alpha})(s) \leq \min_{a_2 \in A_2} [\overline{r} + \gamma \alpha(s)] = \overline{r} + \gamma (U+\varepsilon) = U + \gamma \varepsilon < U + \varepsilon = \alpha(s) \ \text{.}
    \end{equation}
    By Equation~\eqref{eq:thm:osposg:max-justified-bounded}, no value composition can satisfy $\mathsf{valcomp}(\pi_1,\overline{\alpha})(b_s) \geq \alpha(b_s)$ where $b_s(s)=1$ and $b_s(s')=0$ otherwise.
    Consequently, no value composition can satisfy $\mathsf{valcomp}(\pi_1,\overline{\alpha})(b) \geq \alpha(b)$ for every belief $b \in \Delta(S)$ as required by Definition~\ref{def:osposg:max-justified}.
    This contradicts our assumption and concludes the proof.
\end{proof}

\ConvMaxJustified*
\begin{proof}
  Recall that $V$ is max-justified by $\Omega$ if 1) $V(b) = \sup_{\alpha \in \Omega} \alpha(b)$ and 2) for every $\alpha \in \Omega$ there exists $\pi_1 \in \Pi_1$ and $\overline{\alpha} \in \Omega^{A_1 \times O}$ such that $\mathsf{valcomp}(\pi_1,\overline{\alpha}) \geq \alpha$.
  Let $V$ be a value function and suppose that $\Gamma$ satisfies 1) and 2).
  We will now verify that these properties hold for $\mathsf{Conv}(\Gamma)$ as well.
  By Proposition~\ref{thm:osposg:cvx-convexification}, we have that $\sup_{\alpha \in \mathsf{Conv}(\Gamma)} \alpha(b) = \sup_{\alpha \in \Gamma} \alpha(b)$.
  Since the property 1) holds for $\Gamma$ and the value of $V$ remains unchanged, 1) holds for $\mathsf{Conv}(\Gamma)$ as well.
  We will now prove 2) by showing that for every $\alpha \in \mathsf{Conv}(\Gamma)$, there exists $\pi_1 \in \Pi_1$ and $\overline{\alpha} \in \mathsf{Conv}(\Gamma)^{A_1 \times O}$ such that $\mathsf{valcomp}(\pi_1,\overline{\alpha}) \geq \alpha$.
  
  First of all, let us write $\alpha \in \mathsf{Conv}(\Gamma)$ as a finite convex combination $\sum_{i=1}^k \lambda_i \alpha^i$ of $\alpha$-vectors $\alpha^i \in \Gamma$.
  Using the assumption that $V$ is max-justified by $\Gamma$, we have that for every $\alpha^i$ there exists $\pi_1^{(i)} \in \Pi_1$ and $\overline{\alpha}^{(i)} \in \Gamma$, such that $\mathsf{valcomp}(\pi_1^{i},\overline{\alpha}^{i}) \geq \alpha^i$.
  Denote $\pi_1(a_1) := \sum_{i=1}^k \lambda_i \pi_1^i(a_1)$ and $\alpha_{a_1,o} := \sum_{i=1}^k \lambda_i \pi_1^i(a_1) \alpha^i_{a_1,o} / \pi_1(a_1)$.\footnote{Observe that $\alpha_{a_1,o} \in \mathsf{Conv}(\Gamma)$ since $\pi_1(a_1) = \sum_{i=1}^k \lambda_i \pi_1^i(a_1)$ and the coefficients $\lambda_i \pi_1^i(a_1) / \pi_1(a_1)$ thus sum to 1.}
  We claim that $\pi_1$ and $\overline{\alpha}$ witness that $V$ is max-justified by $\mathsf{Conv}(\Gamma)$.
  Since $\pi_1^i$ and $\overline{\alpha}^i$ were chosen s.t. $\mathsf{valcomp}(\pi_1^i,\overline{\alpha}^i) \geq \alpha^i$, we have $\sum_{i=1}^k \lambda_i \mathsf{valcomp}(\pi_1^i,\overline{\alpha}^i) \geq \sum_{i=1}^k \lambda_i \alpha^i = \alpha \in \mathsf{Conv}(\Gamma)$.
  To finish the proof, we show that $\mathsf{valcomp}(\pi_1,\overline{\alpha}) \geq \sum_{i=1}^k \lambda_i \mathsf{valcomp}(\pi_1^i,\overline{\alpha}^i)$.
  By Definition~\ref{def:osposg:val-composition}, we have
  \begin{align*}
      &\mathsf{valcomp}(\pi_1,\overline{\alpha}) = \\
      & = \min_{a_2 \in A_2} \Big[ \sum_{a_1 \in A_1} \pi_1(a_1) R(s,a_1,a_2) \\
      & \hspace{10em} + \gamma \!\!\!\!\!\!\!\!\!\!\!\!\!\!\!\! \sum_{(a_1,o,s') \in A_1 \times O \times S} \!\!\!\!\!\!\!\!\!\!\!\!\!\!\!\! \pi_1(a_1) T(o,s' \,|\, s,a_1,a_2) \alpha_{a_1,o}(s') \Big] \\
      &= \min_{a_2 \in A_2} \Big[ \sum_{i=1}^k \sum_{a_1 \in A_1} \lambda_i \pi_1^i(a_1) R(s,a_1,a_2) + \\
      & \hspace{10em} + \gamma \!\!\!\!\!\!\!\!\!\!\!\!\!\!\!\! \sum_{(a_1,o,s') \in A_1 \times O \times S} \!\!\!\!\!\!\!\!\!\!\!\!\!\!\!\! T(o,s' \,|\, s,a_1,a_2) \sum_{i=1}^k \lambda_i \pi_1^i(a_1) \alpha^i_{a_1,o}(s') \Big] \\
      &= \min_{a_2 \in A_2} \sum_{i=1}^k \lambda_i \Big[ \sum_{a_1 \in A_1} \pi_1^i(a_1) R(s,a_1,a_2) \\
      & \hspace{10em} + \gamma \!\!\!\!\!\!\!\!\!\!\!\!\!\!\!\! \sum_{(a_1,o,s') \in A_1 \times O \times S} \!\!\!\!\!\!\!\!\!\!\!\!\!\!\!\! \pi_1^i(a_1) T(o,s' \,|\, s,a_1,a_2) \alpha^i_{a_1,o}(s') \Big] \\
      &\geq \sum_{i=1}^k \lambda_i \min_{a_2 \in A_2} \Big[ \sum_{a_1 \in A_1} \pi_1^i(a_1) R(s,a_1,a_2) \\
      & \hspace{10em} + \gamma \!\!\!\!\!\!\!\!\!\!\!\!\!\!\!\! \sum_{(a_1,o,s') \in A_1 \times O \times S} \!\!\!\!\!\!\!\!\!\!\!\!\!\!\!\! \pi_1^i(a_1) T(o,s' \,|\, s,a_1,a_2) \alpha^i_{a_1,o}(s') \Big] \\
  &= \sum_{i=1}^k \lambda_i \mathsf{valcomp}(\pi_1^i, \overline{\alpha}^i) \ \text{.}
  \end{align*}
\end{proof}

\PlOneStrategy*
\begin{proof}
    Let $b^{\mathrm{init}}$ and $\rho^{\mathrm{init}}$ be as in the proposition and assume that player~1 follows $\mathtt{Act}(b,\rho)$ for the first $K$ stages and then follows the uniformly-random strategy $\sigma_1^{\mathrm{unif}}$.
    We denote this strategy as $\sigma_1^{b,\rho,K}$.
    To get to our result, we will first consider an arbitrary belief $b \in \Delta(S)$ and gadget $\rho \in \Gamma$.
    We will use induction to prove that the value of $\sigma_1^{b,\rho,K}$ satisfies $\val^{\sigma_1^{b,\rho,K}} \geq \rho - \gamma^K \cdot (U-L)$.
    
    First, assume that $K=0$, i.e., player~1 plays the uniform strategy $\sigma_1^{\mathrm{unif}}$ immediately.
    Value of the uniform strategy $\sigma_1^{\mathrm{unif}}$ is at least $\val^{\sigma_1^{\mathrm{unif}}} \geq L$ (Proposition~\ref{thm:osposg:bounded}) while $\rho \leq U$ (Lemma~\ref{thm:osposg:max-justified-bounded}).
    Hence $\val^{\sigma_1^{b,\rho,0}} \geq L \geq L - (U - \rho) = \rho - \gamma^0 (U-L)$.
    
    Let $K \geq 1$ and assume that $\val^{\sigma_1^{b',\rho',K-1}} \geq \rho' - \gamma^{K-1} (U-L)$ for every belief $b' \in \Delta(S)$ and gadget $\rho' \in \Gamma$.
    Observe that due to the recursive nature of the $\mathtt{Act}$ method, we can represent the strategy $\sigma_1^{b,\rho,K}$ as a composite strategy $\sigma_1^{b,\rho,K} = \mathsf{comp}(\pi_1^*, \overline{\zeta})$, where $\zeta_{a_1,o} = \sigma_1^{\tau(b,a_1,\pi_2,o), \alpha^*_{a_1,o},K-1}$ and $\pi^*_1$ comes from line~\ref{alg:osposg:cr:resolve} of Algorithm~\ref{alg:osposg:cr}.
    (To ensure that $\overline \alpha^*$ and $\pi^*_1$ are correctly defined, the algorithm requires the existence of a value composition satisfying $\mathsf{valcomp}(\pi_1,\overline{\alpha}) \geq \rho$. This requirement holds since $V$ is max-justified by the set $\Gamma$ and $\rho \in \Gamma$.)
    Applying Lemma~\ref{thm:osposg:composition}, the induction hypothesis, and  Definition~\ref{def:osposg:val-composition} (in this order), we have $\val^{\mathsf{comp}(\pi_1^*,\overline{\zeta})}(s) = $
    \begin{align*}
        &= \min_{a_2 \in A_2} \sum_{a_1 \in A_1} \pi_1^*(a_1) \Big[ R(s,a_1,a_2) + \gamma \!\!\!\!\!\! \sum_{(o,s') \in O \times S} \!\!\!\!\!\! T(o,s' \mid s, a_1, a_2) \val^{\zeta_{a_1,o}}(s') \Big] \\
        &\geq \min_{a_2 \in A_2} \sum_{a_1 \in A_1} \pi_1^*(a_1) \Big[ R(s,a_1,a_2) \ + \\
        & \qquad\qquad + \gamma \!\!\!\!\!\! \sum_{(o,s') \in O \times S} \!\!\!\!\!\! T(o,s' \mid s, a_1, a_2) [ \alpha^*_{a_1,o}(s')- \gamma^{K-1}(U-L) ] \Big] \\
        &= \min_{a_2 \in A_2} \sum_{a_1 \in A_1} \pi_1^*(a_1) \Big[ R(s,a_1,a_2) \ + \\
        & \qquad\qquad + \gamma \!\!\!\!\!\! \sum_{(o,s') \in O \times S} \!\!\!\!\!\! T(o,s' \mid s, a_1, a_2) \alpha^*_{a_1,o}(s') \Big] - \gamma^K (U-L) \\
        &= \mathsf{valcomp}(\pi_1^*,\overline{\alpha}^*) - \gamma^K (U-L) \ \text{.}
    \end{align*}
    We thus have $\val^{\sigma_1^{b,\rho,K}} = \val^{\mathsf{comp}(\pi_1^*,\overline{\zeta})} \geq \mathsf{valcomp}(\pi_1^*,\overline{\alpha}^*) - \gamma^K (U-L)$.
    Moreover, according to constraint on line~\ref{alg:osposg:cr:resolve} of Algorithm~\ref{alg:osposg:cr}, we also have $\mathsf{valcomp}(\pi_1^*,\overline{\alpha}^*) \geq \rho$.
    As a result, we also have $\val^{\sigma_1^{b,\rho,K}} \geq \rho - \gamma^K (U-L)$. This completes the induction step.
    
    Denote by $\sigma_1$ the strategy where player 1 follows $\mathtt{Act}(b^{\mathrm{init}},\rho^{\mathrm{init}})$ for \emph{infinite} period of time (i.e., as $K \rightarrow \infty$).
    We then have
    \begin{equation*}
        \val^{\sigma_1} = \lim_{K \rightarrow \infty} \val^{\sigma_1^{b^{\mathrm{init}},\rho^{\mathrm{init}},K}} \geq \lim_{K \rightarrow \infty} [\rho^{\mathrm{init}} - \gamma^K (U-L)] = \rho^{\mathrm{init}}
    \end{equation*}
    which completes the proof.
\end{proof}

\PlTwoStrategy*
\begin{proof}
    For the purposes of this proof, we will use
    \begin{equation*}
    \val_2(\sigma_2', b) = \sup_{\sigma_1 \in \Sigma_1} \E_{b,\sigma_1,\sigma_2'}[\Disc^\gamma]    
    \end{equation*}
    to denote the value a strategy $\sigma_2'$ of player~2 guarantees when the belief of player~1 is $b$.
    Similarly to the proof of Proposition~\ref{thm:osposg:p1-strategy}, we will first consider strategies $\sigma_2^{b,K}$ where player~2 plays according to $\mathtt{Act}(b)$ for $K$ steps, and then follows an arbitrary (e.g., uniform) strategy in the rest of the game, and we show that $\val_2(\sigma_2^{b,K}, b) \leq V(b) + \gamma^K (U-L)$.
    
    First, let $K=0$ and $b \in \Delta(S)$ be the belief of player~1.
    By Proposition~\ref{thm:osposg:bounded}, player~1 cannot achieve higher utility than $U$.
    Moreover, $V$ is min-justified, so we have $V(b) \geq L$ by Lemma~\ref{thm:osposg:min-justified-bounded}.
    Therefore, player~1 cannot achieve higher utility than $\val_2(\sigma_2^{b,0},b) \leq U \leq U + V(b) - L = V(b) + \gamma^0 (U-L)$ when his belief is $b$.
    
    Now let $K \geq 1$ be arbitrary.
    By the induction hypothesis, we have that strategy $\sigma_2^{b',K-1}$ guarantees that the utility is at most $\val_2(\sigma_2^{b',K-1},b') \leq V(b') + \gamma^{K-1} (U-L)$ when the belief of player~1 is $b'$.
    Let us evaluate the utility that $\sigma_2^{b,K}$ guarantees against arbitrary strategy $\sigma_1$ of player~1 in belief $b$.
    In the first stage of the game, player~2 plays according to $\pi_2^*$ obtained on line~\ref{alg:osposg:cr2:resolve} of Algorithm~\ref{alg:osposg:cr2}, and the expected reward from the first stage is $\E_{b,\sigma_1,\pi_2^*}[R(s,a_1,a_2)]$.
    If player~1 plays $a_1$ and observes $o$, he reaches an $(a_1,o)$-subgame where the belief of player~1 is $\tau(b,a_1,\pi_2^*,o)$ and player~2 plays $\sigma_2^{\tau(b,a_1,\pi_2^*,o),K-1}$.
    Using the induction hypothesis, we know that player~1 is able to achieve utility of at most $\val_2(\sigma_2^{\tau(b,a_1,\pi_2^*,o),K-1}, \tau(b,a_1,\pi_2^*,o)) \leq V(\tau(b,a_1,\pi_2^*,o)) + \gamma^{K-1} (U-L)$.
    This implies that an upper bound on the utility that $\sigma_1$ achieves against $\sigma_2^{b,K}$ (i.e., the strategy corresponding to player~2 following $\mathtt{Act}(b)$ for $K$ stages) is
    \begin{align*}
        &\E_{b,\sigma_1,\pi_2^*}[R(s,a_1,a_2)] + \gamma \E_{b,\sigma_1,\pi_2^*}[ V(\tau(b,a_1,\pi_2^*,o)) + \gamma^{K-1} (U-L) ] \\
        &\quad = \E_{b,\sigma_1,\pi_2^*}[R(s,a_1,a_2)] + \\
        &\quad\quad\quad + \gamma \!\!\!\!\!\!\!\! \sum_{(a_1,o) \in A_1 \times O} \!\!\!\!\!\!\!\! \Pr_{b,\sigma_1,\pi_2^*}[a_1,o] \cdot [ V(\tau(b,a_1,\pi_2^*,o)) + \gamma^{K-1} (U-L) ] \ \text{.}
    \end{align*}
    By allowing player~1 to maximize over $\sigma_1$, we get an upper bound on the value $\val_2(\sigma_2^{b,K},b)$ strategy $\sigma_2^{b,K}$ guarantees when the belief of player~1 is $b$.
    \begin{align*}
        &\val_2(\sigma_2^{b,K},b) \leq \\
        &\quad \leq \sup_{\sigma_1 \in \Sigma_1} \Big[ \E_{b,\sigma_1,\pi_2^*}[R(s,a_1,a_2)] \ + \\
        & \qquad\qquad\qquad\qquad + \gamma \!\!\!\!\!\!\!\!\! \sum_{(a_1,o) \in A_1 \times O} \!\!\!\!\!\!\!\!\! \Pr_{b,\sigma_1,\pi_2^*}[a_1,o] \cdot [ V(\tau(b,a_1,\pi_2^*,o)) + \gamma^{K-1} (U-L) ] \Big] \\
        &\quad = \max_{\pi_1 \in \Pi_1} \Big[ \E_{b,\pi_1,\pi_2^*}[R(s,a_1,a_2)] + \\
        &\quad\quad\quad + \gamma \!\!\!\!\!\!\!\!\! \sum_{(a_1,o) \in A_1 \times O} \!\!\!\!\!\!\!\!\! \Pr_{b,\pi_1,\pi_2^*}[a_1,o] \cdot V(\tau(b,a_1,\pi_2^*,o)) \Big] + \gamma^K (U-L) \\
        &\quad = \max_{\pi_1 \in \Pi_1} u^{V,b}(\pi_1, \pi_2^*) + \gamma^K (U-L)
    \end{align*}
    Using the fact that $\pi_2^*$ is the optimal strategy in the stage game $[HV](b)$, the definition of the stage game's value, and the fact that $V$ is min-justified, we get 
    \begin{align*}
        &\max_{\pi_1 \in \Pi_1} u^{V,b}(\pi_1, \pi_2^*) + \gamma^K (U-L) = \min_{\pi_2 \in \Pi_2} \max_{\pi_1 \in \Pi_2} u^{V,b}(\pi_1,\pi_2) + \gamma^K (U-L) \\
        &\qquad\qquad = [HV](b) + \gamma^K (U-L) \leq V(b) + \gamma^K (U-L) \ \text{.}
    \end{align*}
    Hence, the utility player~1 with belief $b$ can achieve against player~2 who follows strategy $\sigma_2^{b,K}$ is at most $V(b) + \gamma^K (U-L)$, and we have $\val_2(\sigma_2^{b,K},b) \leq V(b) + \gamma^K (U-L)$ which completes the induction step.
    
    Now, similarly to the proof of Proposition~\ref{thm:osposg:p1-strategy}, when player~2 follows $\mathtt{Act}(b^{\mathrm{init}})$ for \emph{infinitely} many stages (i.e., plays strategy $\sigma_2$ from the theorem), player~1 is able to achieve utility at most
    \begin{equation*}
        \val_2(\sigma_2, b^{\mathrm{init}}) = \lim_{K \rightarrow \infty} \val_2(\sigma_2^{b^{\mathrm{init}},K}, b^{\mathrm{init}}) \leq \lim_{K \rightarrow \infty} [ V(b^{\mathrm{init}}) + \gamma^K (U-L) ] = V(b^{\mathrm{init}})
    \end{equation*}
    which completes the proof.
\end{proof}

\LBmaxJustByConv*
\begin{proof}
    Observe that during the execution of Algorithm~\ref{alg:osposg:hsvi} the set $\Gamma$ is modified only by the point-based updates on lines~\ref{alg:osposg:hsvi:pb-update-1} and~\ref{alg:osposg:hsvi:pb-update-2} of Algorithm~\ref{alg:osposg:hsvi}.
    To prove the result, it thus suffices to show that (1) the initial lower bound $\uv$ is max-justified by the set $\mathsf{Conv}(\Gamma) = \Gamma = \lbrace \val^{\sigma_1^{\mathrm{unif}}} \rbrace$ and that (2) if $\uv$ is max-justified by $\mathsf{Conv}(\Gamma)$ then any point-based update results in a value function $V_{\mathrm{LB}}^{\Gamma'}$ that is max-justified by the set $\mathsf{Conv}(\Gamma')$.
    
    First, let us show that the initial lower bound $\uv$ is max-justified by the initial set of $\alpha$-vectors $\Gamma=\lbrace \val^{\sigma_1^{\mathrm{unif}}} \rbrace$ (and therefore also by $\mathsf{Conv}(\Gamma) = \Gamma$).
    Clearly, $\sigma_1^{\mathrm{unif}} = \mathsf{comp}(\pi_1^{\mathrm{unif}}, \zeta^{\mathrm{unif}})$, i.e., the uniform strategy $\sigma_1^{\mathrm{unif}}$ can be composed from a uniform stage strategy $\pi_1^{\mathrm{unif}}$ for the first stage of the game, and playing uniform strategy $\zeta^{\mathrm{unif}}_{a_1,o}=\sigma_1^{\mathrm{unif}}$ in every $(a_1,o)$-subgame after playing and observing $(a_1,o)$.
    Hence, $\val^{\sigma_1^{\mathrm{unif}}}=\mathsf{valcomp}(\pi_1^{\mathrm{unif}}, \overline{\alpha}^{\mathrm{unif}})$ for $\alpha^{\mathrm{unif}}_{a_1,o}=\val^{\sigma_1^{\mathrm{unif}}}$ and the initial $\uv$ is therefore max-justified by the set $\mathsf{Conv}(\Gamma) = \Gamma = \lbrace \val^{\sigma_1^{\mathrm{unif}}} \rbrace$.

    Next, consider a lower bound $\uv$ from Algorithm~\ref{alg:osposg:hsvi} and assume that it is max-justified by a set $\mathsf{Conv}(\Gamma)$.
    The point-based update constructs a set $\Gamma' = \Gamma \cup \lbrace \mathsf{valcomp}(\pi_1, \overline{\alpha}) \rbrace$ for some $\pi_1 \in \Pi_1$ and $\overline{\alpha} \in \mathsf{Conv}(\Gamma)^{A_1 \times O}$, see Equation~\eqref{eq:osposg:pbupdate-lb}.
    Since $\uv$ was max-justified by $\mathsf{Conv}(\Gamma)$, we know that for every $\alpha \in \mathsf{Conv}(\Gamma)$ there exists $\pi_1' \in \Pi_1$, $\overline{\alpha}' \in \mathsf{Conv}(\Gamma)^{A_1 \times O}$ such that $\mathsf{valcomp}(\pi_1',\overline{\alpha}') \geq \alpha$.
    The same holds for the newly constructed $\alpha$ vector $\mathsf{valcomp}(\pi_1, \overline{\alpha})$, and $V_{\mathrm{LB}}^{\Gamma'}$ is therefore max-justified by $\mathsf{Conv}(\Gamma) \cup \lbrace \mathsf{valcomp}(\pi_1, \overline{\alpha}) \rbrace$.
    By Lemma~\ref{thm:osposg:conv-max-justified}, we also have that $V_{\mathrm{LB}}^{\Gamma'}$ is max-justified by $\mathsf{Conv}(\mathsf{Conv}(\Gamma) \cup \lbrace \mathsf{valcomp}(\pi_1, \overline{\alpha}) \rbrace) = \mathsf{Conv}(\Gamma')$.
    Every point-based update thus results in a value function $V_{\mathrm{LB}}^{\Gamma'}$ which is max-justified by $\mathsf{Conv}(\Gamma')$ which completes the proof.
\end{proof}